\documentclass[traditabstract]{aa}
\usepackage{graphicx}
\usepackage{color}
\usepackage{natbib}
\usepackage{epsfig}
\usepackage{hyperref}
\usepackage{sidecap}
\usepackage{float}
\usepackage[varg]{txfonts}
\bibpunct{(}{)}{;}{a}{}{,}

\begin{document}

\title{Eruption of a plasma blob, associated M-class flare, and large-scale EUV wave observed by SDO}

\author{Pankaj Kumar \inst{1} \and P.K. Manoharan \inst{2}}
\institute{Korea Astronomy and Space Science Institute (KASI), Daejeon, 305-348, Republic of Korea \\
\email{pankaj@kasi.re.kr}
\and
Radio Astronomy Centre, National Centre for Radio Astrophysics, Tata Institute of Fundamental Research, Udhagamandalam (Ooty) 643 001, India \\
\email{mano@ncra.tifr.res.in}}
%*****************************************************************************
\abstract
{We present a multiwavelength study of the formation and ejection of a plasma blob and associated extreme ultraviolet (EUV) waves in active region (AR) NOAA 11176, observed by SDO/AIA and STEREO on 25 March 2011. The EUV images observed with the AIA instrument clearly show the formation and ejection of a plasma blob from the lower atmosphere of the Sun at $\sim$9 min prior to the onset of the M1.0 flare.
This onset of the M-class flare happened at the site of the blob formation, while the blob was rising in a parabolic path with
an average speed of $\sim$300 km s$^{-1}$.
The blob also showed twisting and de-twisting motion in the lower corona, and the blob speed varied from $\sim$10-540 km s$^{-1}$. The faster and slower EUV wavefronts were observed in front of the plasma blob during its impulsive acceleration phase. The faster EUV wave propagated with a speed of $\sim$785 to 1020 km s$^{-1}$, whereas the slower wavefront speed varied in between $\sim$245 and 465 km s$^{-1}$. The timing and speed of the faster wave match the shock speed estimated from the drift rate of the associated type II radio burst. The faster wave experiences a reflection by the nearby AR NOAA 11177. In addition, secondary waves were observed (only in the 171 \AA~ channel), when the primary fast wave and plasma blob impacted the funnel-shaped coronal loops. The  Helioseismic Magnetic Imager (HMI) magnetograms revealed the continuous emergence of new magnetic flux along with shear flows at the site of the blob formation. It is inferred that the emergence of twisted magnetic fields in the form of arch-filaments/``anemone-type" loops is the likely cause for the plasma blob formation and associated eruption along with the triggering of M-class flare. Furthermore, the faster EUV wave formed ahead of the blob shows the signature of fast-mode MHD wave, whereas the slower wave seems to be generated by the field line compression by the plasma blob. The secondary wave trains originated from the funnel-shaped loops are probably the fast magnetoacoustic waves.}
 
 \keywords{Sun: flares---Sun: magnetic topology---Sun: filaments, prominences---sunspots---Sun: coronal mass ejections (CMEs)}
%*****************************************************************************

\authorrunning{Kumar \& Manoharan}
\titlerunning{Eruption of a plasma blob and associated large-scale EUV wave}

\maketitle
%*****************************************************************************
%%%%%% Section 1 %%%%%%   %%%%%%%%%%%%%%%%%%%%%%%%%%%%%%%%%%%%%%%%%%%%%%%%%%%%
\section{Introduction}
It is widely accepted that magnetic reconnection plays an important role in releasing 
the energy stored in sheared magnetic fields on the solar surface. In a magnetic
reconnection process, oppositely directed field lines come closer and join, resulting 
in the release of magnetic energy in the form of thermal energy and particle 
acceleration (e.g., \citealt{sweet1958}, \citealt{parker1963}, \citealt{petschek1964}). The standard solar flare model, known
as the CSHKP model, explains the energy release process \citep{carm1964,stur1966,hirayama1974,kopp1976}, and it is supported by various observational findings, such as 
cusp-shaped loops \citep{tsuneta1992}, inflows \citep{yokoyama2001}, down-flow signatures \citep{mckenzie2000,mckenzie2001,asai2004}, plasmoid ejections \citep{shibata1995}, loop-top hard X-ray sources \citep{masuda1994,sui2003}, X-ray jets \citep{shibata1992,shimojo1996}, and flux rope/loop interactions \citep{mano1996,kumar2010a,kumar2010b,kumar2010c,torok2011}. 

The exact triggering mechanisms of solar flares/CMEs are however not well understood. A filament in the solar active region often show slow rising motions before the onset of the eruption \citep{kahler1988,Schmieder2008,Sch2008,Cheng2010}. This suggests that a global magnetohydrodynamics (MHD) instability (e.g., kink instability) can trigger an AR filament eruption, leading to two-ribbon flares and an associated CME \citep{moore1988,kleim2004,torok2005, kumar2012a}. A strong correlation between the magnetic flux emergence and filament eruption has been observed \citep{feynman1995}. Therefore, based on the flux rope 2D MHD simulation model proposed by \citet{chen2000}, it is suggested that the flux emergence is an efficient mechanism to trigger the onset of CMEs/flares. The ``tether-cutting" magnetic reconnection, which occurs beneath the erupting filament, has also been employed to explain several eruptions \citep{moore1992,moore2001}. Using SDO/AIA observations, \citet{sterling2011} recently studied a filament eruption that was broadly consistent with the flux cancellation leading to the formation of a helical flux rope that subsequently erupted due to the onset of magnetic instability and/or runaway tether cutting. For more details about the triggering of M- and X-class solar flares and associated flux-rope destabilization, see the review by \citet{sch2009}.
  
According to the standard flare model (e.g., CSHKP), the magnetic reconnection occurs in a vertical current sheet above a closed underlying loop, and the filament/prominence eruption plays an important role in the triggering of fast 
reconnection \citep{forbes2000}. \citet{shibata1995} and \citet{shibata2001} extended the CSHKP model by unifying  reconnection and plasmoid ejection and emphasized the importance of a plasmoid ejection in the reconnection process, which is called the ``plasmoid-induced-reconnection" model.
In this model, the bursty reconnection is initiated by the ejection of a plasmoid and leads to the build-up of magnetic energy in the vertical current sheet. When a plasmoid is ejected, the inflow is induced due to the conservation of mass, which results in enhancing the reconnection rate. In consequence, the plasmoid formed above the current sheet can also become accelerated by the faster reconnection outflow. \citet{nishida2009} performed MHD simulations of the solar flares by changing the values of resistivity and plasmoid velocity and found that the reconnection rate positively correlated with the plasmoid velocity. Therefore, plasmoid ejection plays a key role in the triggering of fast magnetic reconnection and is an observational support for magnetic reconnection in a solar flare. Since plasmoid ejections have been observed in both long-duration events and compact flares (e.g., \citealt{shibata1995}), the magnetic reconnection model may be applicable even for compact flares, which often do not show the other typical features of magnetic reconnection, such as particle acceleration or loop-top source.
 For example, in the soft X-ray image, upward-moving blobs of hot plasma have often been observed \citep{shibata1995,ohyama1998,kim2005}, and the white-light coronagraph observations have also recorded rising blob-like features in the wake of CMEs \citep{ko2003,lin2005}. Slowly drifting pulsating structures (DPSs), occasionally observed in the decimetric frequency range, are also interpreted as the signature of upward or downward moving plasmoid \citep{kliem2000,karl2004,karl2007,ning2007}.

Another important characteristic associated with the eruption are the large-scale coronal disturbances that are often observed. For example, the Extreme-ultraviolet Imaging
 Telescope (EIT) onboard the Solar and Heliospheric
 Observatory (SOHO) mission \citep{thompson1998} has recorded EIT waves in a number of eruptive events. There has been a long debate on the nature of coronal waves (for more detail see the recent reviews by \citealt{warmuth2010}, \citealt{chen2011}). According to the `fast-mode coronal magnetohydrodynamical (MHD) wave model', developed by
 \citet{uchida1968}, 
EIT waves are thought to be the coronal counterparts of
 the H$\alpha$ Moreton waves. However, the speed of a Moreton wave ($\sim$500-1500 km s$^{-1}$) is found to be about three times faster than the typical speed of an EIT wave ($\sim$200-400 km s$^{-1}$) \citep{klassen2000, thompson2009}.
Therefore, the interpretation of
 the EIT wave is divided into two categories, i.e., wave versus non-wave models. According to the non-wave scenario,
 the EIT wave can be explained by the large-scale coronal magnetic restructuring (or field-line stretching) due to the eruption of a CME
 that causes the observed successive brightenings either through plasma compression, heating, 
 or localized energy release (e.g. \citealt{delanee1999}, \citealt{chen2002}, \citealt{chen2005}, \citealt{attrill2007}).
 On the other hand, some authors have explained the EIT waves in
terms of fast-mode MHD waves or shocks \citep{warmuth2004,grechnev2008,temmer2009}. Such 
a wave could be launched and/or driven by flares, CMEs,
or small-scale ejecta. Recently, EUV observations from STEREO have made it possible to view the three-dimensional (3D) structure and
 evolution of EUV waves \citep{patsourakos2009,kienreich2009,ma2009,temmer2011}, including an event in which a full 3D coronal wave
 dome structure has been observed \citep{veronig2010}.
The origin of coronal shock waves (usually evident in the form of type II radio bursts) is also under debate. It may be driven by two possible physical mechanisms, (i) a blast wave ignited by the pressure pulse of a flare \citep{khan2002,narukage2002,hudson2003,magdalenic2012}, or (ii) a piston-driven shock generated by a CME \citep{klassen1999,klassen2003,cho2011}. Coronal shock waves may be associated with solar flares, CMEs, or some combination of these phenomena \citep{magara00,mag2008,vrsnak2008}. Recent observations from SDO/AIA suggest the EUV shock is formed (in the low corona) ahead of an erupting flux rope, which was correlated with a type II radio burst \citep{kozarev2011,ma2011,bain2012,gopal2012}.

In this paper, we present the unique multiwavelength observations of the plasma blob ejection that was associated with the triggering of an M-class flare. It is interesting to observe the formation of EUV waves associated with a plasma blob eruption. This type of dynamical evolution of the plasma blob ejection and associated EUV waves have been rarely recorded in earlier observations. However, the recent high-sensitivity and cadence images from the SDO and STEREO missions have made it possible to detect and track the plasma blob and associated EUV waves. Observations presented here support the formation of a fast-mode MHD shock associated with the plasma blob ejection. In section 2, we present the multiwavelength observational data sets, and in the last section, we discuss the results. 

%
%%%%%% Section 2 %%%%%%   %%%%%%%%%%%%%%%%%%%%%%%%%%%%%%%%%%%%%%%%%%%%%%%%%%%%
\section{Observations and data sets}

The active region NOAA 11176 was located at S16E31 on 25 March 2011. On this day,
its magnetic configuration was rather complex (i.e., $\beta\gamma$), which 
produced several C- and M-class flares. We report the analysis of an M1.0 flare 
on 25 March 2011 that originated
near the edge of the leading sunspot group.  According to the GOES 
soft X-ray profile in the wavelength range of 1-8 \AA, the flare started at 23:08 UT, attained the
peak intensity at 23:22 UT, and ended around 23:30 UT. However, the important point to
be mentioned here is that a plasma blob was formed and its ejection was observed 
before the onset of the flare.  

%%%%%%%%%%%%%%%%%%%%%%%%%%%%%%%%%%%%%%%%%%%%%%%%%%%%%%%%%%%%%%%%%%%%%%%%

\subsection{AIA observations of the plasma blob}
The Atmospheric Imaging Assembly (AIA) instrument onboard the Solar Dynamic Observatory (SDO) mission observes the full-disk image of the Sun at a resolution of $\sim$1.5$^{\arcsec}$ (0.6$^{\arcsec}$ pixel$^{-1}$), and its field of view covers up to $\sim$1.3
R$_\odot$. In this study, we employ images from AIA 171 \AA~ (Fe IX, T$\approx$0.6 MK),
193 \AA~ (Fe XII/XXIV, T$\approx$1.2 \& 20 MK), 131 \AA~ (Fe VIII/XXI, T$\approx$0.4 \& 10 MK), 335 \AA~ (Fe XVI, T$\sim$2.5 MK), 
  304 \AA~ (He II, T$\approx$0.05 MK) and 1600 \AA~ (C IV, 0.1 MK). Thus, images of these wavelengths cover from chromospheric to coronal heights \citep{lemen2011}. They are extremely useful for studying the evolutionary aspects of an eruption in different energy bands.

\begin{figure*}
\centering{
\includegraphics[width=5.5cm]{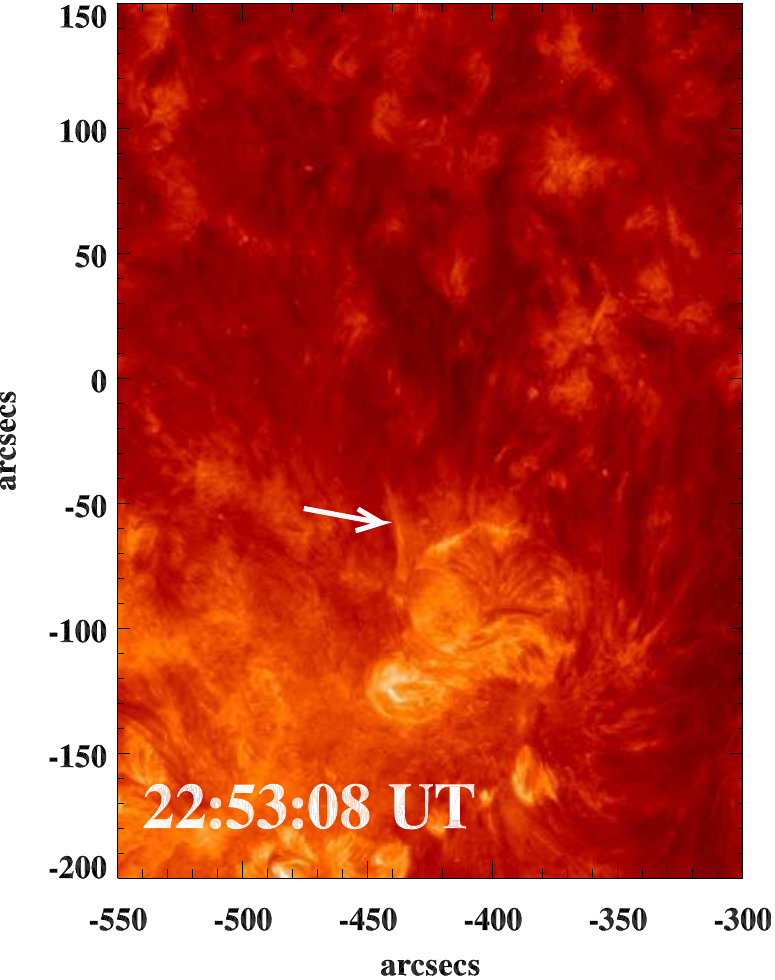}
\includegraphics[width=5.5cm]{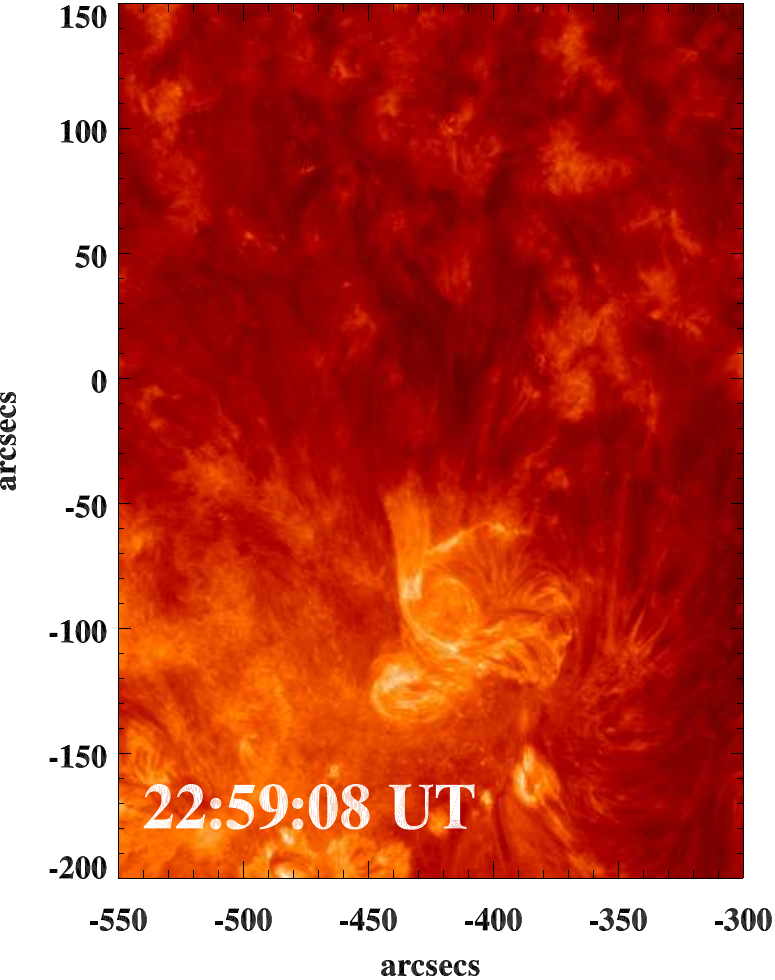}
\includegraphics[width=5.5cm]{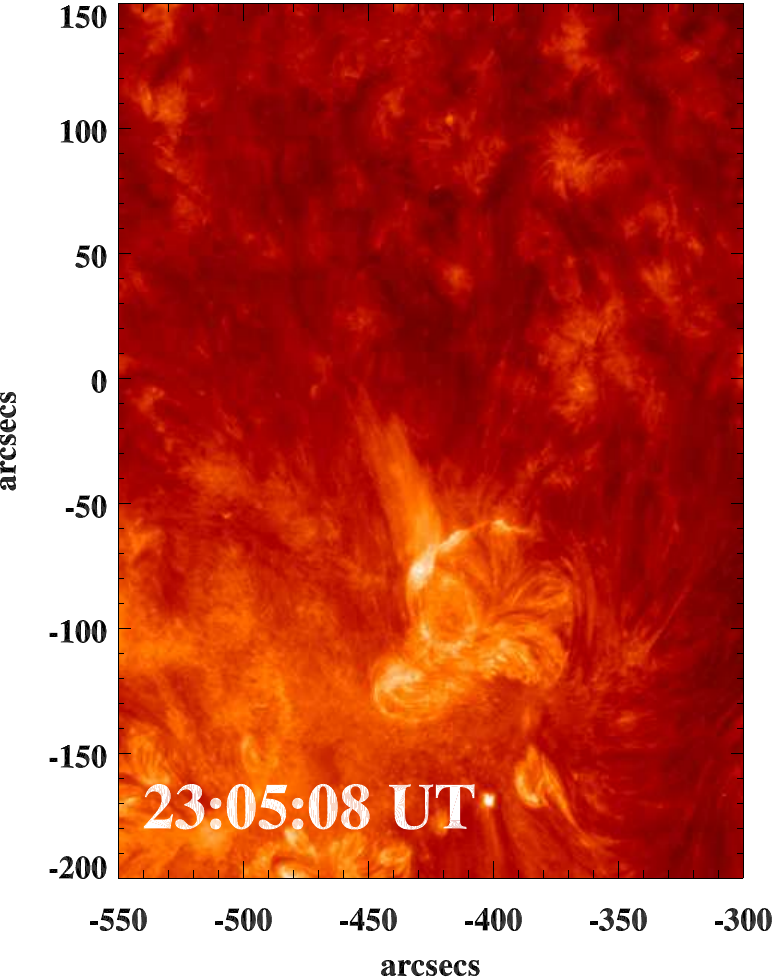}

\includegraphics[width=5.5cm]{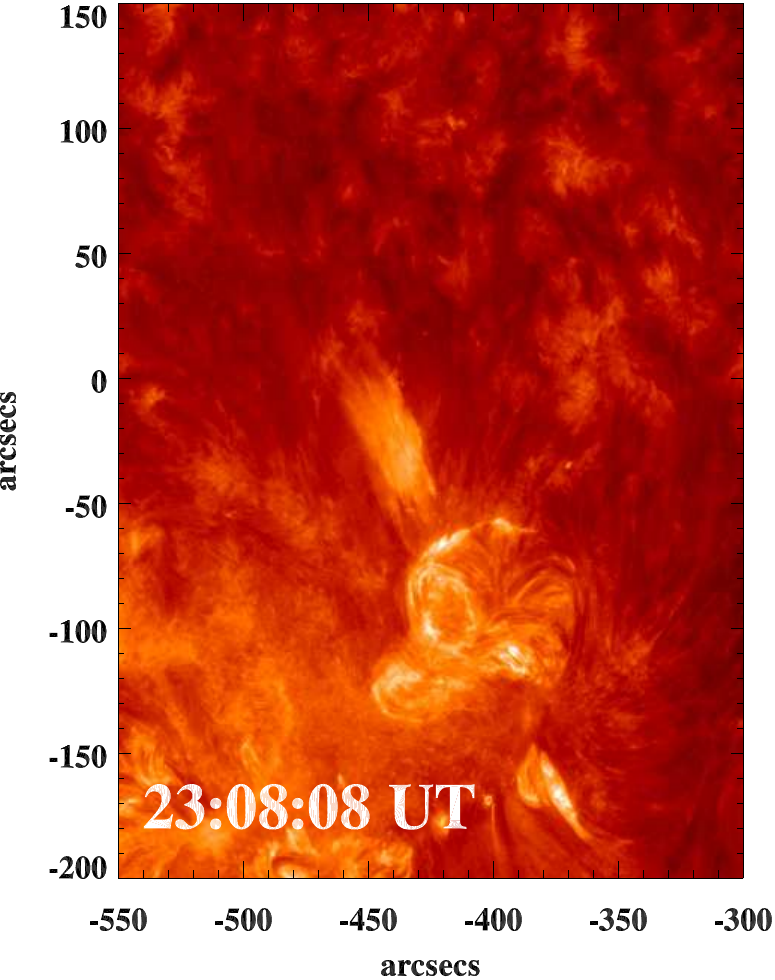}
\includegraphics[width=5.5cm]{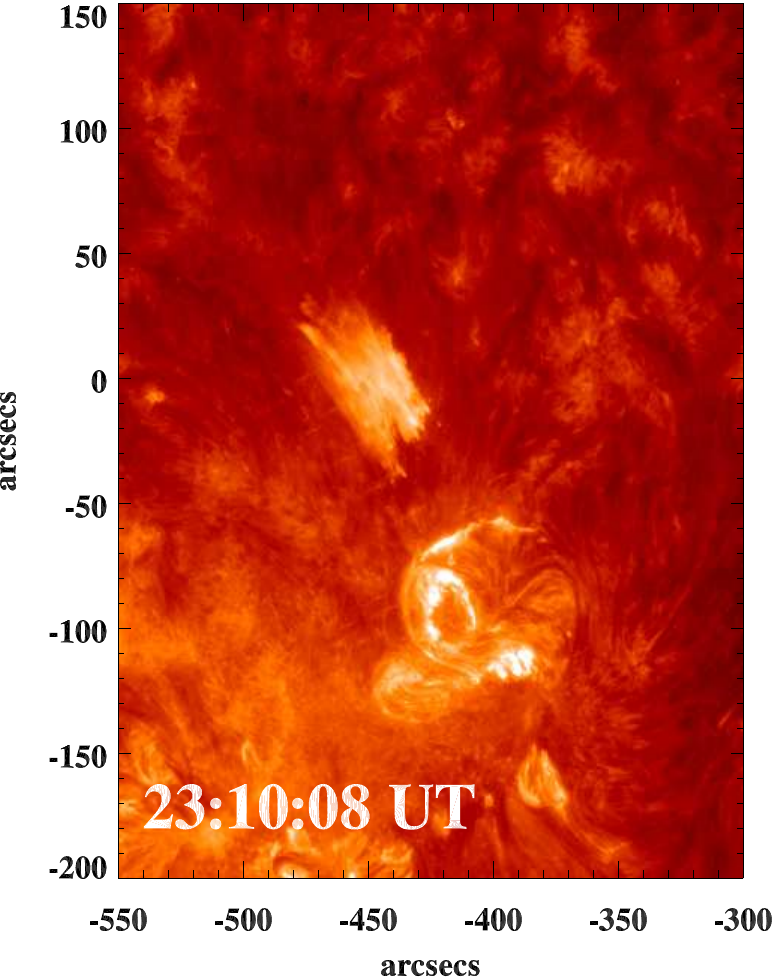}
\includegraphics[width=5.5cm]{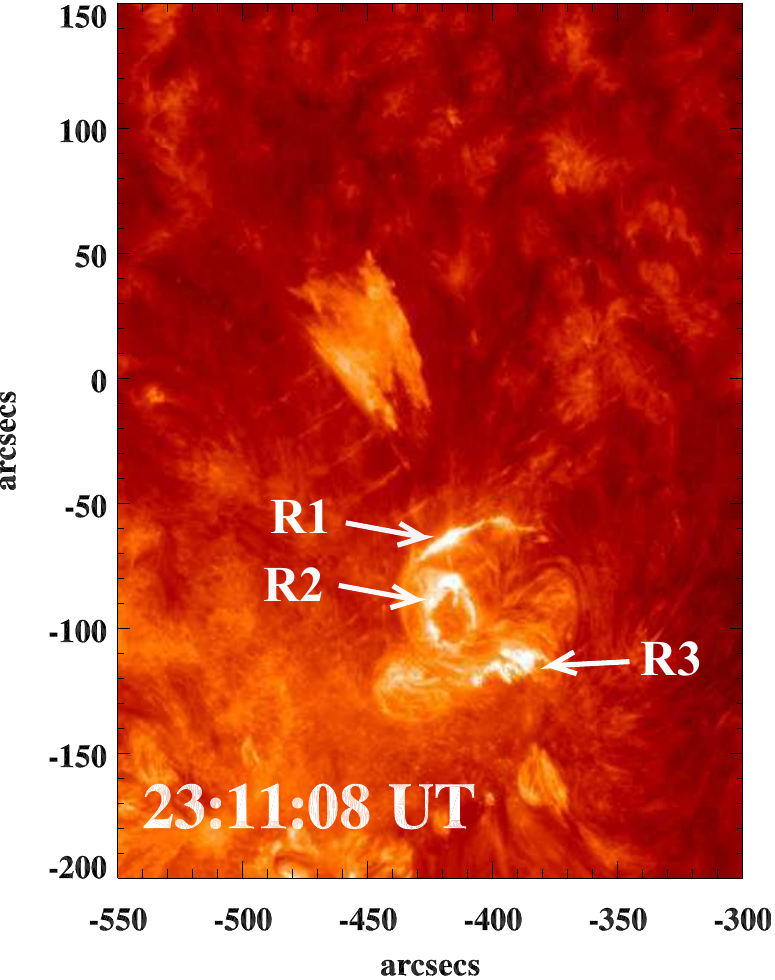}

\includegraphics[width=5.5cm]{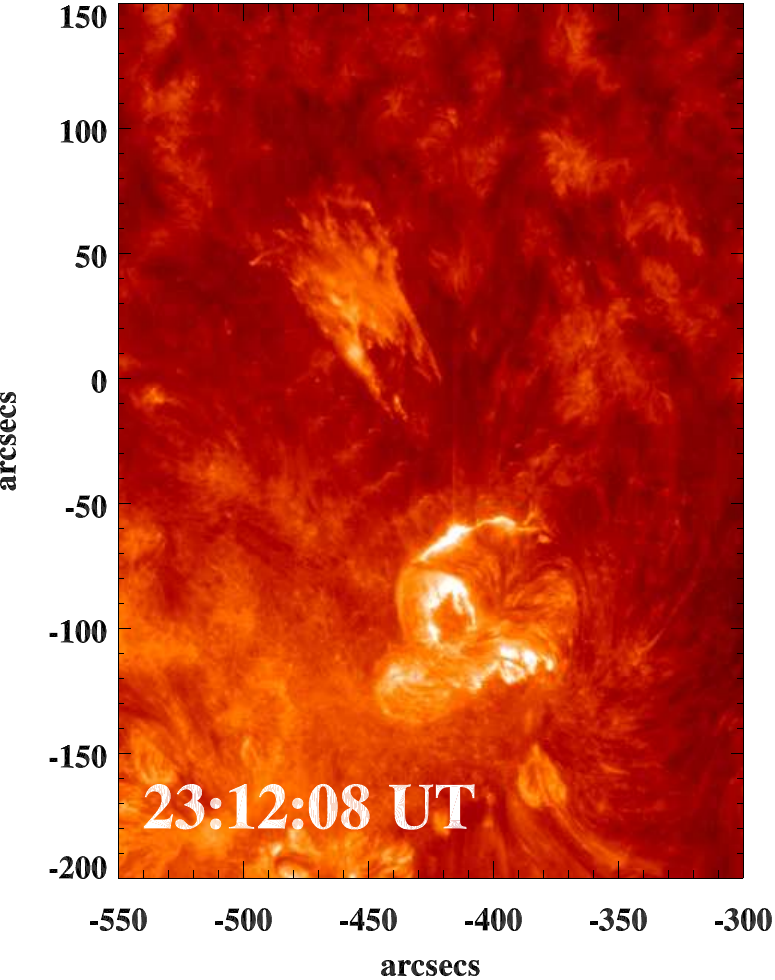}
\includegraphics[width=5.5cm]{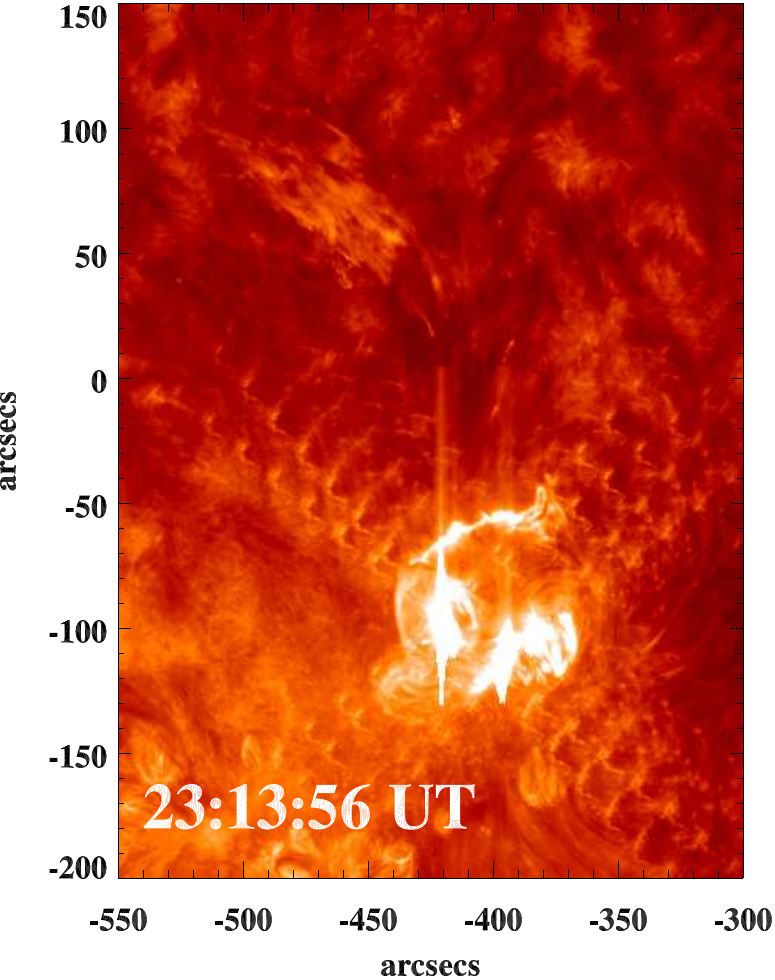}
\includegraphics[width=5.5cm]{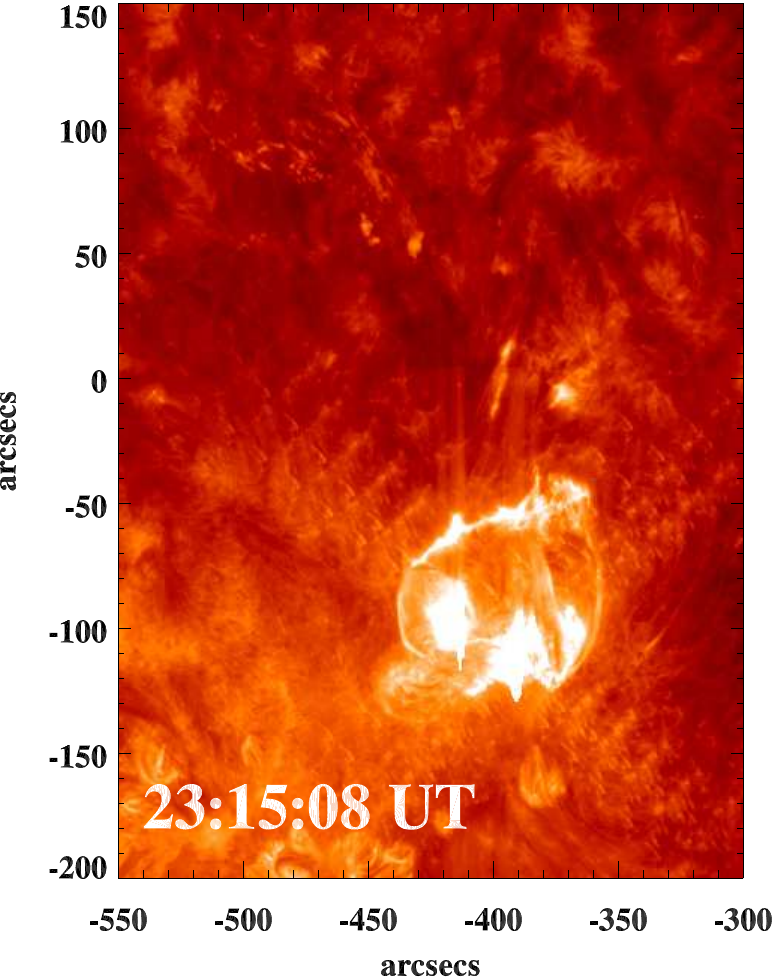}
}
\caption{SDO/AIA EUV images at 304 \AA \ showing the formation and eruption of the plasma blob on 25 March 2011.}
\label{aia304}
\end{figure*}

Figure \ref{aia304} displays some of the selected AIA 304 \AA \ images of the formation/eruption of 
the plasma blob and the associated flare event. These images represent the lower 
solar atmosphere, i.e., the chromosphere and transition region. In these images, the blob formation was observed 
at about 22:53 UT, at the site of a fan-shaped emerging group of a small loop system (marked by the arrow). 
The plasma blob also exhibited a slow upward motion starting from 22:59 UT and became detached from the active 
region at $\sim$23:08 UT. Moreover, at the time of blob separation, the flare brightening started at the site of the blob formation, indicating an increasing reconnection rate.
The length and width of the blob are
$\sim$70$\arcsec$ and $\sim$30$\arcsec$, respectively. The ejection of the plasma blob seems to trigger/enhance the reconnection
processes. The impulsive phase of the flare was recorded at 
23:14 UT, when the blob reached a distance of $\sim$0.21 R$_\odot$ from the flare center.
The blob disappeared at $\sim$23:18 UT and the flare continued to be intensified and peaked at the site of the
blob eruption. The AIA movie taken at 304 \AA~ shows the formation of three ribbon-like structures.
The first ribbon was formed during the slow rising phase of the blob, the second ribbon when the blob started to accelerate.  
The third ribbon was formed in the southward direction. These ribbons are indicated by R1, R2 and R3 in the image taken at 23:11:08 UT (Figure 1).

   Figure \ref{aia131} displays the selected AIA 131 \AA \ EUV base-difference images of the AR site of 
   the plasma blob eruption. This wavelength is sensitive to cool and hot plasma of
   0.4 MK and 10 MK (flaring region). These images show the full episode of the plasma blob formation and its eruption 
   as shown in the AIA 304 \AA \ images (Figure 1). We were able to measure the position of the leading edge of the 
   blob from the flare center (the measurement direction is indicated by the red dotted line). We visually tracked the structure at the leading edge of the blob in these base-difference images. The height-time plot of the blob in the sky plane is shown in the middle panel 
   of Figure \ref{kin}. The top panel shows the profile of GOES soft-X-ray flux, measured in the 1-8 \AA~ channel. To understand the relation between the blob dynamics in producing the non-thermal emission, the hard X-ray flux profile in the energy range of 25-50 keV is also shown in the same panel. The hard X-ray flux profile is obtained from the Fermi Gamma-ray Burst 
   Monitor (GBM) \citep{meegan2009}. 
   The height-time profile reveals the blob that  moves in a parabolic path and the soft X-ray flare seems to start with 
   the acceleration of the plasma blob at 23:08 UT. The hard X-ray emission peaks at 23:14 UT and continues 
   till 23:16 UT, after the acceleration phase of the blob. There is a delay of $\sim$8 min between the peak of hard X-ray emission and the start time of the blob acceleration. This indicates that most of the nonthermal particles 
   are accelerated after the plasma blob acceleration phase. 
    The bottom panel shows the temporal evolution of the blob speed 
  during its eruption, which is computed from the height-time measurements using a numerical differentiation with
 three-point Lagrangian interpolation \citep{zhang2004}. Initially, it shows a slow rise-up with a speed of $\sim$10 km s$^{-1}$ 
   and attains the maximum speed of $\sim$540 km s$^{-1}$ during its acceleration phase. However, the average speed of the blob is $\sim$300 km s$^{-1}$. 
   The movie from AIA 131 \AA~
   shows about three counterclockwise rotations in the blob structure during its eruption, which is probably related to the three peaks in the speed profile. 
During the upward motion, the orientation of the blob changes and is reflected in the speed profile as an increase and decrease in the speed. Very likely the changes in orientation and resulting projection have caused speed variations.
We also plot the mean counts of the AIA 1600 \AA \ images during the blob eruption, which correspond to the emission from the photospheric (5000 K) and transition regions (10000 K). The EUV flux slowly rises at $\sim$23:09 UT and its peak time matches the hard X-ray profile. This suggests that the hard X-ray emission comes from the footpoint sources, which are usually cospatial with flare ribbons. Thus, the above data sets reveal the kinematic evolution of the plasma blob, its association with the flare, and hard X-ray emission.

  Moreover, to investigate the source region of the blob, we checked AIA images before and after the M-class flare. Figure \ref{aia_3} displays selected snapshots of the AIA 304 and 335 \AA~, which provide chromospheric and coronal views of the blob formation site. The emerging ``anemone-type" loops start to grow in size continuously from  $\sim$09:00 UT onward on 25 March 2011. The top panel shows the AIA images before the flare at 22:40 UT. We observe several arch filaments in the emerging fan, in which one arch filament (indicated by arrows) is evident prior to the blob-eruption. The middle panel shows the plasma blob formation most likely above the arch filament, which is visible in AIA hot and cool channels. The bottom panels display the postflare loops and the weakening of the emerging fan at the blob eruption site after the flare.
In the middle panel, the AIA 304 \AA~  image is overlaid with the HMI magnetogram contours. The color codes white and black show positive and negative magnetic polarity regions, respectively. The originating location of the blob is near the small-satellite sunspot of the negative polarity region. The essential point to consider is that the blob originates from the lower solar 
  atmosphere (i.e., chromosphere), which is clearly shown in 304 \AA~ image.   
 
\subsection{Evolution of magnetic fields}
The time sequence of the HMI magnetograms were analyzed to see the evolution of the magnetic field before and after the flare. The HMI has been designed to study the oscillations and the magnetic field at the solar surface or photosphere \citep{schou2012,scherrer2012}. It observes the full disk of the Sun (4096$\times$4096 pixels) at 6173 \AA~ with a resolution of 1$^{\prime\prime}$ (0.5$^{\prime\prime}$ pixel$^{-1}$) and a typical cadence of 45 s. Figure \ref{hmi} displays some of the selected HMI magnetograms of 25 March 2011. The evolution of magnetic field at the blob formation site shows a significant amount of flux emergence. This started at $\sim$4 UT on 25 March 2011, which was nearly $\sim$19 hours before the flare occurrence. These magnetograms also show the emergence of negative flux toward the north of the main positive polarity sunspot P (shown within the ellipse indicated by N at 6:00 UT). Additionally, a small bipole emerged, indicated by P1 and N1 at $\sim$18:00 UT. To show the temporal evolution of the emerging flux, we estimated the positive, unsigned negative and total unsigned magnetic fluxes (in Mx) for the field of view shown in Figure \ref{hmi}. These quantities are plotted in the bottom-left panel of Figure \ref{hmi}. It clearly shows the flux emergence at the northern side of the big positive polarity spot. As revealed by the magnetic-flux profile, the negative flux emergence was rapid after about 5:00 UT and continued till 20:00 UT. Thus, the negative flux increased from 0.8$\times$10$^{21}$ to 1.4$\times$10$^{21}$ Mx ($\sim$75$\%$). However, the positive flux increase was lower by $\sim$15$\%$ than that of the negative flux. In addition, the HMI movie shows the motion of the big positive sunspot (P) toward the west, whereas the emerged negative field region moved toward the east. Therefore, the rapidly emerging/moving fields, particularly negative fields, seem to play the main role in the energy build-up at the site of blob formation and its subsequent eruption that leads to the M-class flare.

It is evident that the main emerging fluxes are N1 and P1, but if we carefully look into the footpoint polarities of the emerged fan shaped loop system/arch filaments, we see the connectivity of the emerged negative fields with the ambient positive field region on both the sides (north and south of the negative field) (refer to Figure \ref{aia_3}). Moreover, the initial phase of the blob eruption reveals two ribbons R1 and R2 in the AIA 1600/304 \AA~ images, and later we see mainly R2 and R3 during the impulsive phase of the flare (in the hard X-ray flux). The polarities of the ribbons R1, R2 and R3 are positive, negative (N1), and positive (P1), respectively. In addition, at R1 we observe only a weak positive flux region, which does not show changes with time.  Therefore, the main emerging fluxes are located at R2 and R3.

%%%%%%%%%%%%%%%%%%%%%%%%%%%%%%%%%%%%%%%%%%%%%%%%%%%%%%%%%%%%%%%%%%%%%%%%%%%%%%%%%%%%%%
\subsection{AIA observations of coronal EUV waves}
The images observed with AIA at 193, 335 and 171 \AA~ are useful for investigating the coronal EUV disturbance/wave associated with the blob eruption. The top panels in Figure \ref{aia193} show the AIA 193 \AA~ intensity image (left) and the HMI magnetogram (right). It is evident that the blob is formed at the leading edge of AR NOAA 11176 and moves in the northeast direction in a parabolic path. As depicted by the magnetogram, the nearby AR 11177 is also located to the north of AR 11176, and the EUV image shows the location of a coronal hole (CH) close to AR 11177. We display these images to show the typical magnetic environment around the blob eruption site and its propagation path. To show more details of the blob eruption site, in Figure \ref{aia193} some of the selected median filtered running difference images obtained from AIA at 193 \AA~ are shown. The running difference image at 23:13:55 UT (middle panel) shows the flare site and erupted plasma blob (indicated by arrows). The flare occurred during the impulsive acceleration phase of the blob and an EUV wave/shock was formed in front of the blob (indicated by arrows). We see a clear expanding wavefront associated with the blob eruption at 23:15:31 UT. The shock was formed at a distance of $\sim$0.23 R$_\odot$ away from the flare center. The accelerating blob seems to be the driver for the generation of the shock in the corona. The shock-wave front expands and moves up as the plasma blob propagates away from the flare site. However, after about 23:18 UT, the blob disappeared. The expansion of the blob very likely tends to decrease the associated density and the blob could not be observed in good contrast. However, the circular wavefront moved toward AR NOAA 11177 and was compressed and deflected at its boundary (AIA images at 23:18:19 UT). Additionally, the CH, located nearby AR 11177, seems to further deflect the path of the wavefront. Thus the locations of AR 11177 and the CH affect the wave propagation in the west and east of the CH (refer to image at 23:19:07 UT).

 Figure \ref{aia171-335} shows the selected running-difference images at AIA 335 and 171 \AA~ to show the formation/propagation of EUV waves in these channels. We have noticed some interesting and novel features in these channels.

(i) The top panels show the formation of a thick wavefront (indicated by `S') ahead of the plasma blob that moves in a northeast direction. Another wavefront (marked by `F') seems to be deflected from AR 11177 and the CH boundary similar to that observed in AIA 193 \AA~(see AIA running-difference composite movie). Therefore, two wavefronts (`F' and `S') are observed in AIA 335 \AA~ running-difference image at 23:17:39 UT. The thick wavefront `S' seems to be the slower and stopped at the boundary of another AR, whereas wavefront `F' was deflected at both sides (east and west) of the AR 11177.

(ii) The middle and bottom panels display the formation of EUV waves in AIA 171 \AA. Surprisingly, some of the wave features are observed only in the AIA 171 \AA~ channel, not in other AIA channels. There is a clear wave signal that appears to be transmitted (indicated by `T') across the AR 11177 (from the funnel-shaped loops) during 23:18-23:21 UT. Before and after the passage from the AR, the wave signal in running difference images turns from dark to bright at the leading edge, indicating a switch from emission reduction to enhancement at the wavefront. In other EUV channels, the wave seems to be deflected from the AR, and turned to the east and west of it.

(iii) Another remarkable feature is observed at 23:18 UT, when the erupting blob impacted the funnel-shaped loops emanating from the trailing (eastern) polarity of this AR. These loops become a new epicenter and emitted a thick wavefront marked by `F1' that moved in a northeast direction. This is most likely a secondary wave train. At  23:21:12 UT, another wave pulse was observed to be ejected from the same funnel-shaped loops (marked by `F2') behind `F1'. Obviously, the funnel-shaped loops are the sources of the secondary waves that move in the northeast direction.
To compare the spatial location of the wavefronts, we overplot the position of slower wavefront (`S') observed in the 335 \AA~ channel (at 23:20:03 UT) at the top of the AIA 171 \AA~ image (23:20:00 UT). The bottom-left panel clearly shows that the slower component (marked by `+' symbol) was separated from the fast components observed in the AIA 193 and 171 \AA~ channels.

To show the kinematic evolution of the plasma blob and formation of the EUV wave, we created space-time plots using the AIA 171, 193, and 335 \AA~ running difference images. The slices used for the space-time plots are shown in the top-right panel of Figure \ref{aia171-335} marked by `S1' to `S5'. We used the straight slices to obtain the stack plots to estimate the speed. We measured the plane-of-sky velocities of the EUV wavefronts, which are lower limits of the wave velocities in real 3D space. In Figure \ref{slice1}, we display the stack plots obtained for slices `S1', `S2', and `S3' using the AIA 171, 193, and 335 \AA~ running-difference images. The first panel shows the stack plot along slice `S1' for 171, 193, and 335 \AA~. The fast wave (`F') was observed in all three channels. From the linear fit to the data points, the mean speed of the wave was $\sim$912$\pm$33 km s$^{-1}$. The error in the position measurement was assumed to be four pixels (i.e., 2.4$\arcsec$). In the second panel, the stack plot shows the transmitted wave (through AR 11177) observed only in the 171 \AA~ channel at 23:18 UT. The speed of the transmitted wave (`T') was $\sim$1245$\pm$28 km s$^{-1}$, which is faster than the speed of the primary wave. The stack plot along slice `S3' shows the slowly rising plasma blob in a parabolic path (indicated by the arrow) during 23:00-23:18 UT. The onset of the EUV wave was observed in front of the plasma blob at 23:12-23:13 UT, which is indicated by a vertical line. Additionally, the coronal dimming was also observed behind the moving plasma blob. The apparent dimming in the running-difference images could indicate a recovery to the pre-event level of the emission after the passage of a bright wave front. To confirm the true dimming signatures, we also investigated the base-difference images. We observed clear dimming signatures behind the slow wavefront `S'. However, the apparent dimming behind the plasma blob and fast wavefronts in the running-difference images is mainly due to a recovery to the pre-event level of the emission. The impulsive phase of the flare is evident at 23:13-23:14 UT.

In Figure \ref{slice2}, the bottom panel of slice `S4', we track the wavefront (`S') ahead of the plasma blob (335 \AA). The wavefront propagates with a speed of $\sim$341$\pm$11 km s$^{-1}$ during 23:13--23:18 UT and finally it seems to be stopped at the boundary of another AR. The AIA 171 \AA~ stack plot shows the two fast wavefronts `F1' and `F2' (at 23:18 and 23:21 UT, respectively), which propagate beyond the limb across the existing AR. The speeds of `F1' and `F2' are $\sim$1173$\pm$15 and 1211$\pm$17 km s$^{-1}$, respectively. These two wavefronts originate from the funnel-shaped loops. The initial path of these wavefronts is shown in the slice `S5'. The typical speed of these wavefronts are $\sim$1296$\pm$25 and $\sim$1011$\pm$32 km s$^{-1}$, respectively.    

The comparison of the wavefronts in AIA 335 and 193 \AA \ reveals two main wavefronts, moving with different speeds in different directions. The faster wavefront (indicated by `F') is observed clearly in AIA 193 \AA \ images, whereas the slower one is observed in the AIA 335 \AA \ images. Both  wavefronts initially originate at the front of the plasma blob. However, it is observed that initially the faster wavefront is generated by the blob during its impulsive acceleration phase, later the faster wavefront arrives at the boundary of the other AR and then is deflected toward the west from the CH site. The AIA 335 \AA~ running difference movie and space-time plot clearly show the plasma blob running behind the slower wavefront till 23:18 UT.

On the other hand, the transmitted wavefront (`T') and wave trains (`F1' and `F2') observed in 171 \AA~ originate from the funnel-shaped loops. These wavefronts may be interpreted as secondary waves generated by the impact of the primary fast wave and the plasma blob.

Using AIA 171 \AA~ observations, \citet{liu2011} observed the quasi-periodic fast-wave trains (phase speed$\sim$2200 $\pm$130 km s$^{-1}$, period$\sim$3 minute) that emanate near the flare kernel and propagate outward along a funnel of coronal loops. These fast-wave trains temporally coincided with quasi-periodic pulsations of the flare (hard X-ray flux), suggesting a common origin. Liu and collaborators concluded that these are fast-mode magnetosonic waves that are excited by quasi-periodic magnetic reconnection and energy release. Recently, \citet{liu2012} reported quasi-periodic fast wave trains (period$\sim$2 minute) both ahead (within the broad global EUV wave pulse) and behind the expanding CME bubble.
In our observation, the quasi-periodic wavefronts (`F1' and `F2') are observed (i) away from the flare site, i.e., not originating exactly from the flare energy release site, as observed in \citet{liu2011}, (ii) and `F1' appeared exactly at the same time (at 23:18 UT) as the plasma blob impacted the funnel-shaped loops (moving in the northeast direction). For more details, we refer to AIA running difference multi-panel movie. However, the period of the wave trains in our observation is $\sim$3 minute, and the waves move across the AR, beyond the eastern limb, suggesting a true wave nature. Therefore, these waves may be interpreted as the fast magnetoacoustic waves. But in our case, their origin seems to be different from the site reported in \citet{liu2011} (i.e., plasma blob).

As suggested in the numerical simulation by \citet{ofman2011}, the quasi-periodic-wave trains are driven by motions at lower coronal boundary that propagate outward in a magnetic funnel and are evident through density fluctuations caused by the compressibility. The waves are best observed in 171 \AA~, which is associated with the cool coronal temperature of 0.6 MK, which results in stronger gravitational stratification of the density compared to hotter ARs, leading to the faster magnetosonic speed that depends on density in low-$\beta$ plasma \citep{ofman2011}.

\subsection{Plasma blob and EUV waves observed by STEREO}
We used STEREO (Solar TErrestrial RElations Observatory, \citealt{kaiser2008}) observations at 195 \AA~ to investigate the coronal EUV wave associated with the eruption of the plasma blob. In the STEREO-B view of the Sun on 25 March 2011, the active region was located close to the solar western limb. Figure \ref{st-b} displays running-difference EUV images taken at 195 \AA \ of the eruption site as seen in the STEREO-B view. These images have a typical cadence of $\sim$5 min and a spatial resolution of 1.6$\arcsec$ pixel$^{-1}$. The top-left image recorded at 23:05 UT shows the plasma blob close to the west limb of the Sun. This then moves up in a parabolic path, as observed in the AIA images. The shock formation can be also seen in the bottom-right image at 23:15:30 UT. The structure is like a bow-shock formed in front of the blob. The faster shock front (`F') is indicated by arrows. The next image at 23:18 UT shows the formation of a bright wavefront that moves toward the east. This is the slower wavefront as observed in AIA 335 \AA \ (indicated by `S') running ahead of the plasma blob.

A comparison between AIA and STEREO running difference images confirms the shock formation ahead of the blob.
The eastward deflection of the shock wave, observed as the bright front, is clearly visible in the images at 23:18 and 23:20 UT and is consistent with the findings obtained from the AIA images. However, we were unable to see the westward-propagating wavefront in the STEREO images, probably because of the limb side projection.  From these images,
we also measured the propagation of the leading edge of the blob in the sky plane
 (from the flare center, indicated by red dotted line) as a function of time (shown in Figure \ref{wave}) along with the GOES soft X-ray flux profile in 1-8 \AA \ (top panel). The soft X-ray flare started at 23:08 UT, when the blob reached a projected height of $\sim$5$\times$10$^{4}$ km away from the flare center. The soft X-ray flux enhanced with the acceleration of blob. The blob disappeared at $\sim$0.35 R$_\odot$ from the center of the flare (AIA running-difference images). A second-order polynomial fit to these measurements yields the relationship
 h=(2.7$\times$10$^4$)t$^2$+(4.1)t+0.2, where `h' is the height (in km) and `t' is the time (in sec) from 23:03 UT. This suggests the motion of the blob in a typical parabolic path. The middle panel shows the blob speed profile derived from the height-time measurements using a three-point Lagrangian interpolation method. The blob slowly rises with an initial speed of $\sim$70 km s$^{-1}$ and  accelerates to a maximum speed of $\sim$345 km s$^{-1}$. 
Another important point to note is that we observe no deceleration here as seen in AIA images, probably because of the limitation of the low-cadence STEREO images.

\subsection{Type II radio-burst and CME} 
The top panel of Figure \ref{spectrum} shows the radio dynamic spectrum from the Learmonth observatory, Australia, in the frequency range of 25-180 MHz. The GOES soft X-ray flux in the wavelength band of 1-8 \AA \ is also included in the plot. This radio spectrum shows a narrow type III radio burst at 23:14-23:15 UT during the impulsive phase of the flare, which suggests a narrow acceleration region along the opening of field lines. A fast-drifting type II radio burst observed in the frequency range of $\sim$40-180 MHz, near the flare maximum between 23:16 and 23:25 UT, nicely matches the timing and location of the shock formation ahead of the blob as observed in the AIA images.  We estimated the speed of the type II radio burst source using its frequency-drift rate and compared it with the speed of the EUV wave observed in the corona. The bottom panel shows the height of the type II radio burst (using the fundamental band) in the corona. The shock height (speed also) depends on the choice of the coronal density model. Therefore, we used Newkirk one-fold and two-fold density models \citep{newkirk1970} to estimate the source height. First of all, we converted the plasma frequency of the type II burst into the corresponding electron density (f$_p$=9$\surd$N$_e$ MHz with the electron density N$_e$ in m$^{-3}$). Then, using the Newkirk coronal electron density models, we obtained the corresponding source heights (i.e., heliocentric distances). The shock heights derived from the Newkirk one-fold (circle) and two-fold (triangle) density models are plotted in the bottom panel of Figure \ref{spectrum}.
The distance-time profile of the faster coronal wave is also included (`+' symbol, blue color) for comparison. Using a linear fit to the height-time data points, we calculated the shock speed.
 The mean speed of the shock wave obtained from the Newkirk one-fold and two-fold density models is $\sim$660 and $\sim$820 km s$^{-1}$. The speed of the faster coronal wave in AIA 193 \AA~ is 785-1020 km s$^{-1}$. These speeds almost agree with each other. The results (height and speed) from the Newkirk two-fold density model fit the observed wave kinematics well. Moreover, the formation time of the shock wave agrees with these measurements and is consistent with the formation signature of the coronal shock at the leading edge of the blob.
 
In general, EUV waves have been observed to interact with coronal
structures during their propagation \citep{thompson1998,thompson1999,wills1999,veronig2006}. \citet{thompson1999} reported EUV waves that were stopped by coronal holes. Furthermore, the above result also was demonstrated by numerical simulations \citep{wang2000,wu2001}.
The slowing-down of the EUV wave by an AR
was also reported by \citet{ofman2002}. In addition, these authors found strong reflection and refraction
 of the primary wave from an AR, as well as the generation of secondary waves by the dynamic distortion of the AR magnetic field.
  In the present study, the high cadence of the AIA images provides the opportunity to track the shock wave and study the deflection of the fast-mode shock wave
in the vicinity of the high Alfv\'en-speed region as well as reflection in the region of the steep gradient of the Alfv\'en speed \citep{uchida1973,uchida1974,wu2001,zhang2011}. We observed the transmission of the wave through AR 11177 observed in AIA 171 \AA. Additionally, we observed the fast secondary wave trains probably generated by the impact of the plasma blob on the funnel-shaped loops. Using STEREO observations, \citet{gopal2009} reported the strong reflection of the EUV wave by the boundary of a CH, which suggested that the nature of the EUV wave is a fast-mode MHD wave. In the present study, the type II radio burst during the EUV wave formation is consistent with the fast-mode MHD wave possibly driven by the ejection of the plasmoid.

In the analysis of the type II radio burst, the Newkirk one-fold and two-fold density models were used to estimate the shock height with respect to the center of the Sun (i.e., in heliocentric distance units), whereas for the EUV images we measured the projected shock height from the center of the flare site. We considered the same reference point for the coronal shock observed in radio and the wave seen in EUV as well as their overlapping time and confirm the detection of the fast-mode MHD wave in the EUV measurements.   
For example, recently \citet{gopal2012} found the shock formation in EUV images in a limb event that coincided with the onset of a type II radio burst. Moreover, the estimated height of the type II source matched the formation height of the EUV wave in the AIA 193 \AA \ images well. In their study, the fundamental band formed at a frequency of 150 MHz, which implied a
 local plasma density of 2.8$\times$10$^8$ cm$^{-3}$, which was consistent with
 the shock formation close to the solar surface (0.19 R$_\odot$ above
 the surface, or 1.19 R$_\odot$ from the Sun center).
  In this study, we observed the fundamental emission at 100 MHz,  which corresponds to a source height of $\sim$1.24-1.36 R$_\odot$ using different density models (from the Sun center, Figure \ref{spectrum}).  The projected height of the shock formation as observed in EUV images is $\sim$0.23 R$_\odot$ above the surface of the Sun. Therefore, the present estimate of shock formation heights in radio and EUV are consistent with the results from \citet{gopal2012}. 
Additionally, the plasma blob follows a parabolic path during its propagation. The shock formed in front of the blob also propagates obliquely near another AR 11177 and CH at the north (i.e., it is deflected by the existing active region and CH).

 No CME was recorded by SOHO/LASCO associated with the plasma blob eruption. In Figure \ref{cor1}, the top panels display the 195 \AA~ EUV images from STEREO A and B. In the first two images (at 23:20 UT), a tip of the EUV ray is marked with a `+' symbol. The coronal ray initially seems to be stretched upward and then some of its threads showed southward deflection (at 23:40 UT) during the passage of the reflected EUV wave from the AR 11177 (see the STEREO movies). In addition, bunches of loop/thread, located south of the EUV ray, are also deflected southward, and showed transverse oscillation. Transverse loop oscillations are usually observed during the passage of the shock through the coronal structures \citep{patsourakos2012,kumar2012b}. This confirms the passage of the shock through the coronal structures, which leads to the onset of oscillations.
The bottom panels display the composite STEREO EUVI and COR1 running-difference images. The first panel (at 23:20 UT) shows a bright streamer (marked by the arrow) above the eruption site. In the next three panels, we observe the passage of a weak CME and associated disturbance through the streamer in the southward direction (the front is marked by `F'). The COR1 movie shows the heating and deflection of the streamer (which persists for 1-2 hours) possibly because of the passage of the disturbance/shock through it in the southward direction.
As we observed the reflection of the fast wave in the AIA images from AR 11177, it is possible that the passage of the reflected fast EUV wave causes the deflection in the EUV ray in the southward direction and the associated heating/deflection of the streamer.

Moreover, we observed the spectral bump in the type II radio burst structure that may be interpreted as the interaction between the shock and streamer. The timings of the type II radio burst bump and the shock-streamer interaction observed in the COR1 images closely match. A similar kind of spectral bump was recently reported by \citet{feng2012}, who provided a schematic cartoon (see Figure 5 of their paper) for the magnetic topology of the streamer structure, the outward propagation of the coronal shock wave, and the location where the type II spectral bump took place during the shock-streamer interaction. The same scenario as for the type II bump was suggested by \citet{kong2012} for the 27 March 2011 eruption that occurred in the same AR, that we studied here on 25 March 2011. Therefore, the above observational findings support our interpretation of the shock-streamer interaction, which leads to the bump in the type II radio burst.

\section{Results and discussion}

We presented a multiwavelength analysis of a plasma blob ejection associated with an M-class flare and EUV waves on 25 March 2011. This event shows observational evidence of the formation of a plasma blob and its ejection followed by an intense flare. The blob formation was observed $\sim$9 min prior to the flare onset. Initially, the blob slowly rises with a mean speed of $\sim$70 km s$^{-1}$ and attains a maximum speed of $\sim$540 km s$^{-1}$. It also shows untwisting motion during the eruption/propagation and both acceleration and deceleration patterns in the low corona. The plasma blob shows impulsive acceleration just before the onset of the flare impulsive phase.

The arch filaments are usually observed in the new emerging flux region (EFR) in chromospheric H$\alpha$ images \citep{chou1988}. This filamentary structure arises
spontaneously from the magnetic Rayleigh-Taylor instability \citep{isobe2005,isobe2007}. The ends of an arch filament are anchored to the opposite polarities of EFR. In the EUV, the $\Omega$-shaped emerged loops are generally observed in the corona above the EFR. In the present study, the HMI observations clearly show the newly emerging flux at the blob formation site and AIA images reveal  expanding emerged anemone-shaped loops/arch filaments. The plasma blob is formed just above the arch filament, which strongly supports the scenario that the emerging flux plays an important role in the formation of the blob. The flare was initiated by the eruption of the blob. \citet{saka2004} also found a tiny two-ribbon flare driven by the emerging flux that accompanied the miniature filament eruption (equivalent to plasmoid). In our analysis, we obtained much clearer observational evidence of the flux emergence and the associated plasma blob formation/ejection.

The plasma blob was observed mostly in all coronal EUV channels (hot and cold), which reveals its multi-thermal nature. \citet{shibata1998} showed the ejection of soft-X-ray reconnection jets (T$\sim$1-10 MK) above the anemone-shaped emerging loops, which have speeds of the order of the Alfv\'en speed ($\sim$1000 km s$^{-1}$). The collimated plasma ejections along the fan-spine topology \citep{antiochos1998,pariat2010} are usually interpreted as jets/surges generated by low-altitude reconnection. On the other hand, the ejection of a detached hot plasma is interpreted as a plasmoid/blob \citep{shibata1998}.
As discussed by \citet{shibata1998}, the plasmoid ejection could play an important role in triggering the main energy release during the impulsive phase. During the pre-flare phase, \citet{ohyama1998} found that the plasmoid ejection started ($\sim$10 km s$^{-1}$) well before the impulsive phase.
However, in \citet{shibata1995}, the plasmoid is the top portion of an erupting (twisted) flux-rope or filament. In our case, the initial morphology looks like a jet with unwinding twists, because of its multiple bright strands elongated along the direction of motion \citep{shimojo1996,patsourakos2008}. In contrast, it lacks the shape of the closed flux-rope/loop with both ends anchored on the solar surface. Its eruption direction is closely aligned with the nearby coronal loops, as expected for a jet erupting along the magnetic field lines. Moreover, it is known that (helical) surges/jets often occur in the emerging flux region and are often associated with flares/CMEs (e.g., \citealt{shibata1986}). Therefore, the eruption studied here may be interpreted as a jet rather than a plasmoid.

The plasma blob acceleration starts with the slow rising of soft-X-ray flux (brightening below the blob, formation of ribbons `R1' and `R2'), which is probably associated with the ``tether-cutting reconnection" usually seen below the filaments in the initial phase of eruptions \citep{moore2001,sterling2004}. After the ejection of the blob, the reconnection rate seems to be enhanced in the current-sheet presumably formed below it, along the fan-spine topology, leading to the abrupt release of magnetic energy and acceleration of the charged particles along the footpoints, forming ribbons `R2' and `R3'.

The EUV images of our study also show the formation of coronal shock wave at the front/leading edge of 
the plasma blob during its impulsive acceleration phase. Additionally, a metric type II radio burst is
observed during the formation of EUV shock in the corona. The speeds of the EUV shock wave and the type II radio burst source are comparable. This provides evidence of 
 shock formation in front of the plasma blob.  
 In the model of solar eruptive flares with the plasmoid ejection \citep{yok2001}, the current sheet is formed below
the ejected plasmoid. As the reconnection progresses,
the plasmoid is simultaneously pushed upward by the plasma reconnection 
outflows. In a uniform medium, the plasmoid would need to have super-Alfv\'enic speed to generate a fast-mode shock at its leading edge. However, for a non-uniform medium, the shock can be formed from an ordinary simple wave and steepen into a shock wave, if there is continuous energy supply \citep{vrsnak2008}. For example, when a CME is suddenly accelerated, it can lead to a simple wave of small-scale sizes
\citep{chen2011a}. Additionally, in the numerical simulation of \citet{chen2000}, it is shown that a plasmoid with a speed of $\sim$300 km s$^{-1}$ in the corona, generates the shock above it. Moreover, \citet{klein1999} suggested that the origin of the type II burst is closely
 related to the dynamics of the X-ray plasma blob moving with a speed of $\sim$770 km s$^{-1}$. In the present study, the fastest sky-plane speed of the plasma blob is $\sim$540 km s$^{-1}$, which may be sufficient to generate a shock in the corona.

We observed both the faster ($\sim$912 km s$^{-1}$ ) and the slower ($\sim$341 km s$^{-1}$) wavefronts formed in front of the moving plasma blob. Initially, both wavefronts are cospatial and later the faster wavefront arrives at the boundary of another AR, is reflected by the AR 11177/CH boundary, and moves toward the west. The slower wavefront moves in a northeast direction, which has been revealed by the AIA 335 and STEREO 195 \AA \ images. The faster wavefront is the coronal Moreton wave as predicted in the numerical simulation of \citet{chen2002,chen2005}, which was recently confirmed from the AIA observations \citep{chen2011,kumar2012b}. Moreover, the speed and timing of the faster wave matches the type II radio burst. Therefore, the coronal Moreton wave (i.e., the fast MHD wave) is the source of the type II radio burst. The faster wavefront is possibly triggered by the impulsive acceleration of the blob, and later this wavefront propagates freely, because the plasma blob has been observed to run behind the slower wavefront in AIA 335 \AA~ and STEREO till 23:18 UT. There may be different possibilities to observe the slower wavefront. (i) This may be the signature of the same shock (fast) propagating in the low corona, (i.e., anisotropy of the wave propagation/different local fast-mode speeds). But it seems to be stopped at the boundary of another AR 11177 and does not propagate further across it. This is in contrast to its wave nature. (ii) This may be the slower EUV wave, running behind the faster MHD shock wave, as predicted from 2D and 3D simulations \citep{chen2005,cohen2009,downs2011,downs2012}, which was recently confirmed by \citet{chen2011} and \citet{kumar2012b}. The slower EUV wave is found to be cospatial with the CME leg because of the field line stretching. However, initially the slower component is probably merged with the faster component, and when the direction of the plasma blob changes to a northeast direction, it decouples (running in front of the blob). In the numerical simulation, the radial expansion speed of the shock is generally about three times the lateral expansion speed, i.e., the speed of the coronal Moreton wave is about three times the speed of the EIT wave \citep{chen2005}, which is consistent with our observations. The lateral expansion of the CME leg may be analogous to the blob propagation in the low corona. Therefore, the slower wavefront indicates stronger heating in the northeast direction caused by the plasma blob impact/compression, which is responsible for the bright emissions observed in the AIA 335 and STEREO 195 \AA~ channels.

Furthermore, we observed the transmitted EUV wave (171 \AA) from the funnel-shaped loops of another AR. We also found the emergence of two fast EUV waves/ripples (`F1' and `F2') from the funnel-shaped loops originating from the trailing (eastern) polarity of the flaring AR, possibly generated by the impact of the primary wave. These two wavefronts move in a northeast direction (limb side) across the existing AR that shows the signatures of fast MHD waves.
Recently, \citet{olmedo2012} reported the reflection/refraction of the EUV wave from the CH that was associated with the X2.2 flare on 15 February 2011. They found for the first time the transmission of part of the primary wave through the CH.
In the 07 June 2011 eruption, \citet{ting2012} observed a 3D dome-shaped EUV wave and found the apparent disappearance of the primary EUV wave upon arrival at an AR on its path.
A secondary wave rapidly re-emerged within the AR boundary at the same speed. Moreover, they also observed the reflection of some part of the primary EUV wave. Most importantly, though, they observed the secondary wave in four EUV channels, i.e., AIA 193,  211, 171, and 335 \AA. However, in our case, we observed the secondary EUV waves only in the 171 \AA~ channel that originates from the funnel-shaped loops, with the speed faster than about 20-30$\%$ the primary wave speed.

The speed of the driver is always slower than the speed of the generated shock in the case of piston-driven shocks, which is the reason why we see the clear shock signature. \citet{veronig2008} studied an eruption and found the filament speed to be faster than the speed of the EUV wave, they proposed that the filament might not be the possible driver of the shock wave. They suggested that the propagating perturbation (disturbance) was powered only temporarily by a source region expansion, which could be generated by the flare-related pressure pulse, by small-scale flare ejecta, or by the CME expanding flanks (which propagate laterally only over a limited distance). Additionally, if the CME speed approaches that of the associated EUV wave, the wave signatures are very likely merged with the associated phenomena of the CME. When the CME propagates much slower than the EUV wave, the wave signatures would show well \citep{zheng2011}. \citet{patsourakos2012} also mentioned that the EUV wave was decoupled from the CME once the CME was sufficiently decelerated.
 Our observational findings are consistent with the numerical simulation results of \citet{chen2002}. In their numerical simulation, the speed of the generated shock (in front of the flux-rope) is about 2-3 times faster than the speed of the piston (P.F. Chen 2012, private communication). 
The present event most likely belongs to the small-scale ejection, i.e., the plasma blob ejection. Moreover, in our case, the blob speed is slower than the speed of the EUV shock wave, and additionally, the shock is formed exactly in front of the blob and moves in the same direction. 

\begin{figure*}
\centering{
\includegraphics[width=5.5cm]{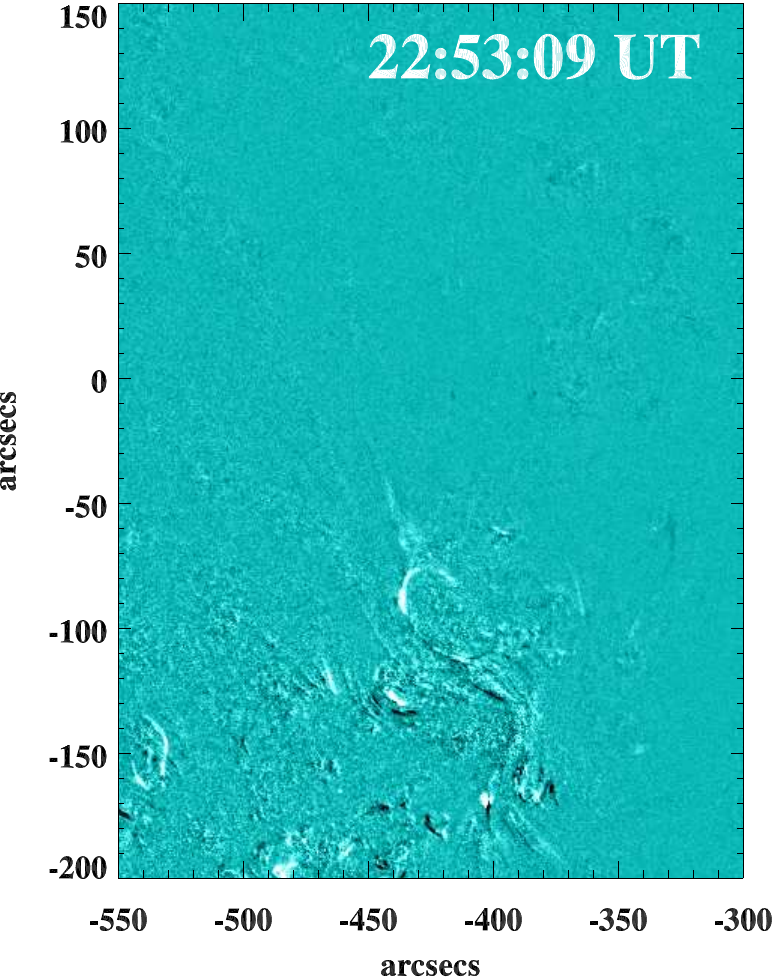}
\includegraphics[width=5.5cm]{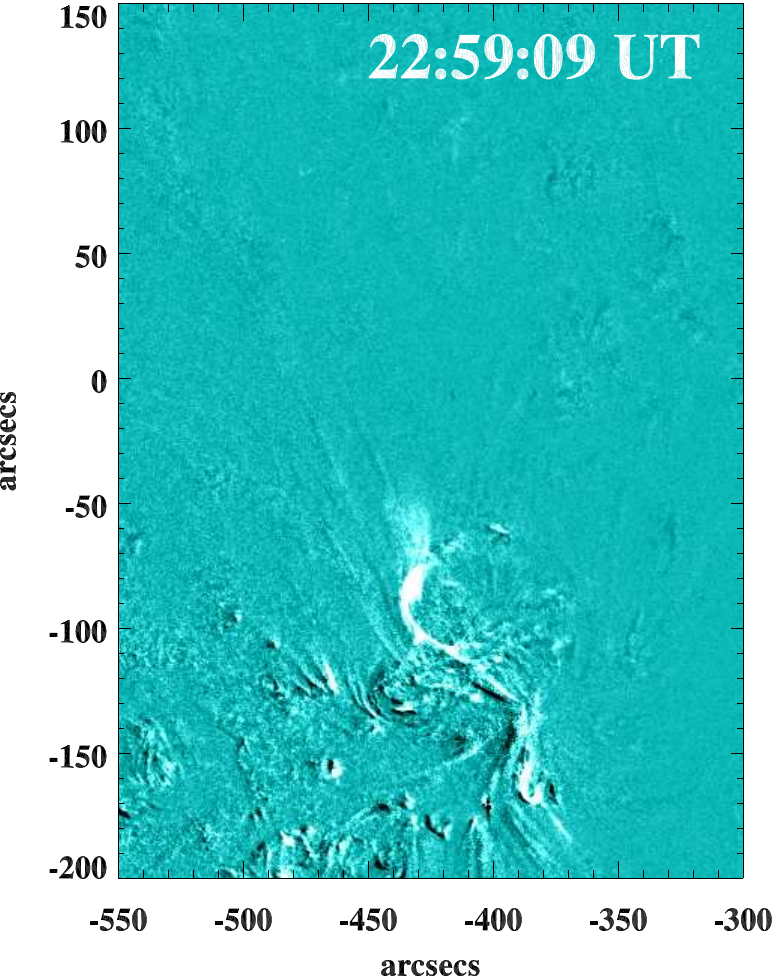}
\includegraphics[width=5.5cm]{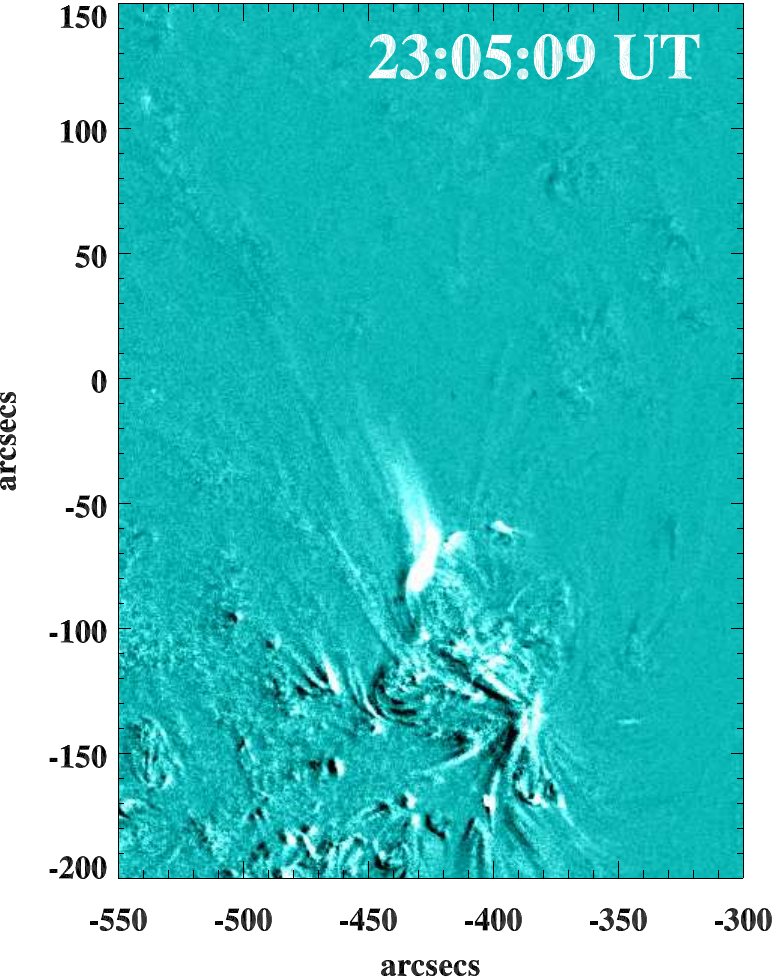}

\includegraphics[width=5.5cm]{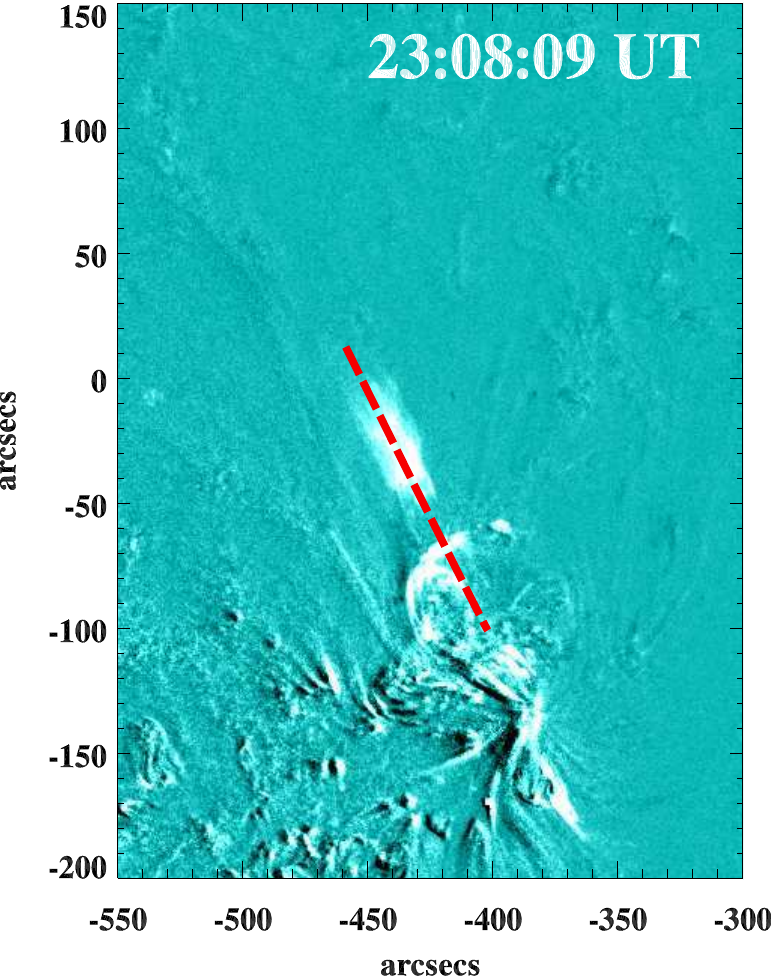}
\includegraphics[width=5.5cm]{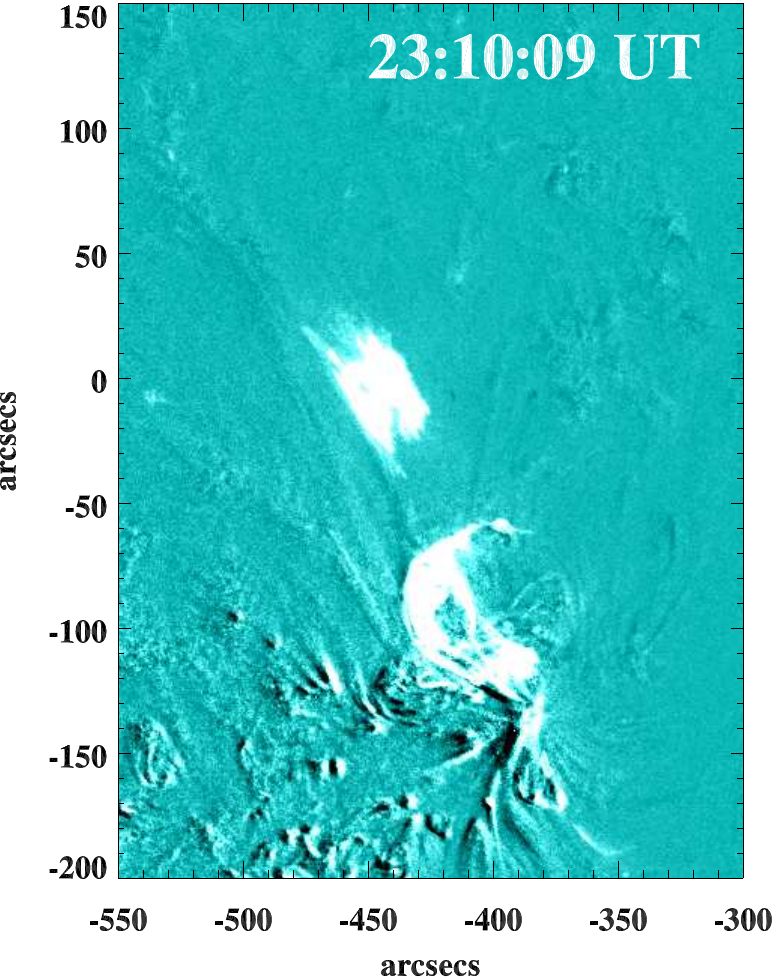}
\includegraphics[width=5.5cm]{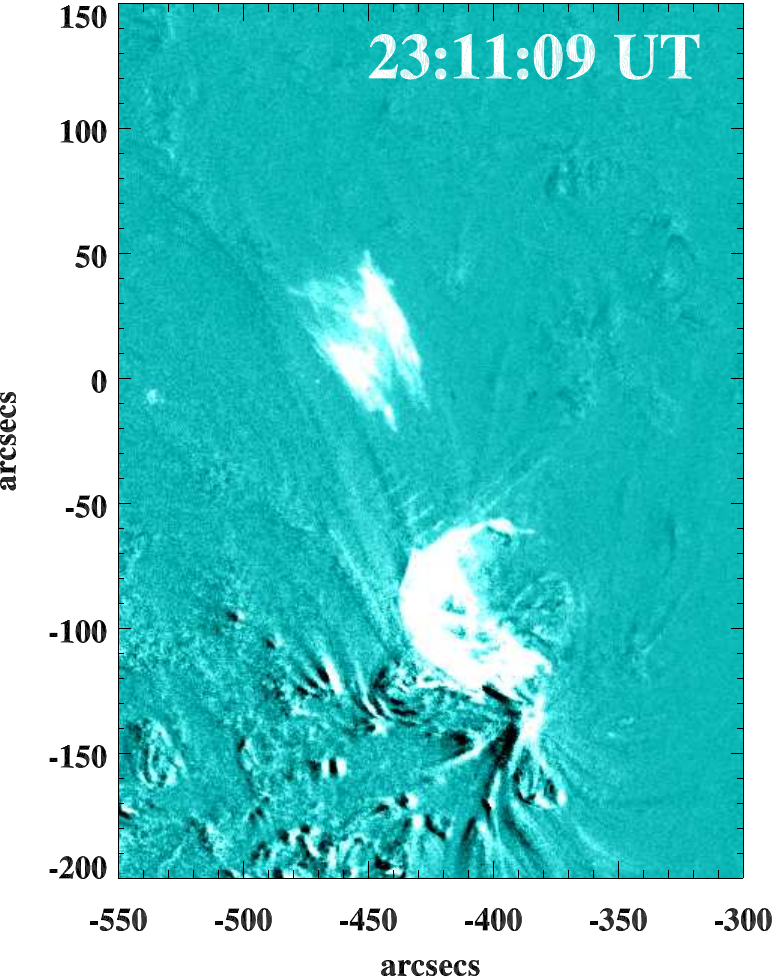}

\includegraphics[width=5.5cm]{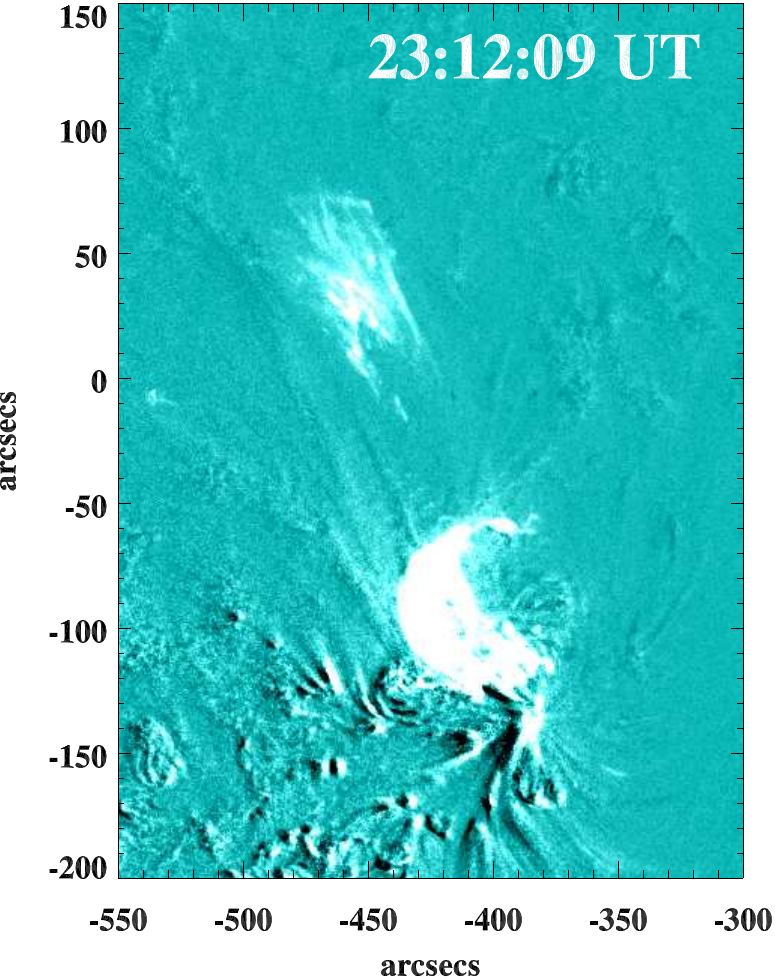}
\includegraphics[width=5.5cm]{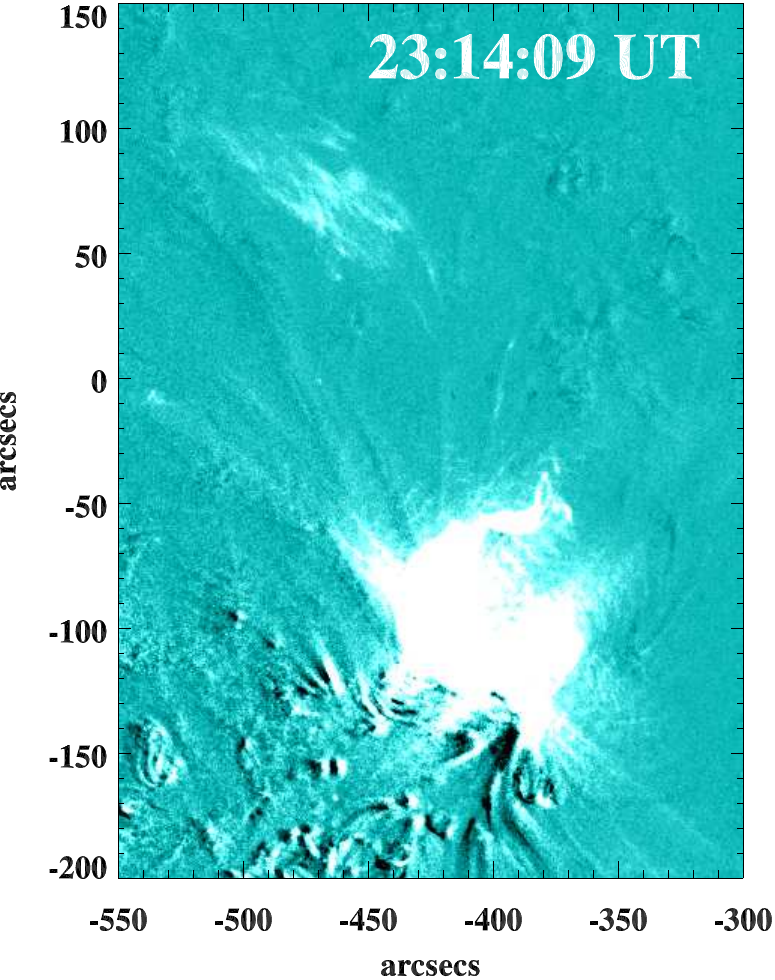}
\includegraphics[width=5.5cm]{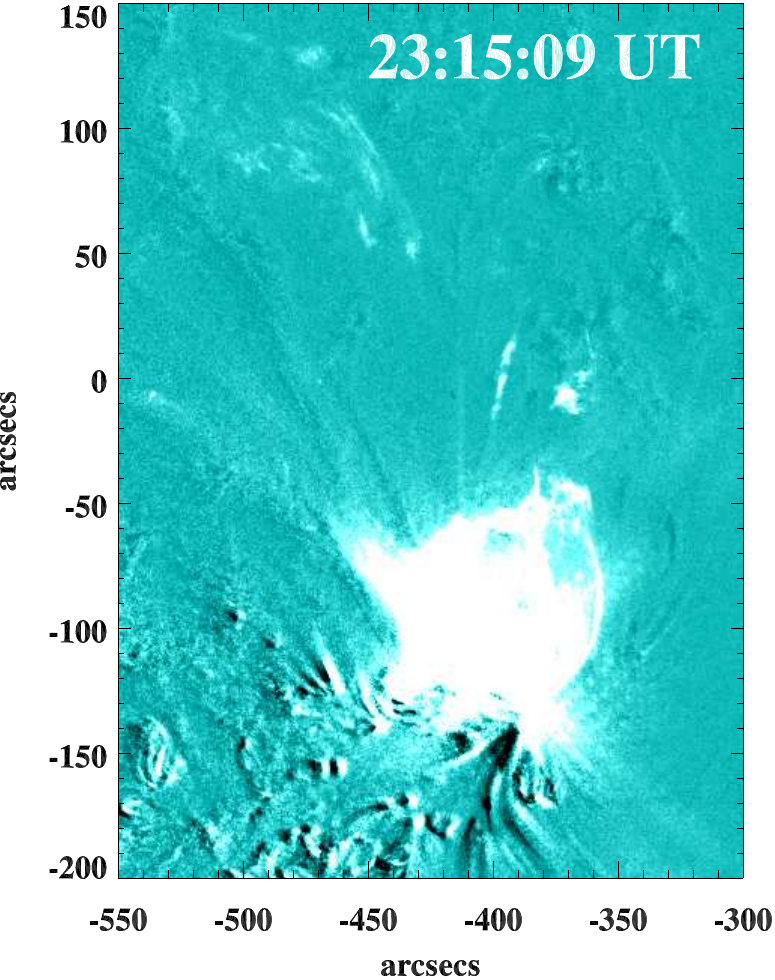}
}
\caption{SDO/AIA 131 \AA \ base-difference images showing the formation and eruption of the plasma blob on 25 March 2011.}
\label{aia131}
\end{figure*}

\begin{figure*}
\centering{
\includegraphics[width=12cm]{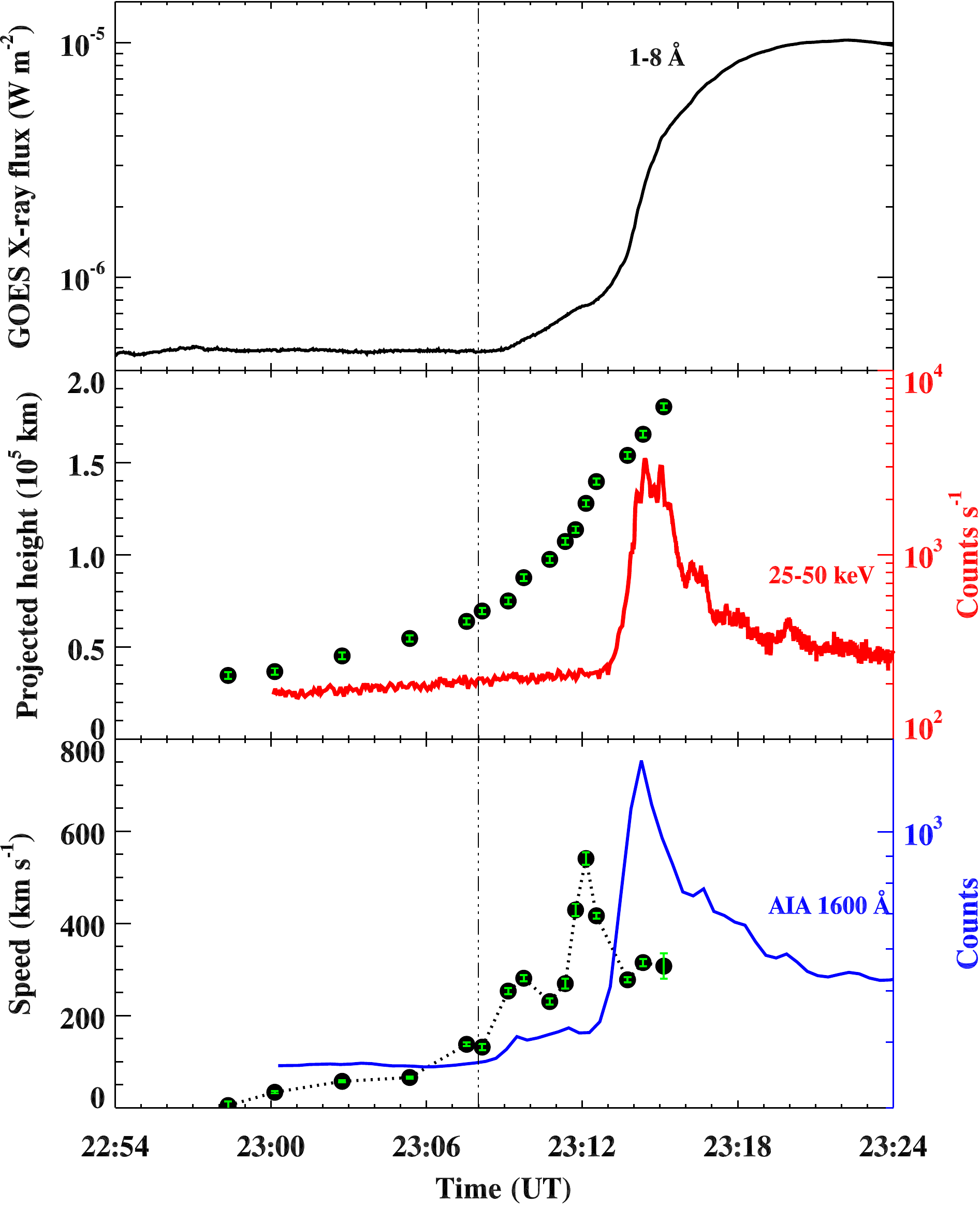}
}
\caption{Top: GOES soft X-ray flux profile in the 1-8 \AA~ wavelength channel. Middle: Projected height-time of the blob plotted with the hard X-ray flux profile in the energy range of 25-50 keV observed with the Fermi Gamma-ray Burst Monitor. Bottom: Temporal evolution of the blob speed derived by the height-time measurements using AIA 131 \AA~ images. The AIA 1600 \AA~ average integrated counts profile is plotted as a blue curve. The uncertainty in the speed estimate is mainly due to the error in the height measurement, which is assumed to be four pixels (i.e., 2.4$\arcsec$). The vertical dotted line shows start time of the flare at 23:08 UT.}
\label{kin}
\end{figure*}

\begin{figure*}
\centering{
\includegraphics[width=6cm]{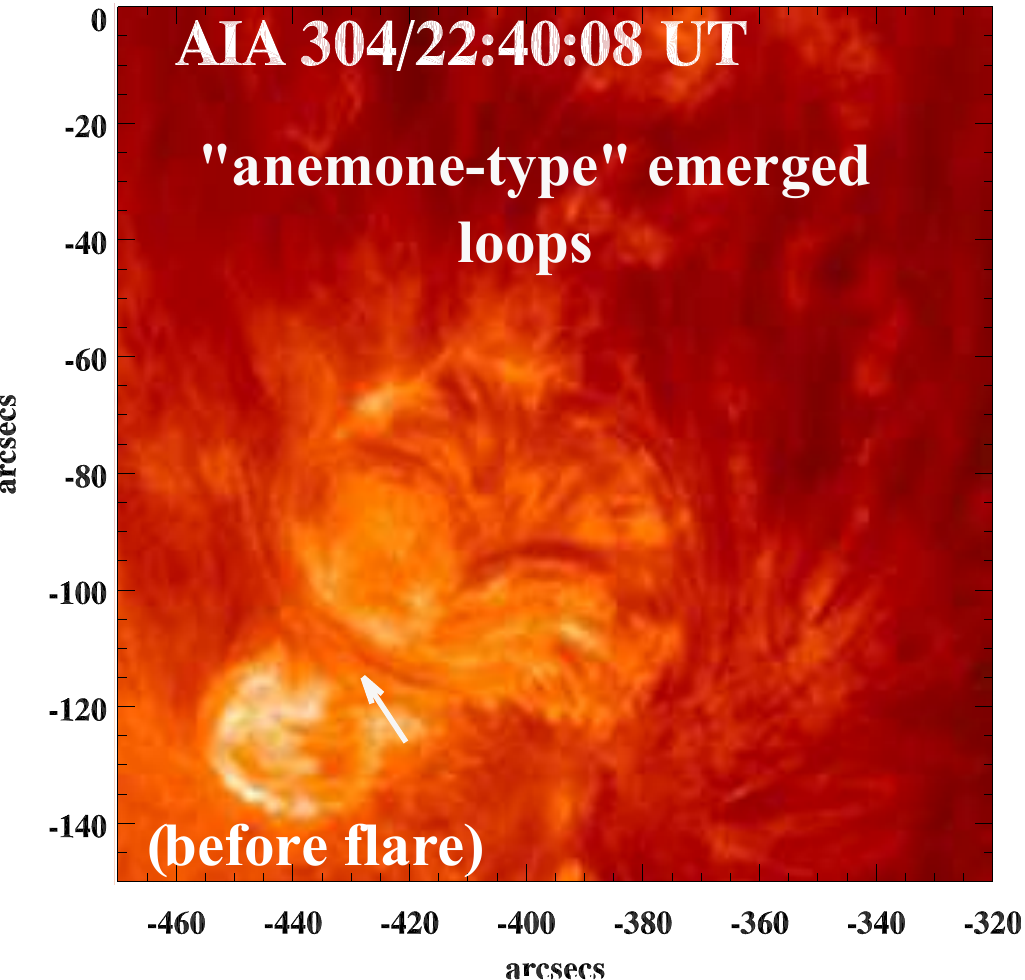}
\includegraphics[width=6cm]{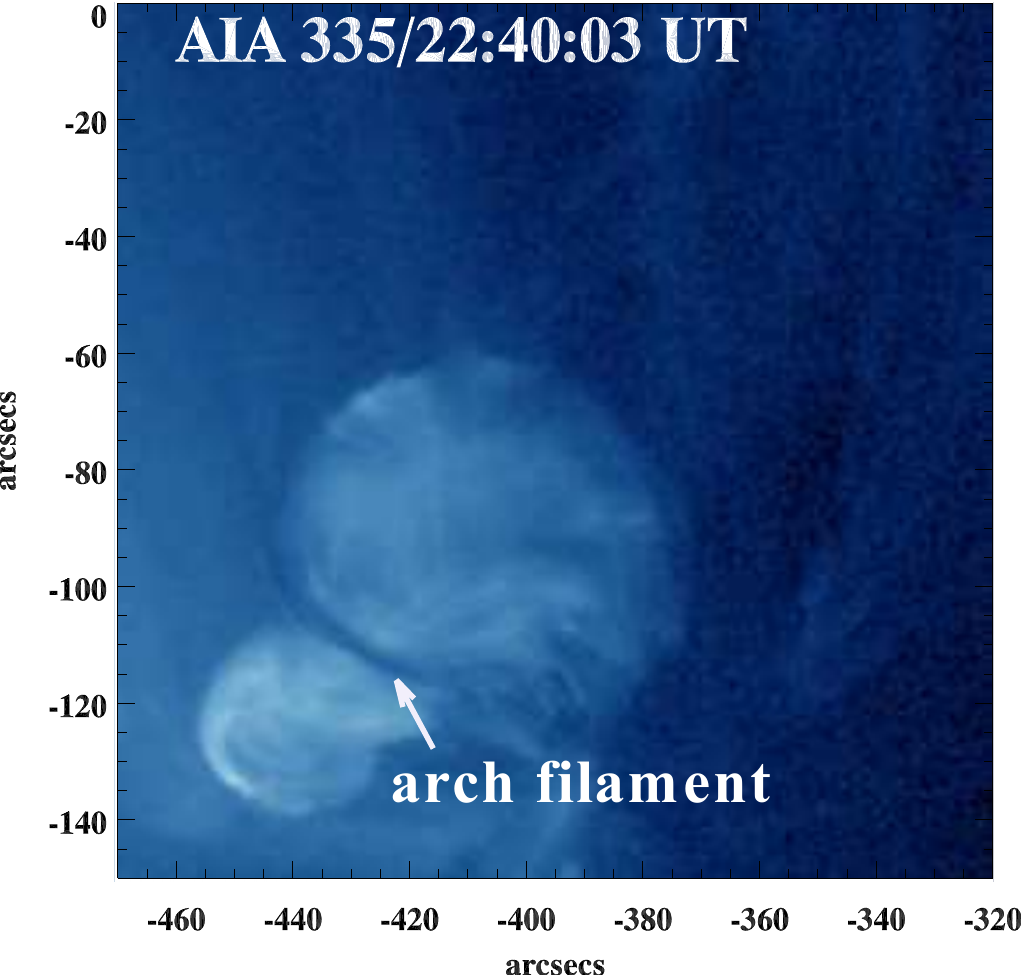}

\includegraphics[width=6cm]{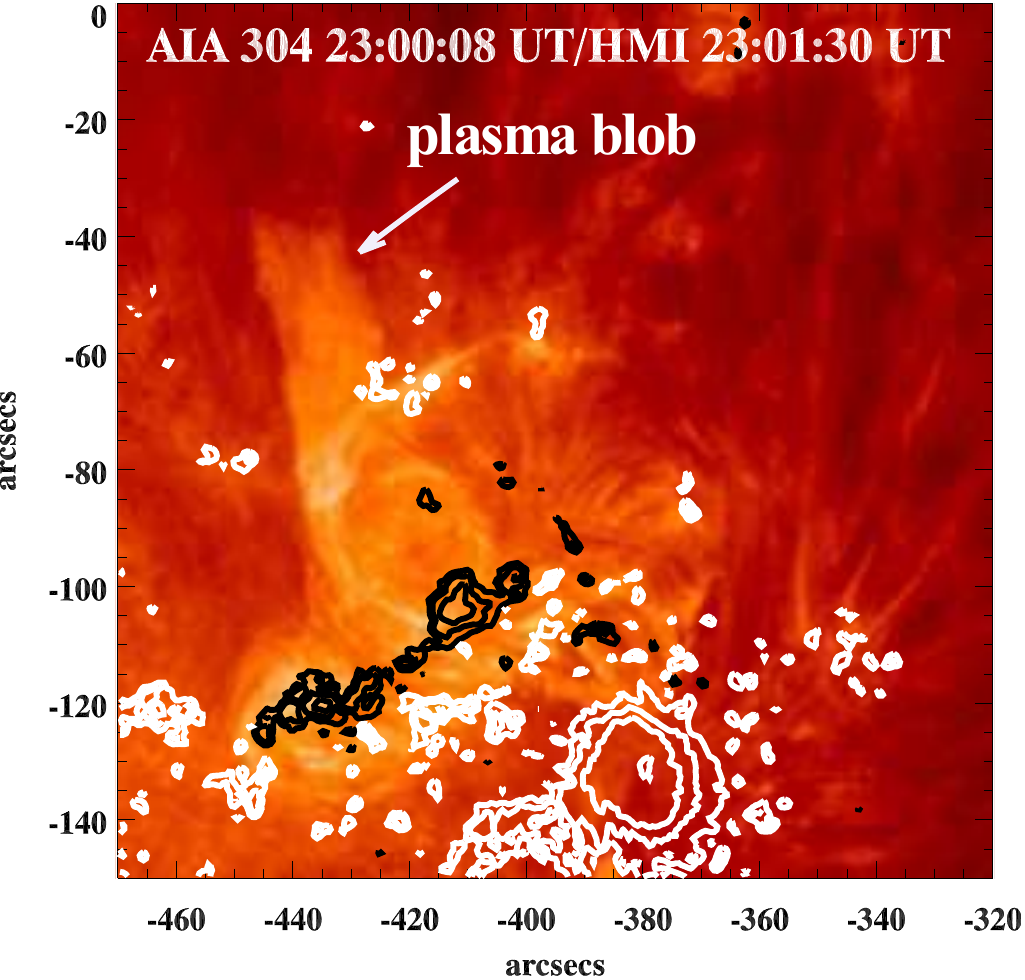}
\includegraphics[width=6cm]{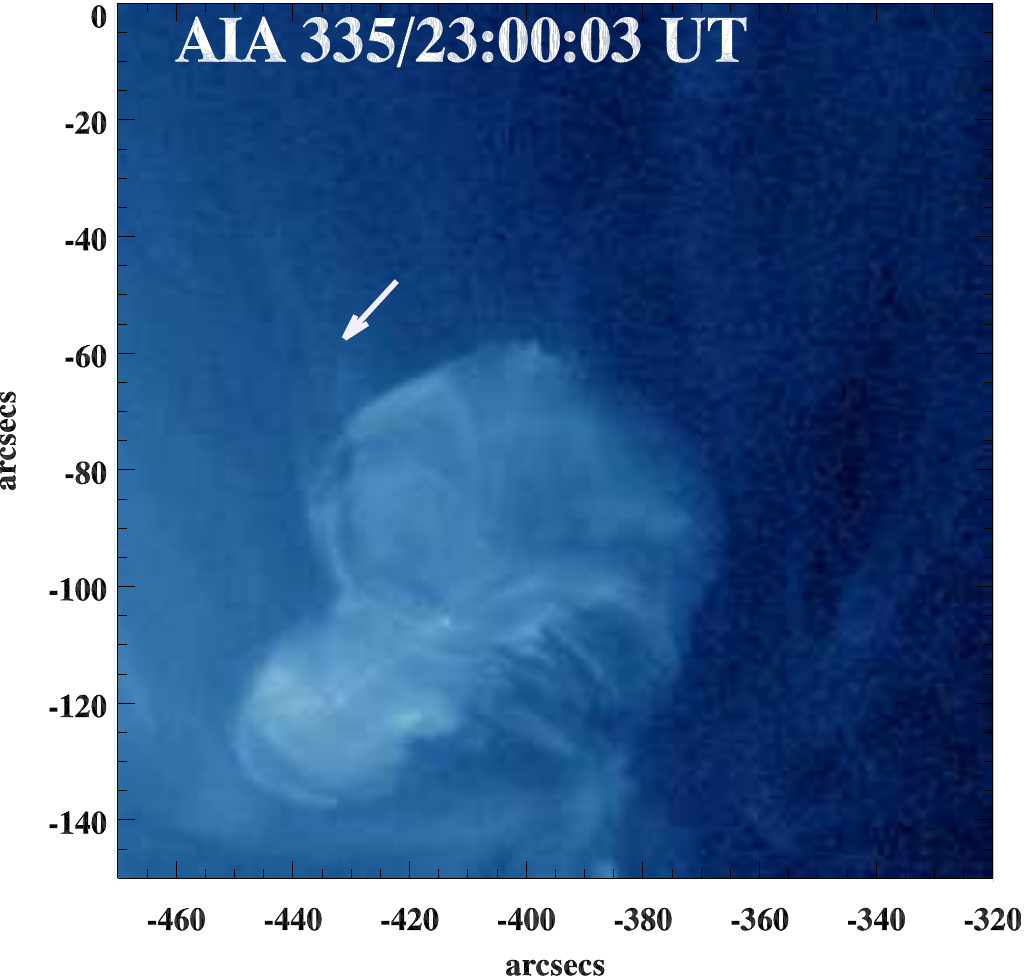}

\includegraphics[width=6cm]{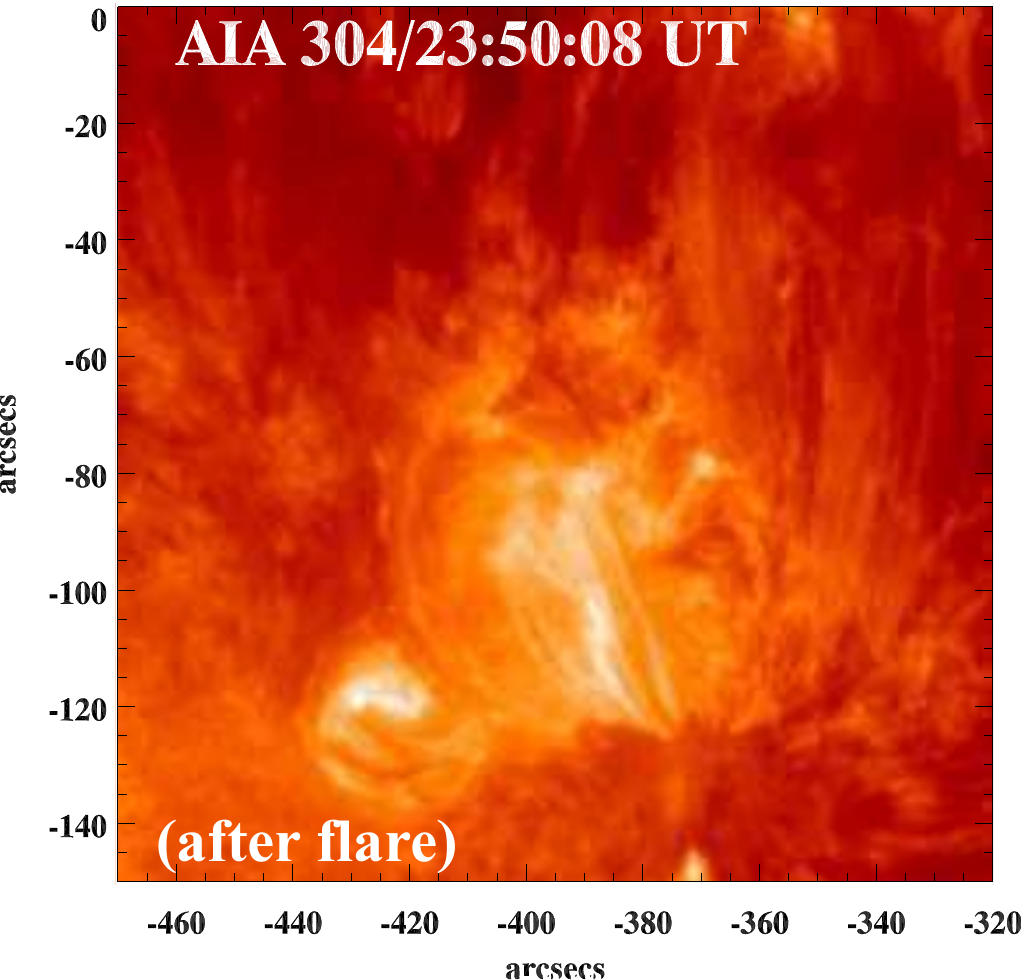}
\includegraphics[width=6cm]{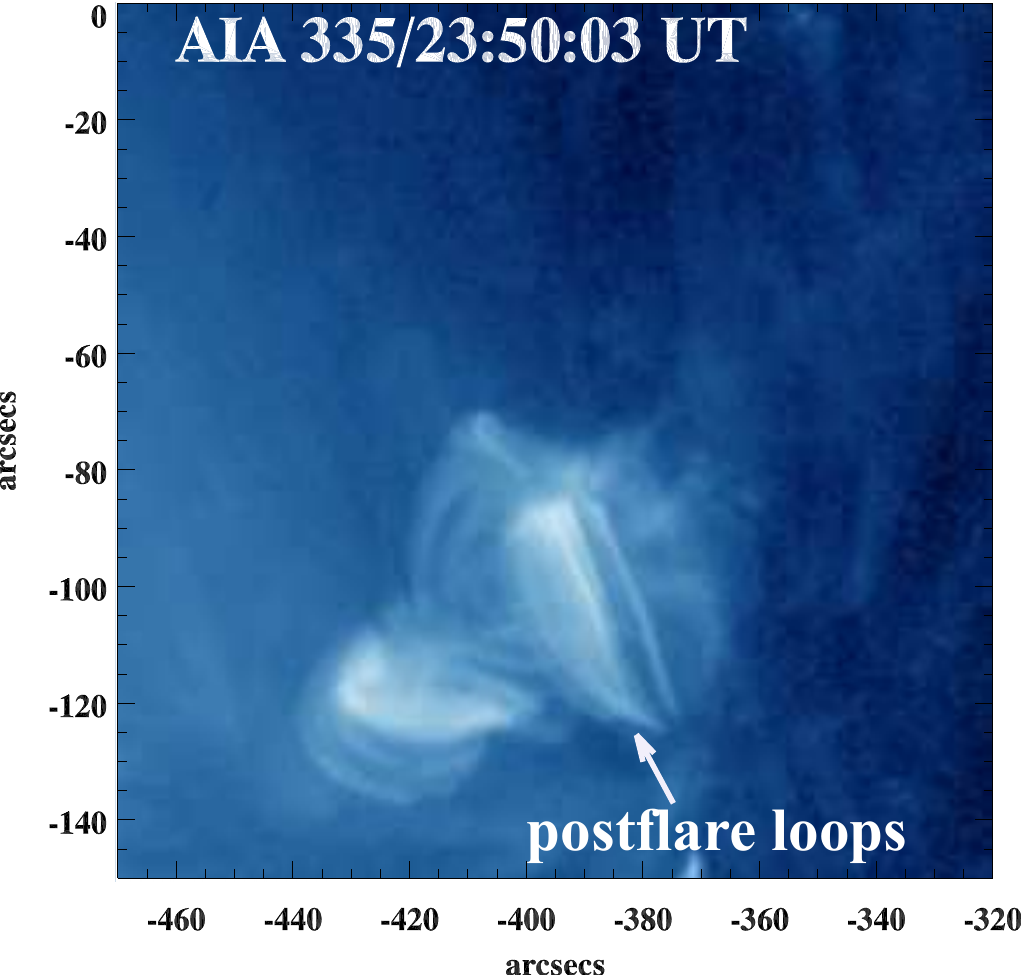}
}
\caption{SDO/AIA 304 and 335 \AA \ images of the blob formation site before and after the flare. The plasma blob was formed above the arch filament (indicated by arrows) at the site of the emerging anemone-type loops. The AIA 304 \AA~ image in the middle panel is overlaid with the HMI magnetogram contours of positive (white) and negative (black) polarities. The contour levels are $\pm$200, $\pm$500, $\pm$1000, and $\pm$2000 G.}
\label{aia_3}
\end{figure*}

\begin{figure*}
\centering{
\includegraphics[width=5cm]{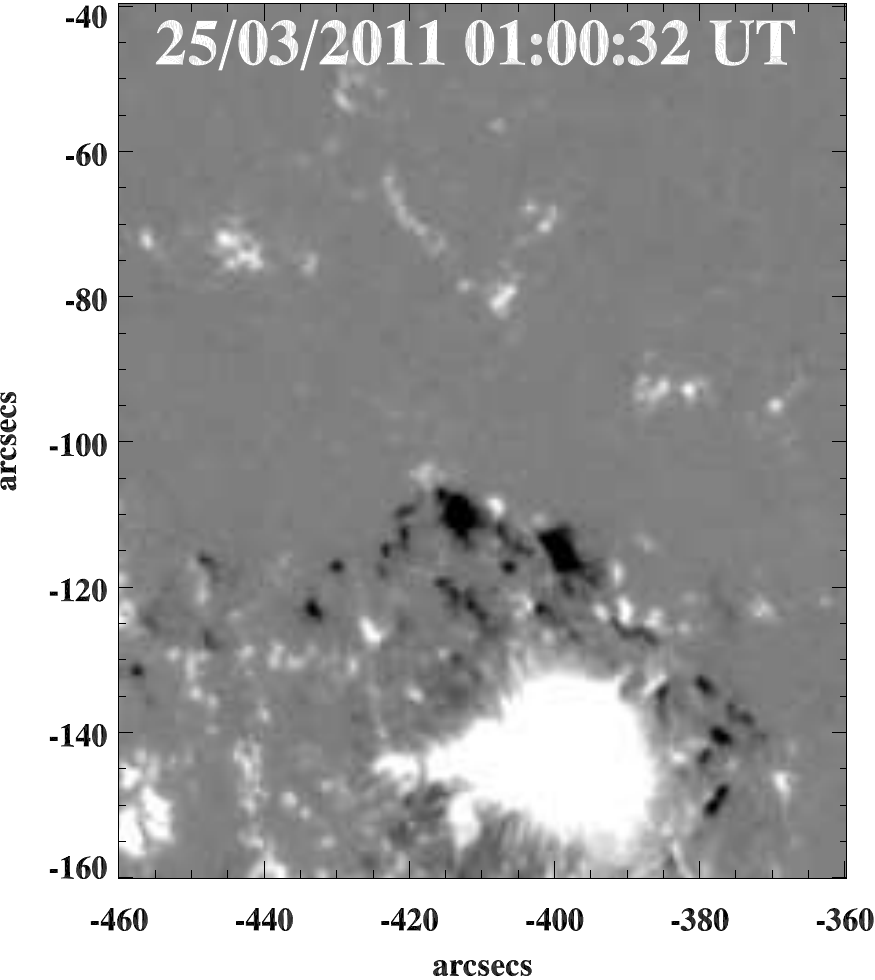}
\includegraphics[width=5cm]{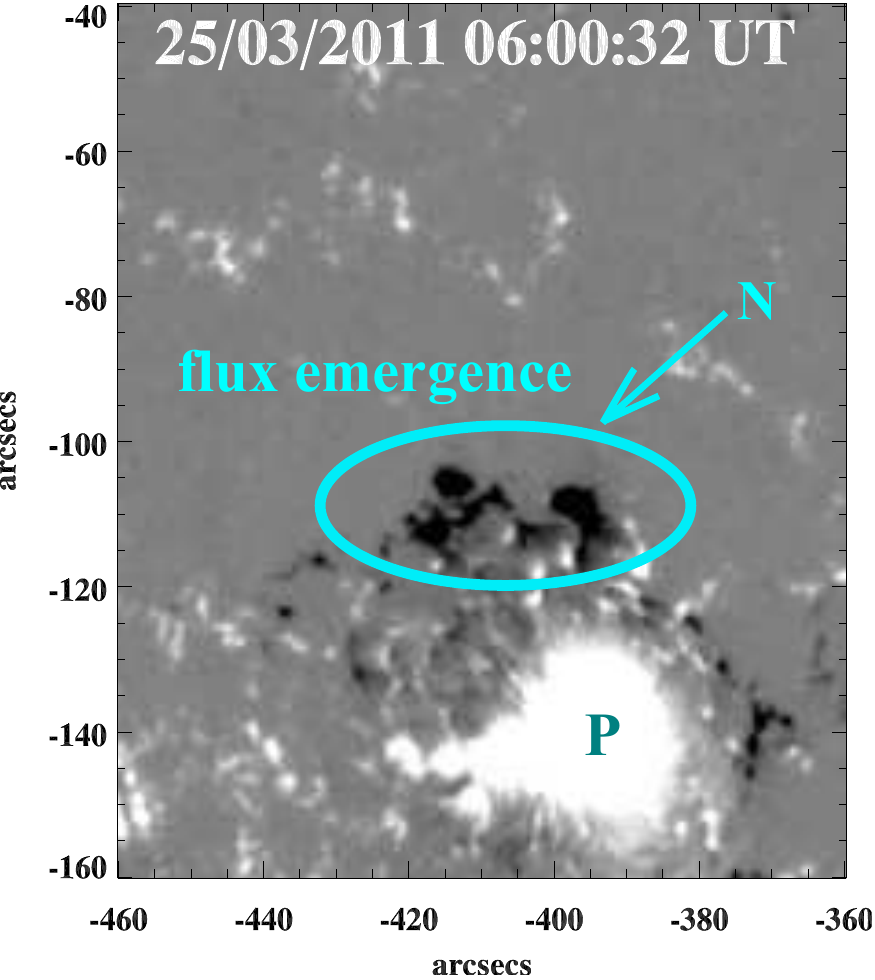}

\includegraphics[width=5cm]{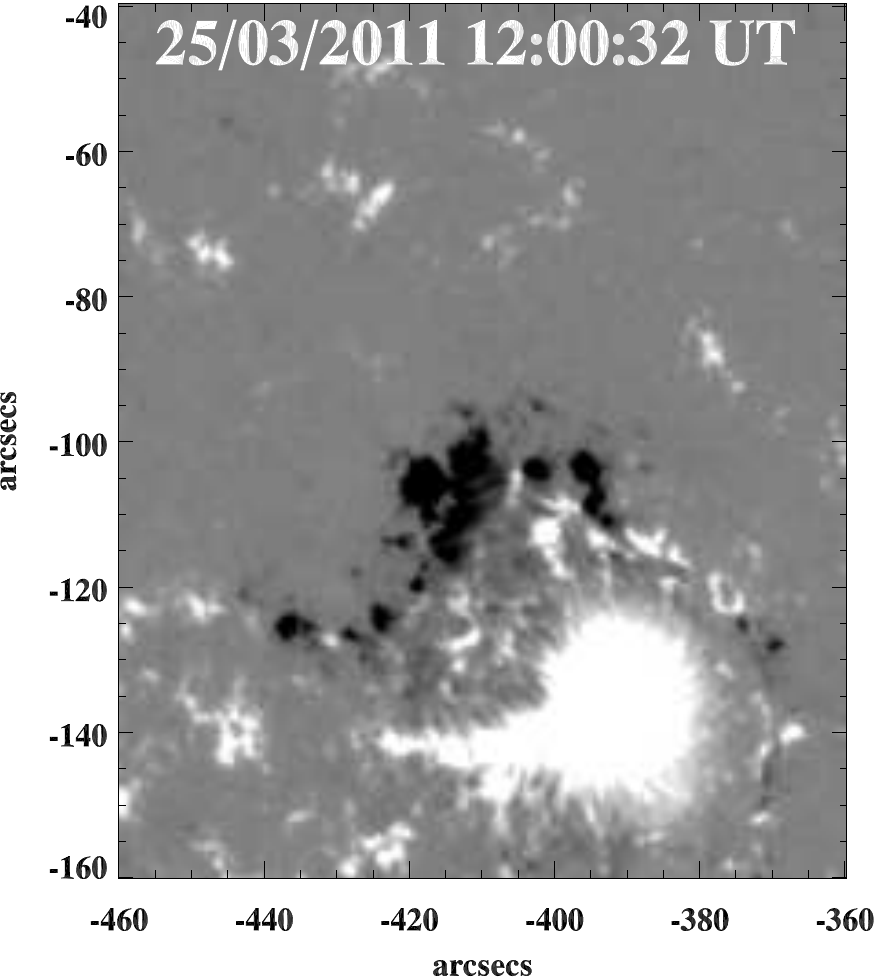}
\includegraphics[width=5cm]{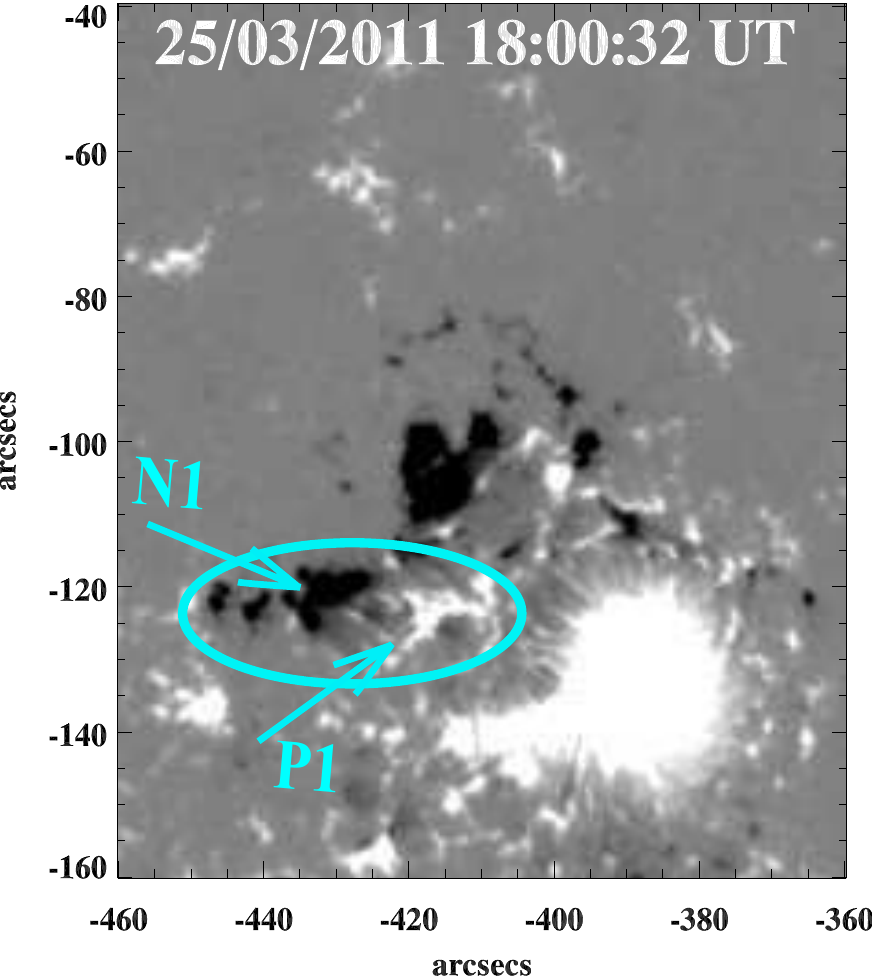}

\includegraphics[width=5cm]{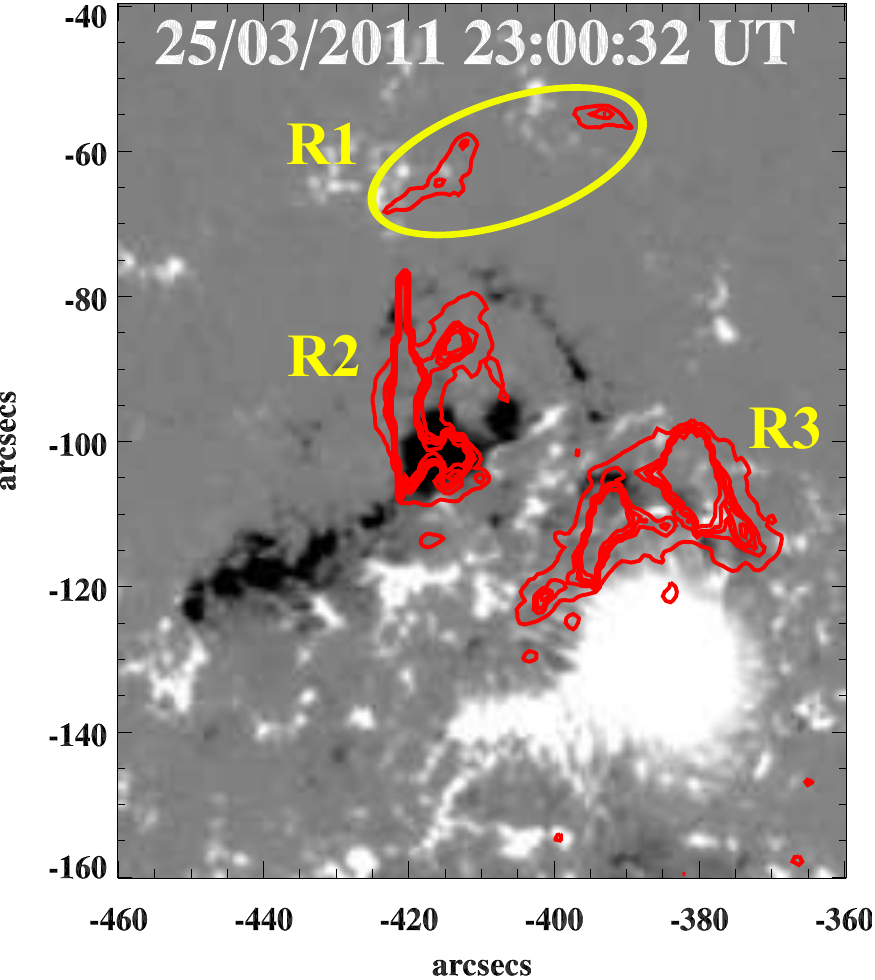}
\includegraphics[width=5cm]{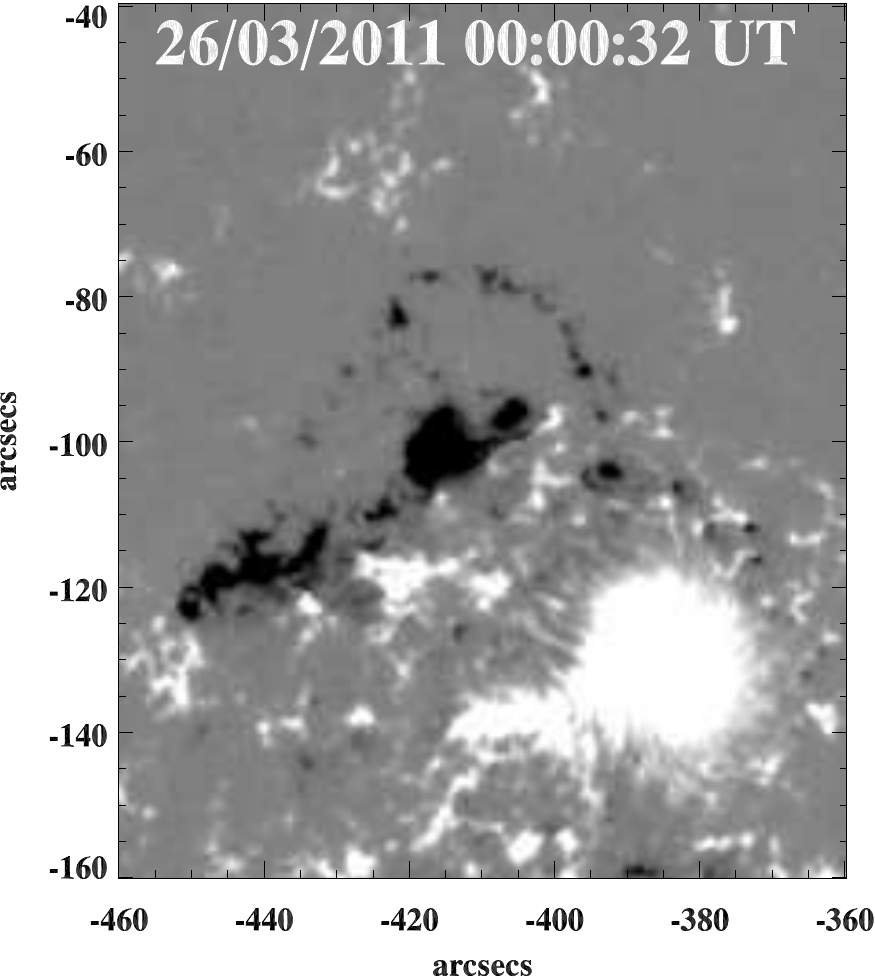}
}
\centering{
\includegraphics[width=10cm]{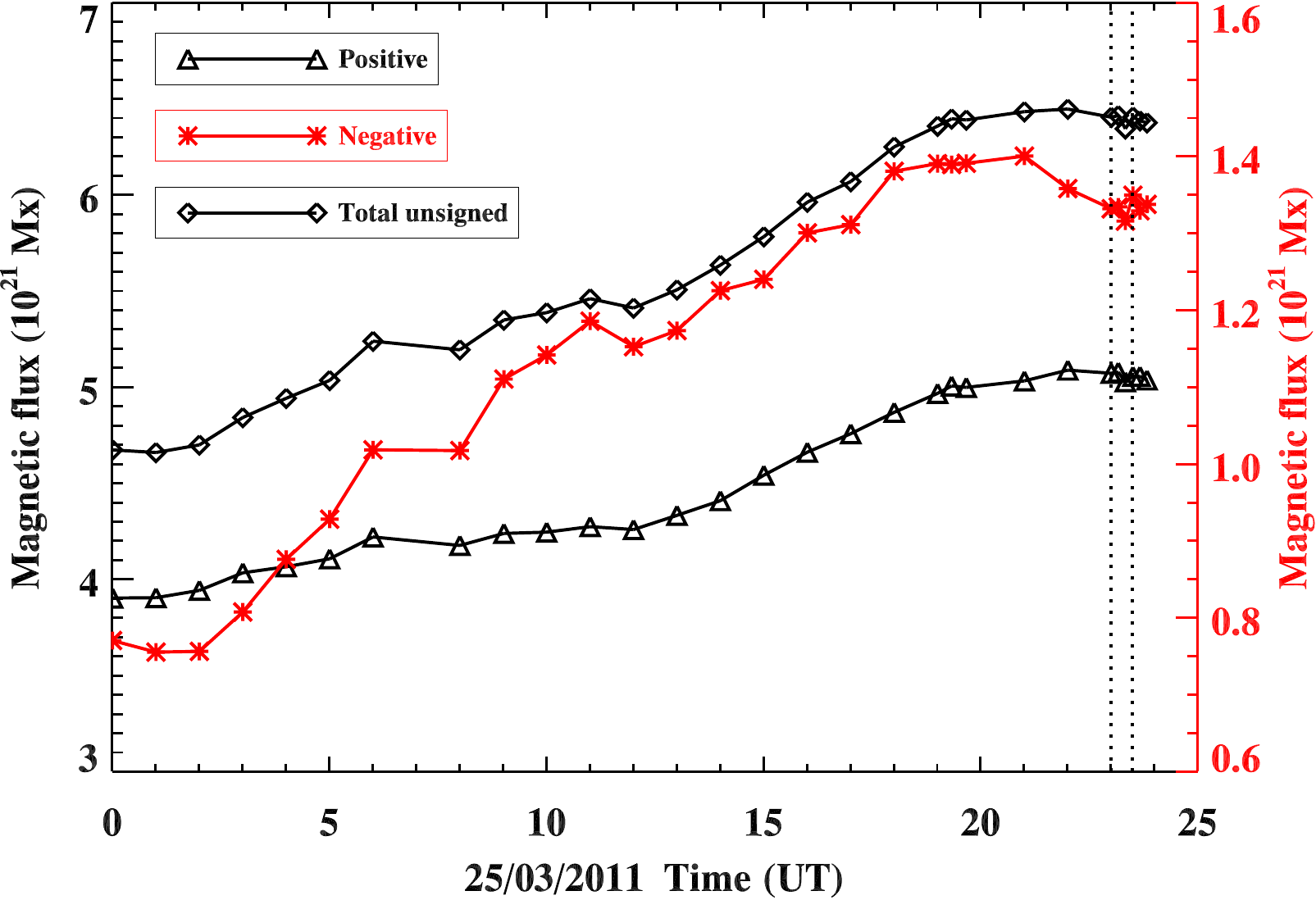}
}

\caption{HMI magnetograms showing the magnetic flux emergence at the site of the blob eruption. The magnetogram at 23:00 UT is overlaid with the AIA 1600 \AA~ EUV flare ribbons near the flux emergence site. Bottom: The temporal evolution of positive, unsigned negative, and total unsigned flux derived from the HMI magnetograms on 25 March 2011. Two vertical dotted lines indicate the duration of blob eruption and the associated flare.}
\label{hmi}
\end{figure*}

\begin{figure*}
\centering{
\includegraphics[width=6cm]{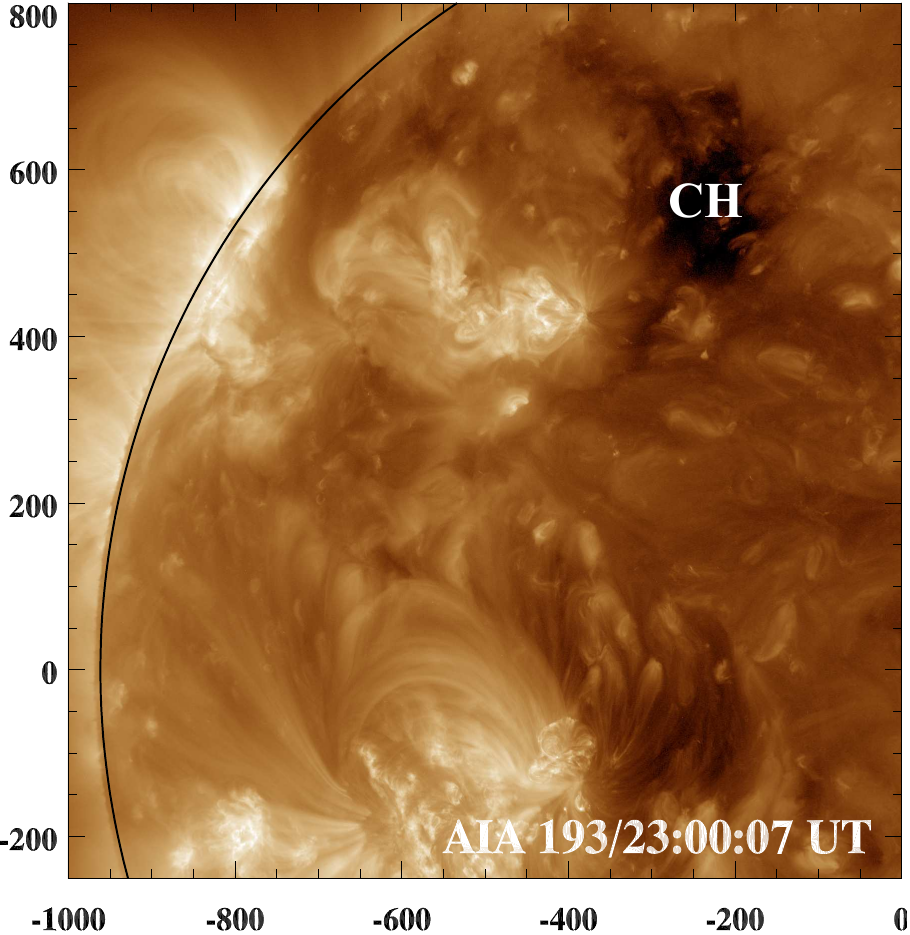}
\includegraphics[width=6cm]{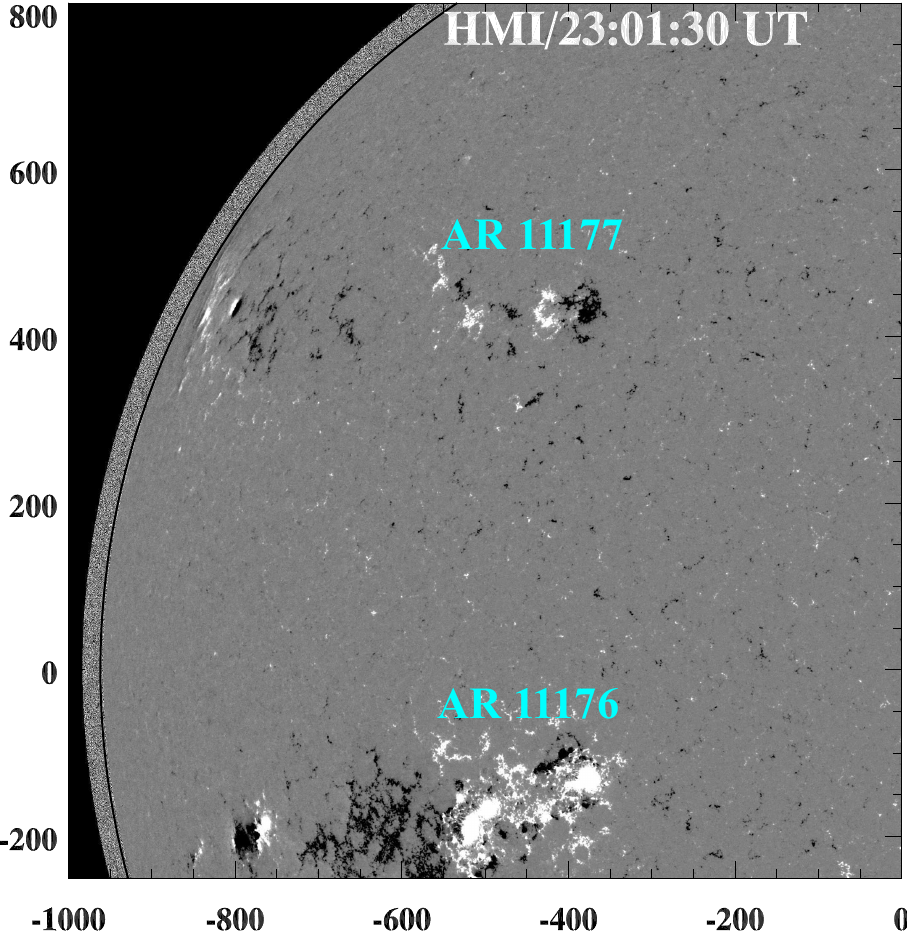}

\includegraphics[width=6cm]{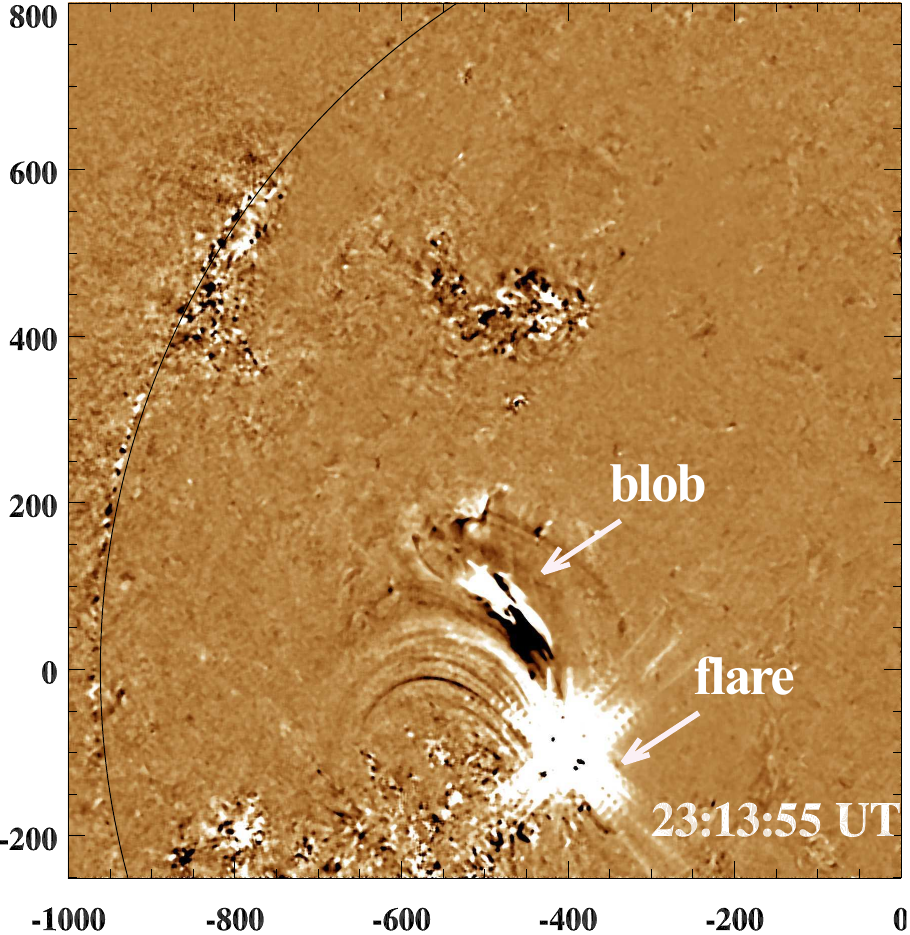}
\includegraphics[width=6cm]{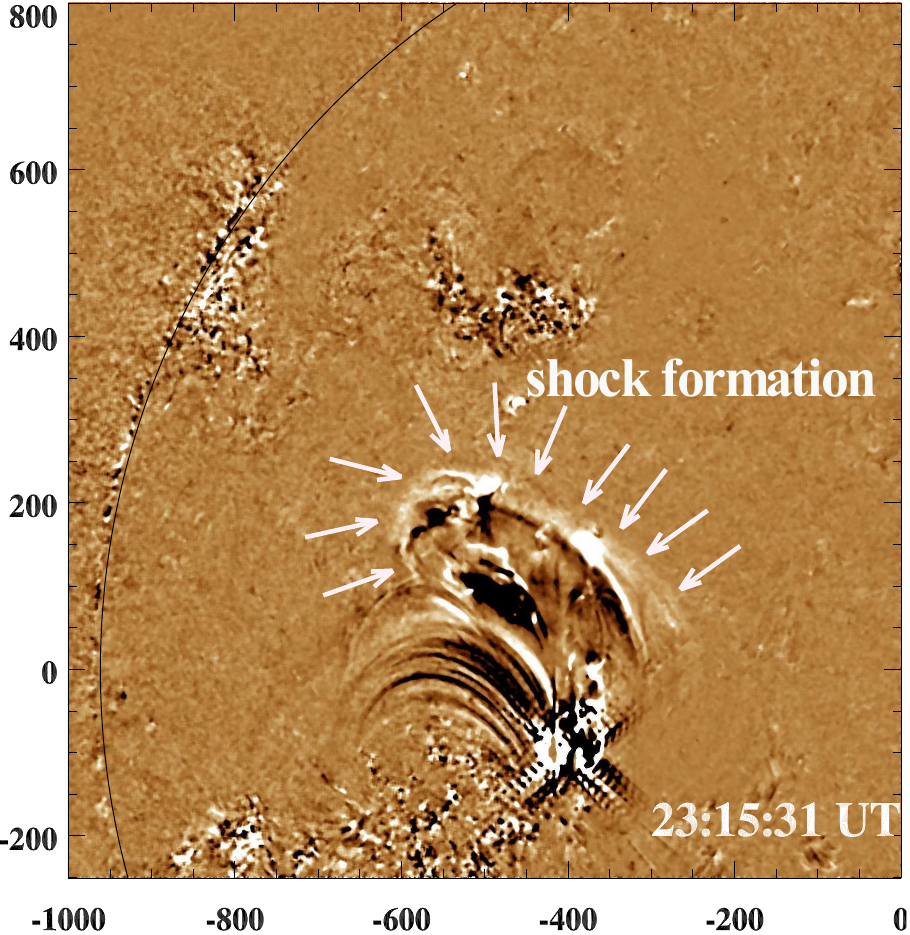}

\includegraphics[width=6cm]{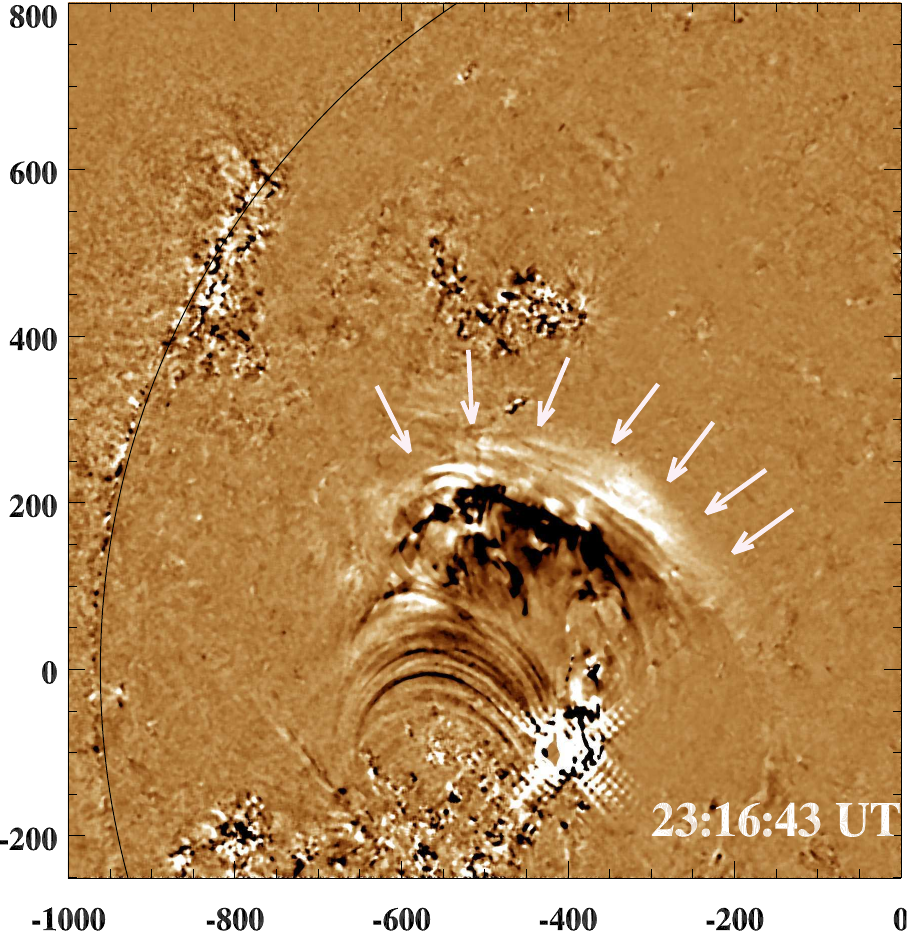}
\includegraphics[width=6cm]{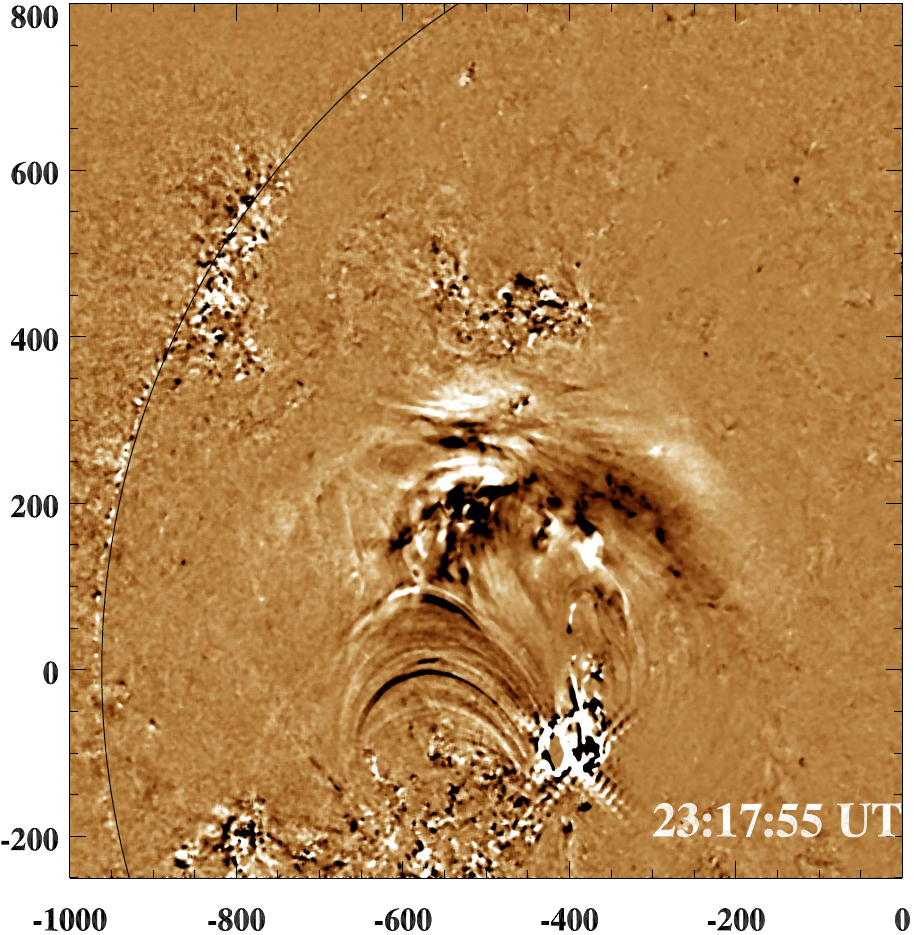}

\includegraphics[width=6cm]{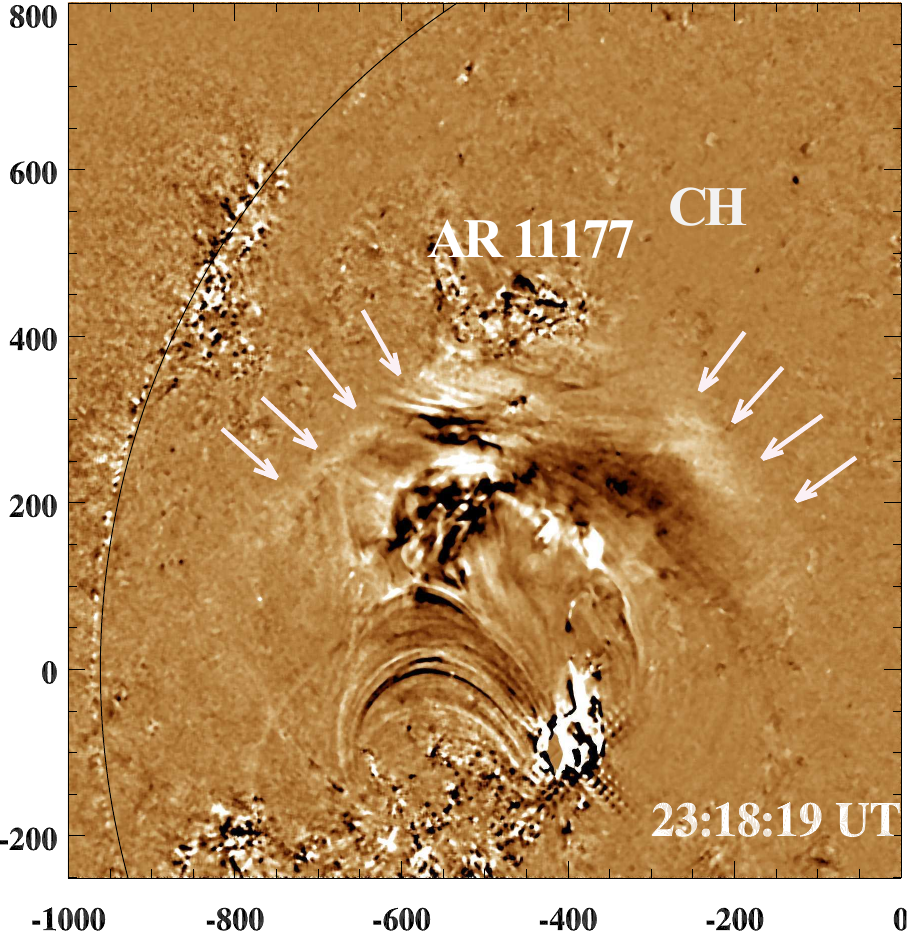}
\includegraphics[width=6cm]{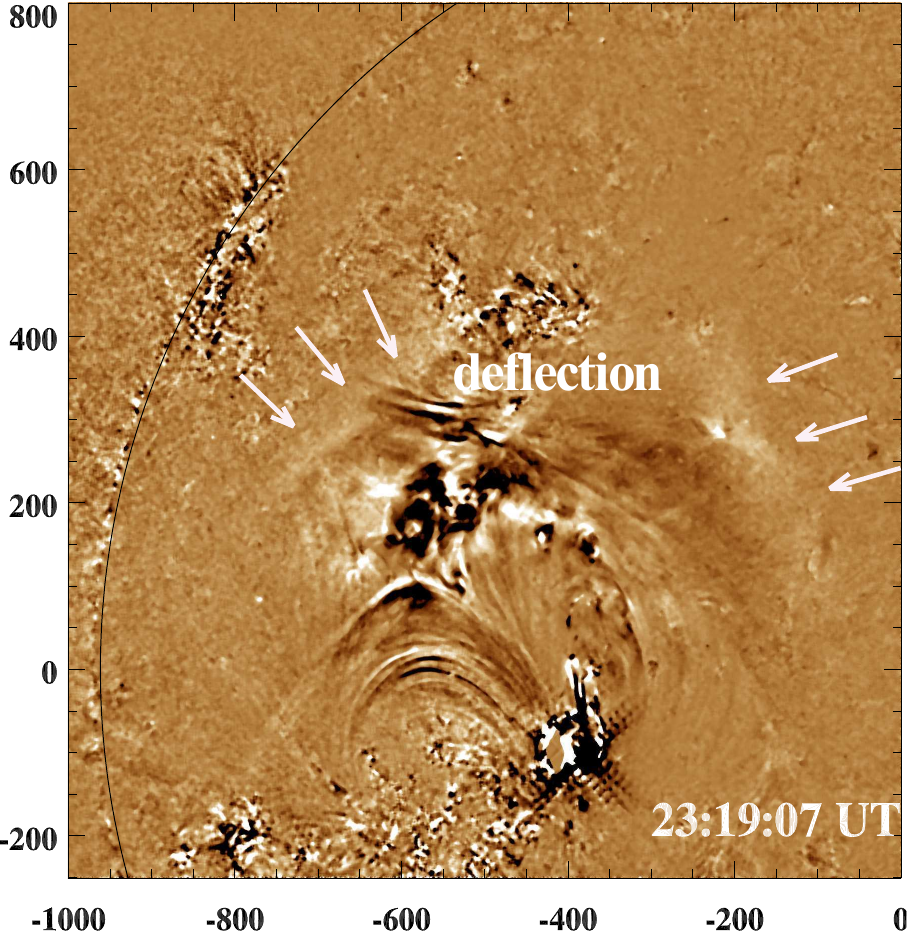}
}

\caption{AIA 193 \AA~ intensity, HMI magnetogram and AIA 193 \AA~ running difference images showing the eruption of the plasma blob and the formation of EUV wave in front of the blob on 25 March 2011. $X$ and $Y$ axes are given in arcsecs.}
\label{aia193}
\end{figure*}

\begin{figure*}
\centering{

\includegraphics[width=6cm]{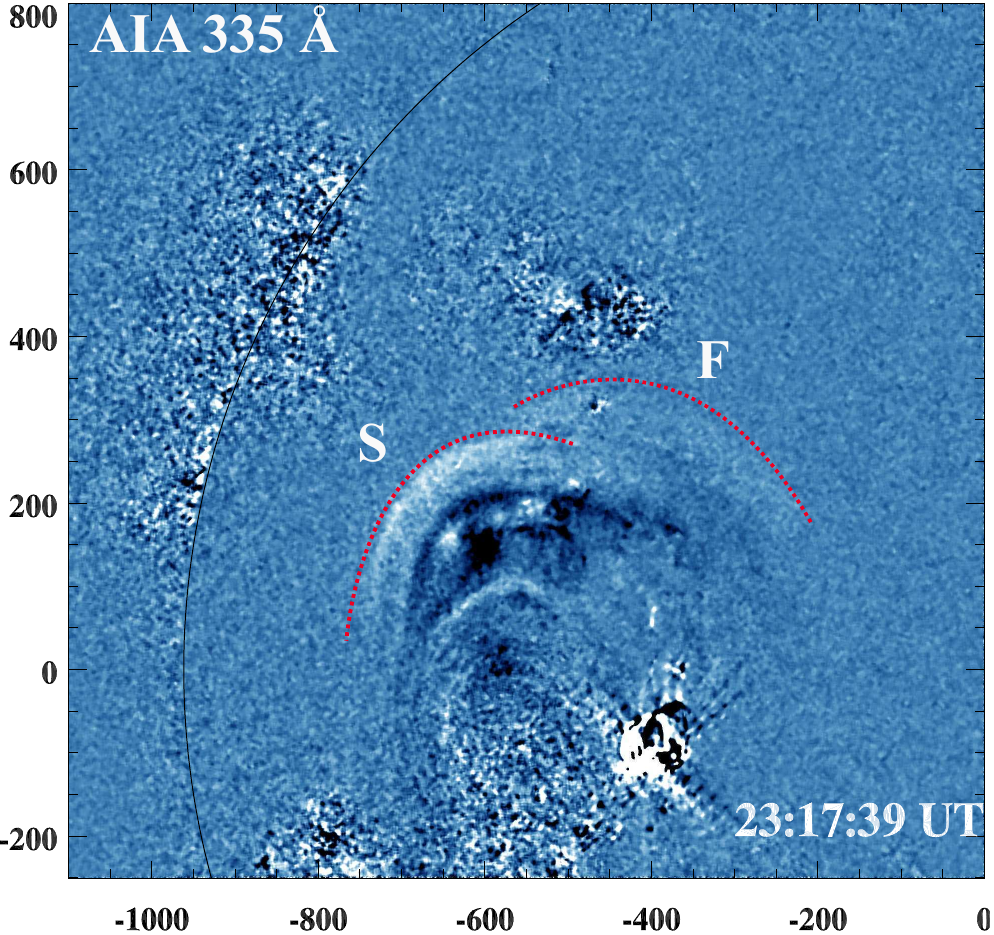}
\includegraphics[width=6cm]{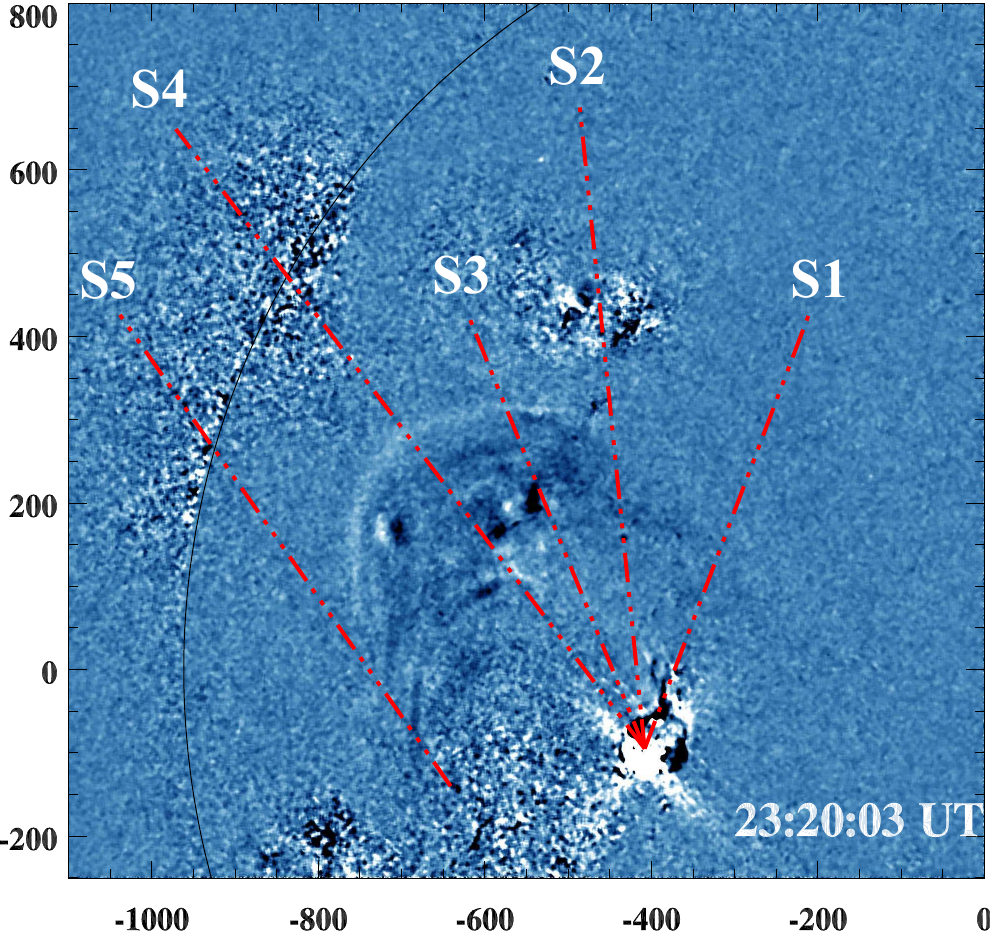}

\includegraphics[width=6cm]{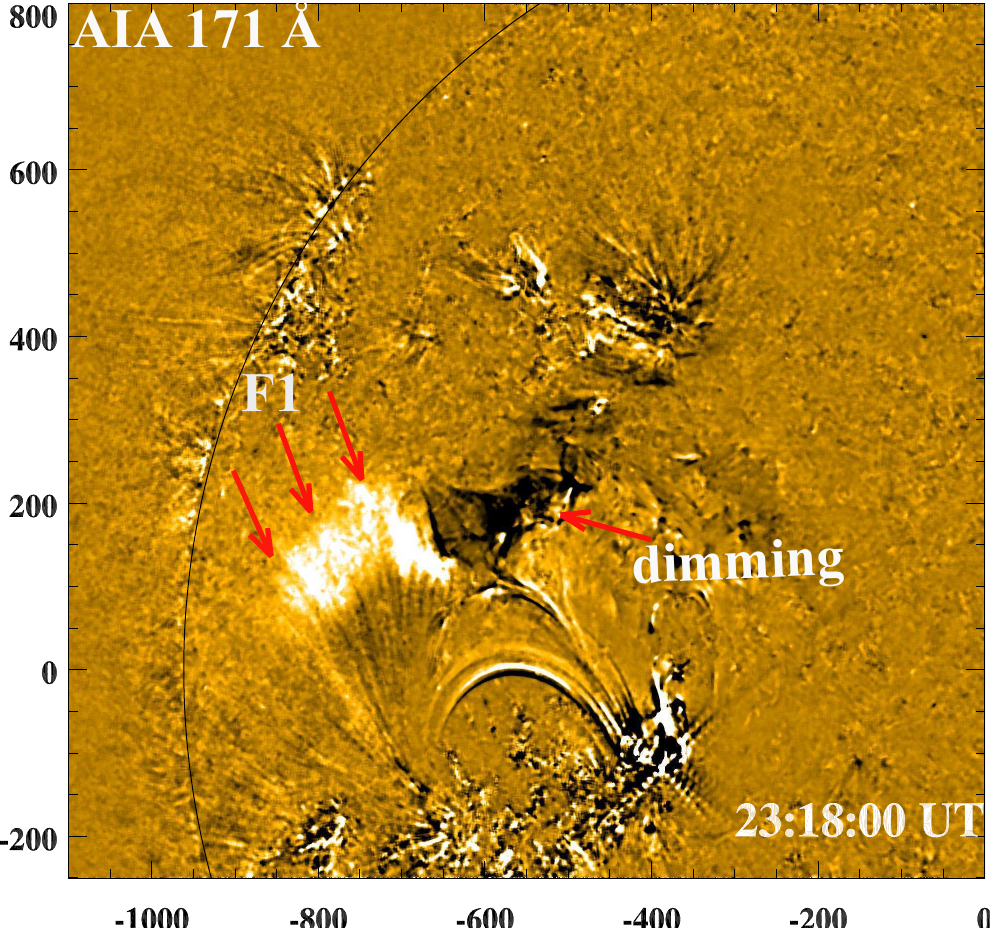}
\includegraphics[width=6cm]{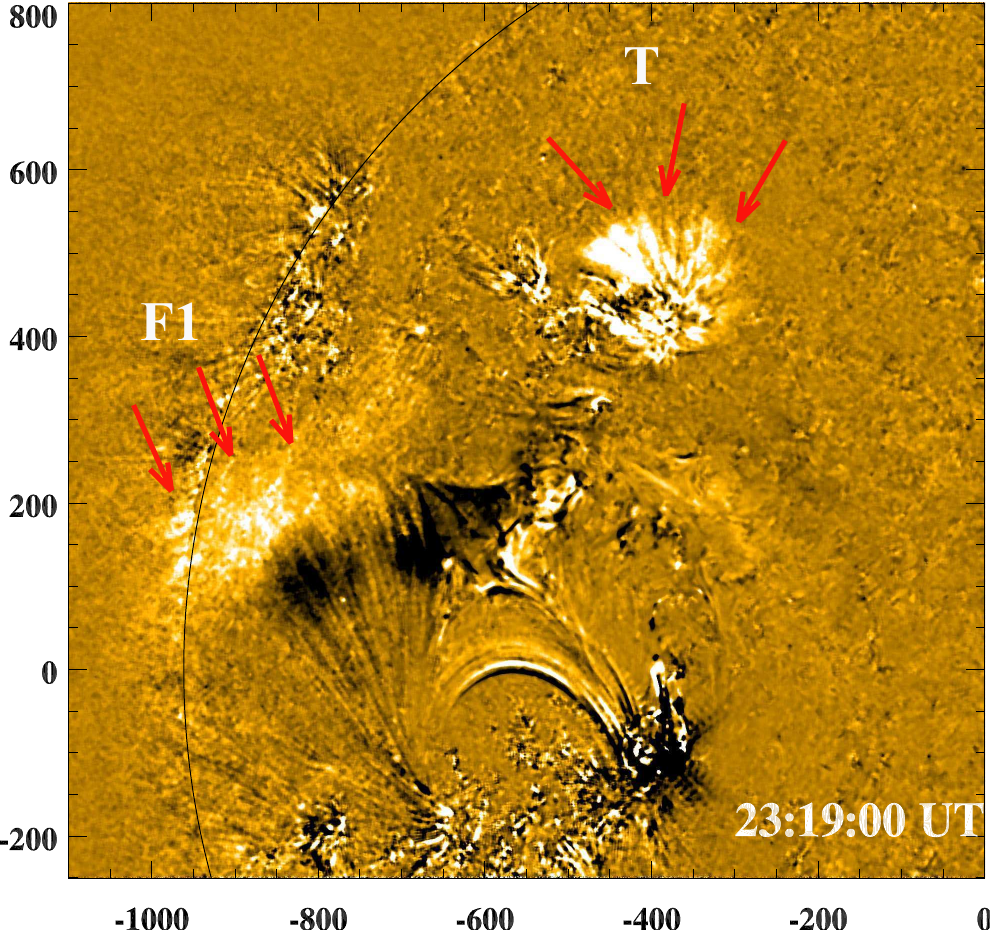}

\includegraphics[width=6cm]{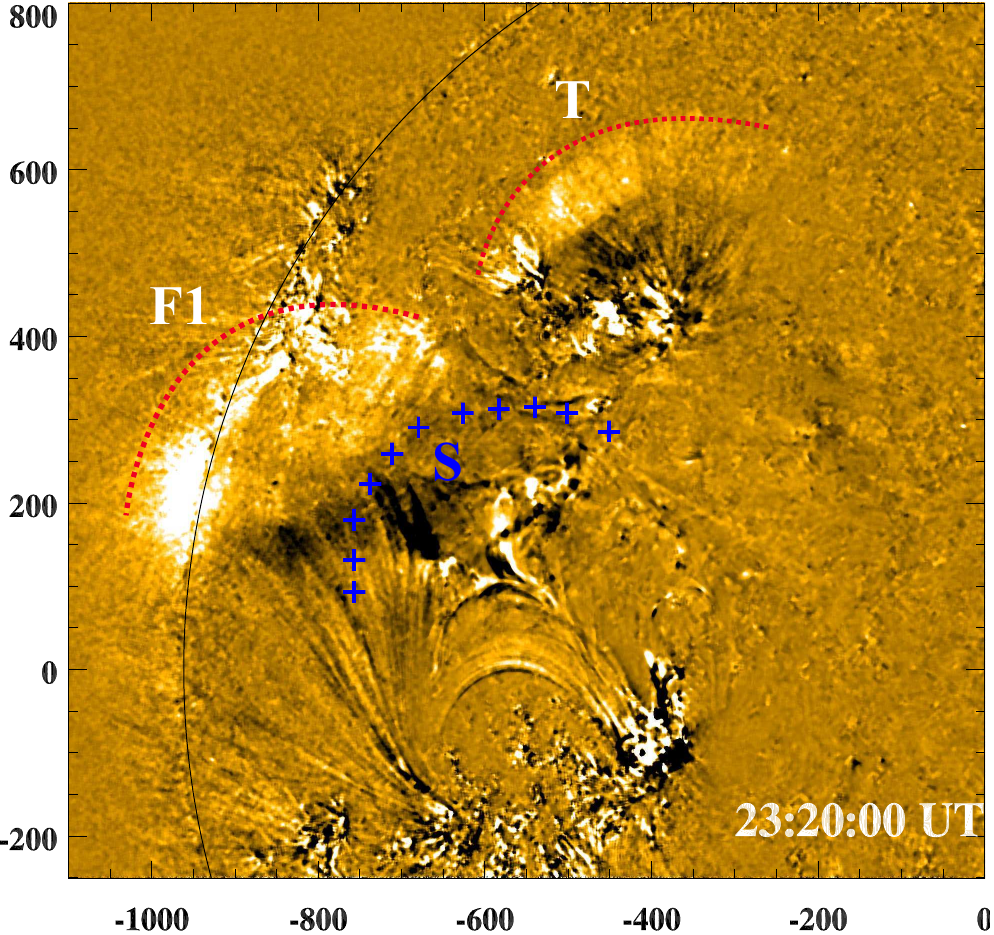}
\includegraphics[width=6cm]{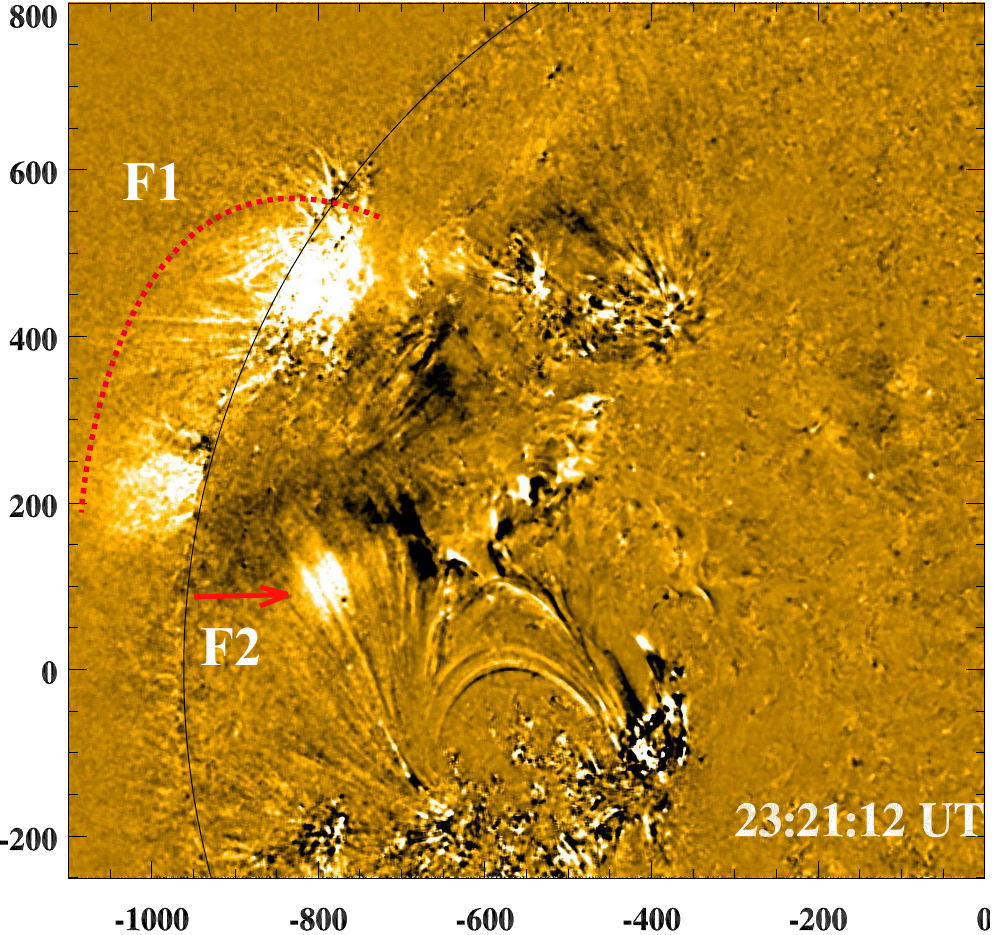}
}
\caption{AIA 335 and 171 \AA \ EUV running-difference images showing the propagation of EUV waves on 25 March 2011. `F' and `S' indicate the faster and slower wavefronts, respectively. The slices selected for the space-time intensity plots are marked by `S1' to `S5' (top-right). `F1' and `F2' also represent the fast wavefronts, while `T' shows the transmitted wavefront from AR NOAA 11177. $X$ and $Y$ axes are given in arcsecs.}
\label{aia171-335}
\end{figure*}

\begin{figure*}
\centering{
\includegraphics[width=6.3cm]{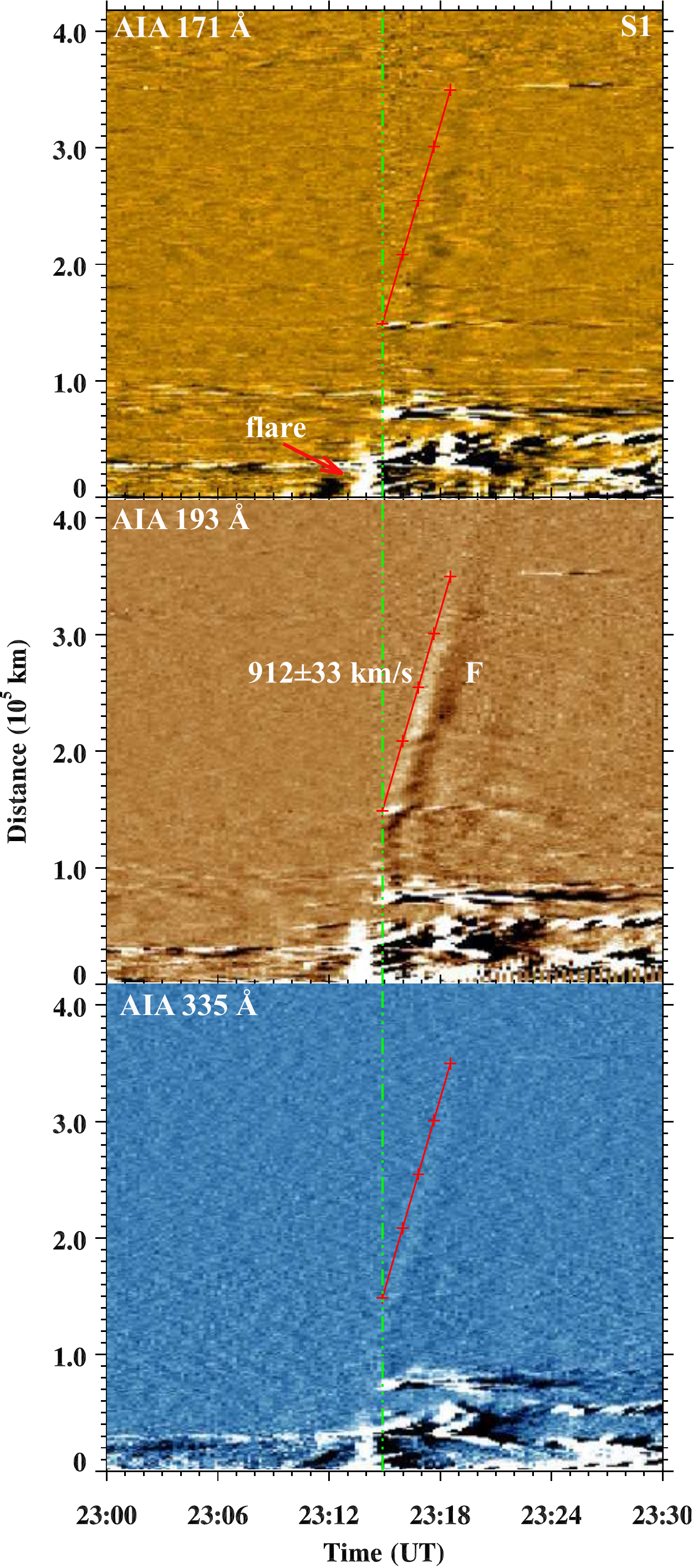}
\includegraphics[width=5.82cm]{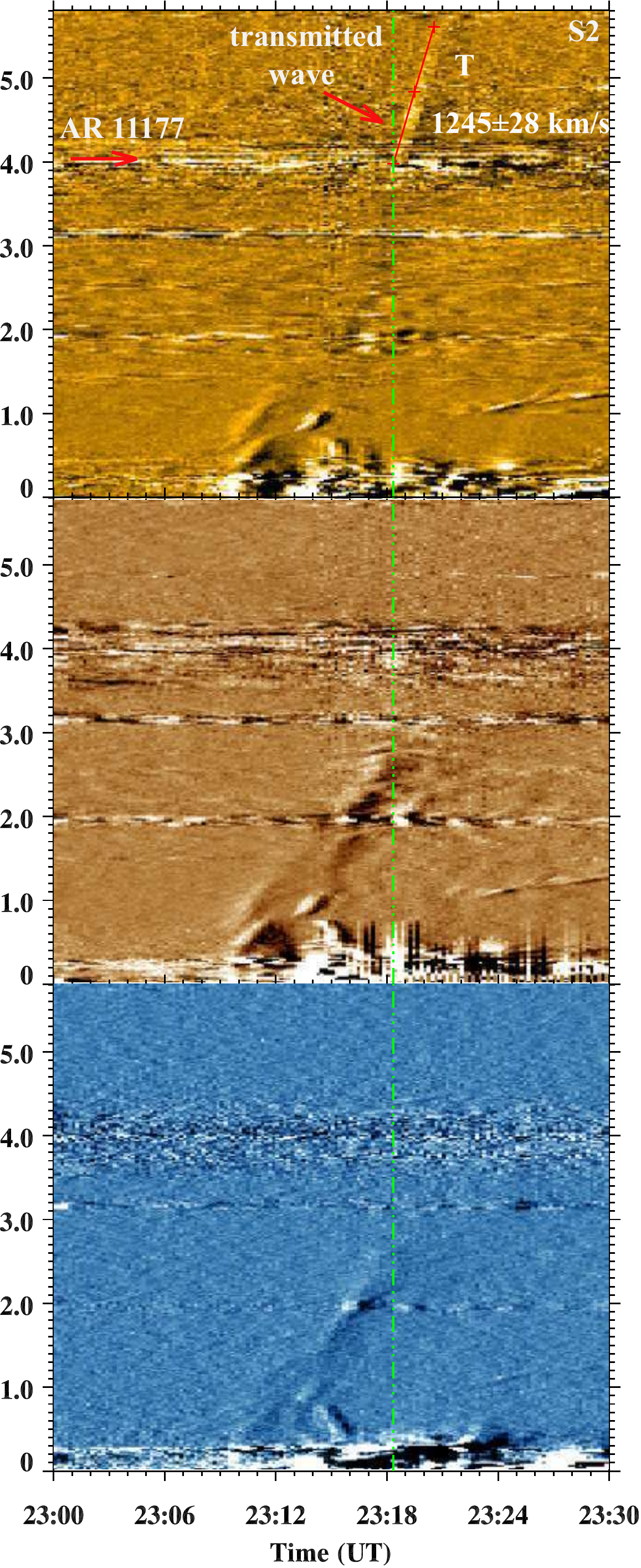}
\includegraphics[width=5.82cm]{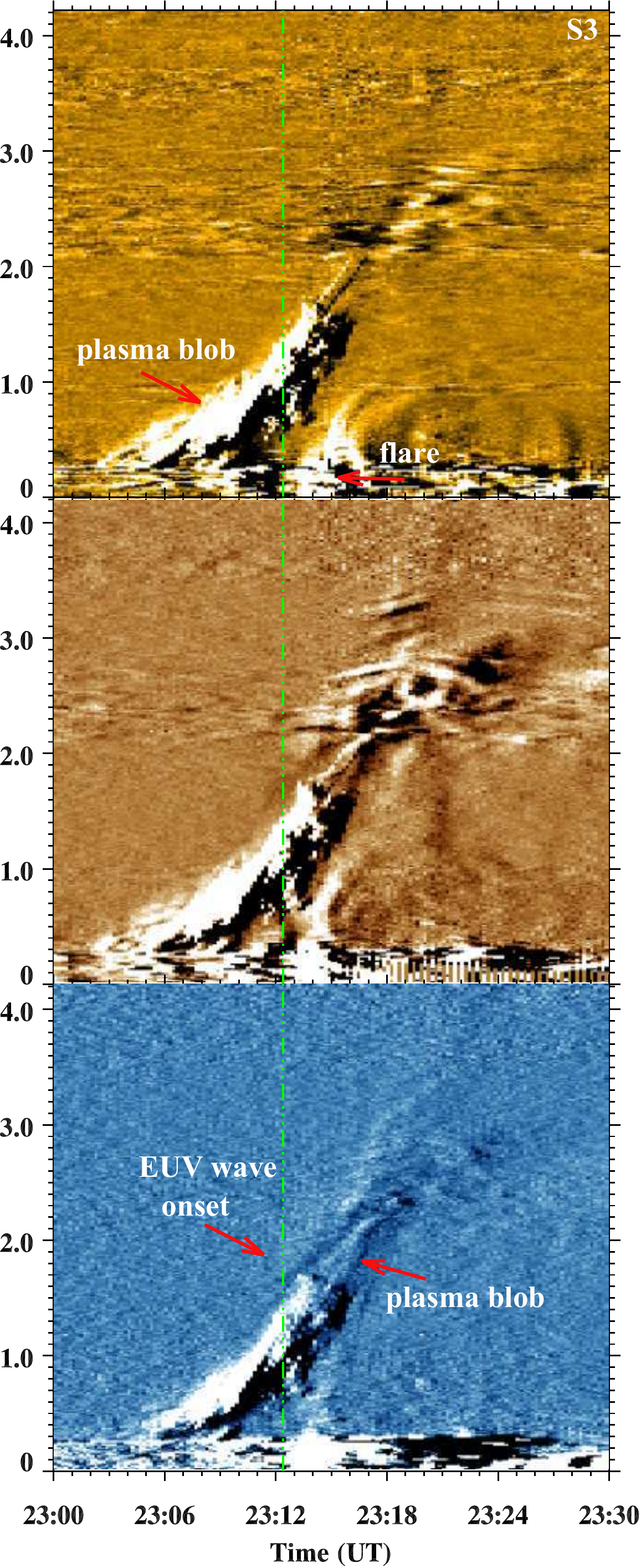}
}
\caption{Temporal evolution of the intensity along slices `S1', `S2', and `S3' for the AIA 171, 193, and 335 \AA~ running-difference images. The faster wave is indicated by `F' while the transmitted wave from AR NOAA 11177 is marked with `T'.}
\label{slice1}
\end{figure*}

\begin{figure*}
\centering{
\includegraphics[width=6.65cm]{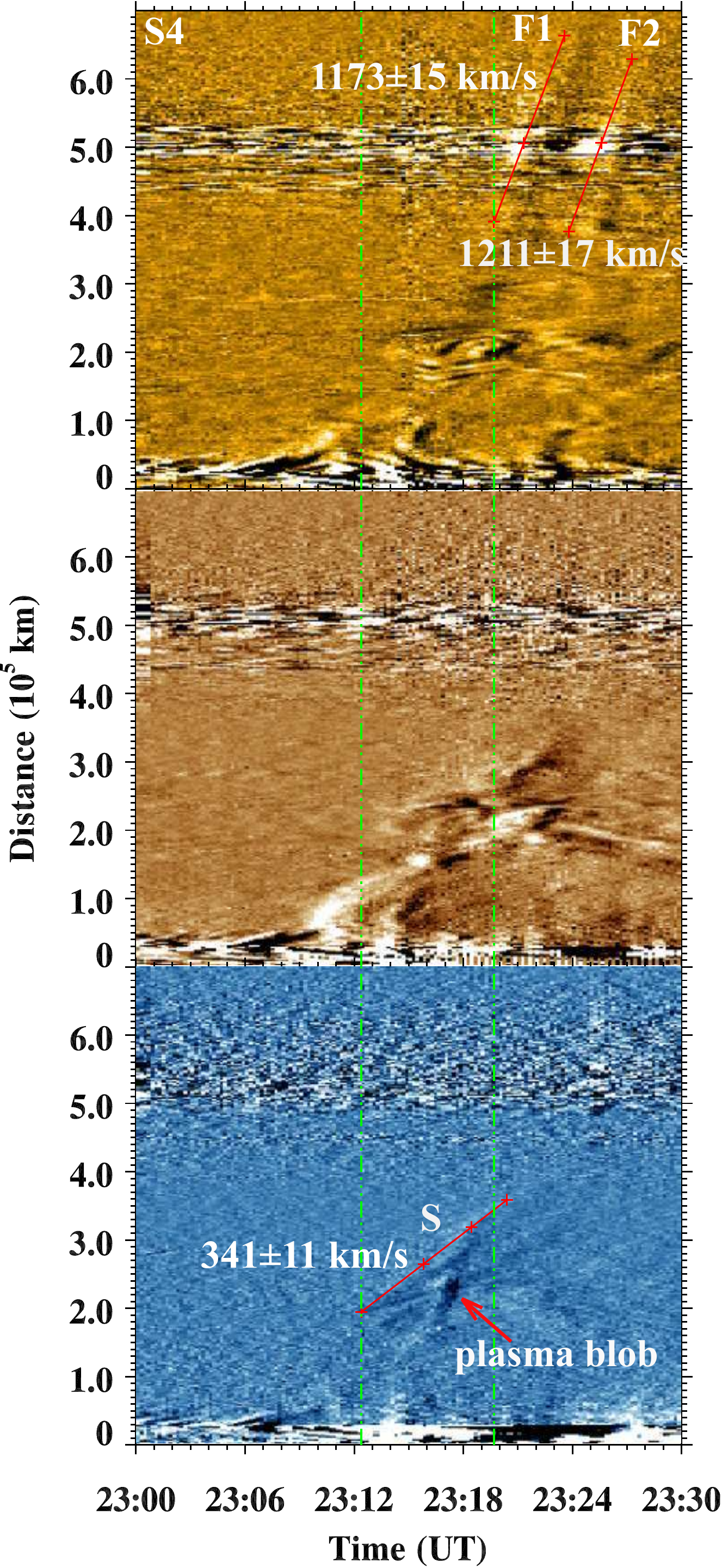}
\includegraphics[width=6.0cm]{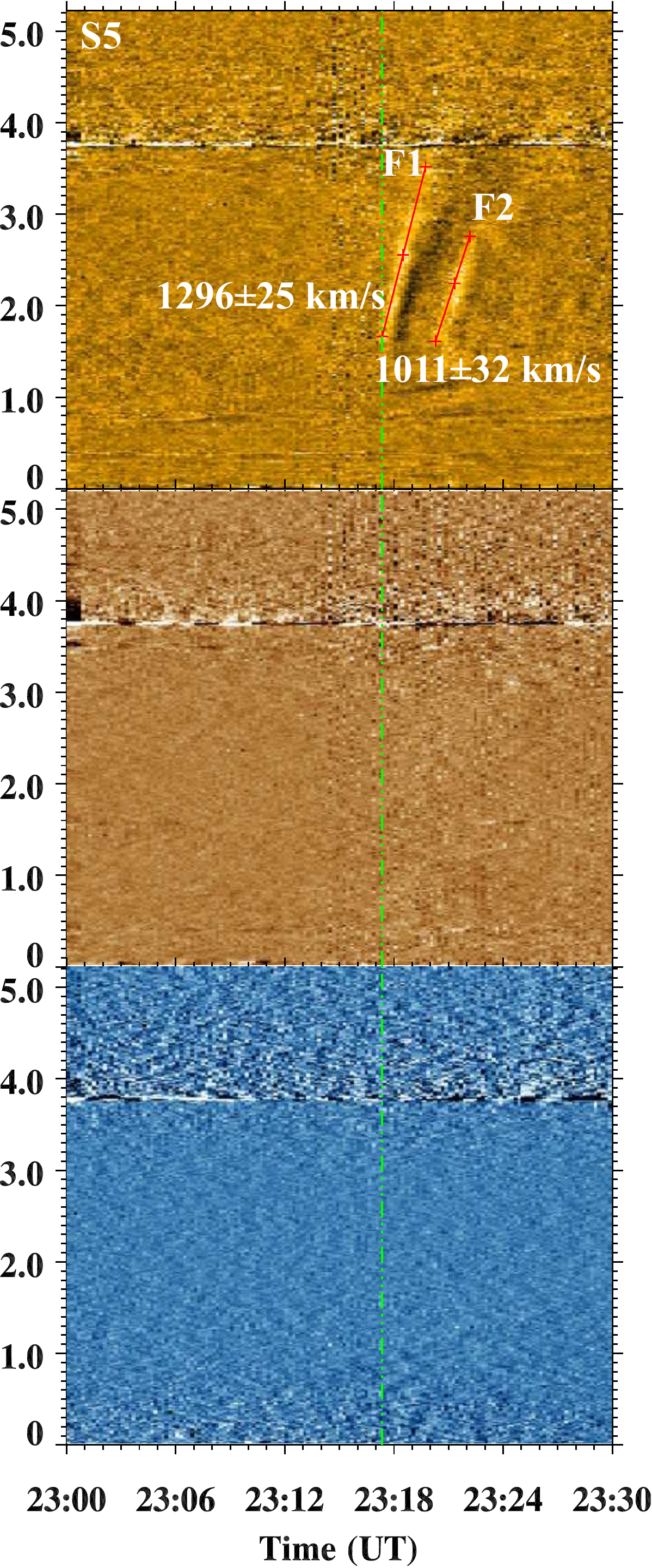}
}
\caption{Temporal evolution of the intensity along slices `S4' and `S5' for the AIA 171, 193, and 335 \AA~ running-difference images. The faster wavefronts are indicated by `F1' and `F2', while the slower wavefront in front of the plasma blob is marked with `S'.}
\label{slice2}
\end{figure*}

\begin{figure*}
\centering{
\includegraphics[width=5cm]{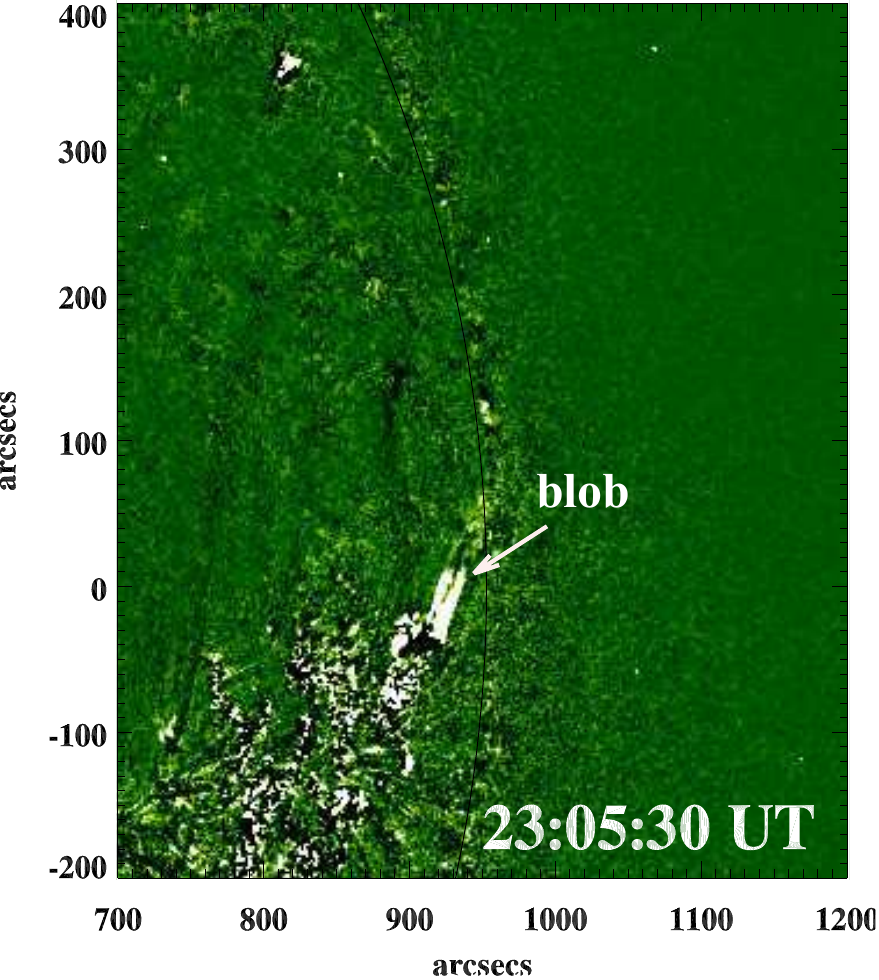}
\includegraphics[width=5cm]{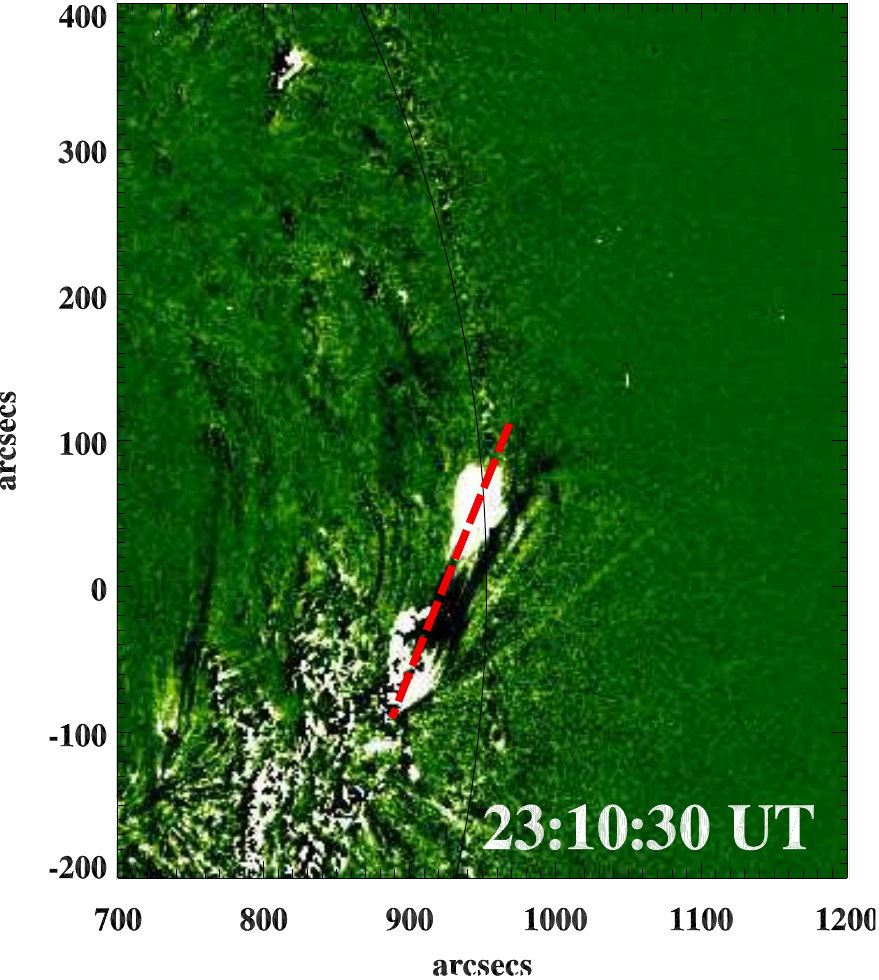}

\includegraphics[width=5cm]{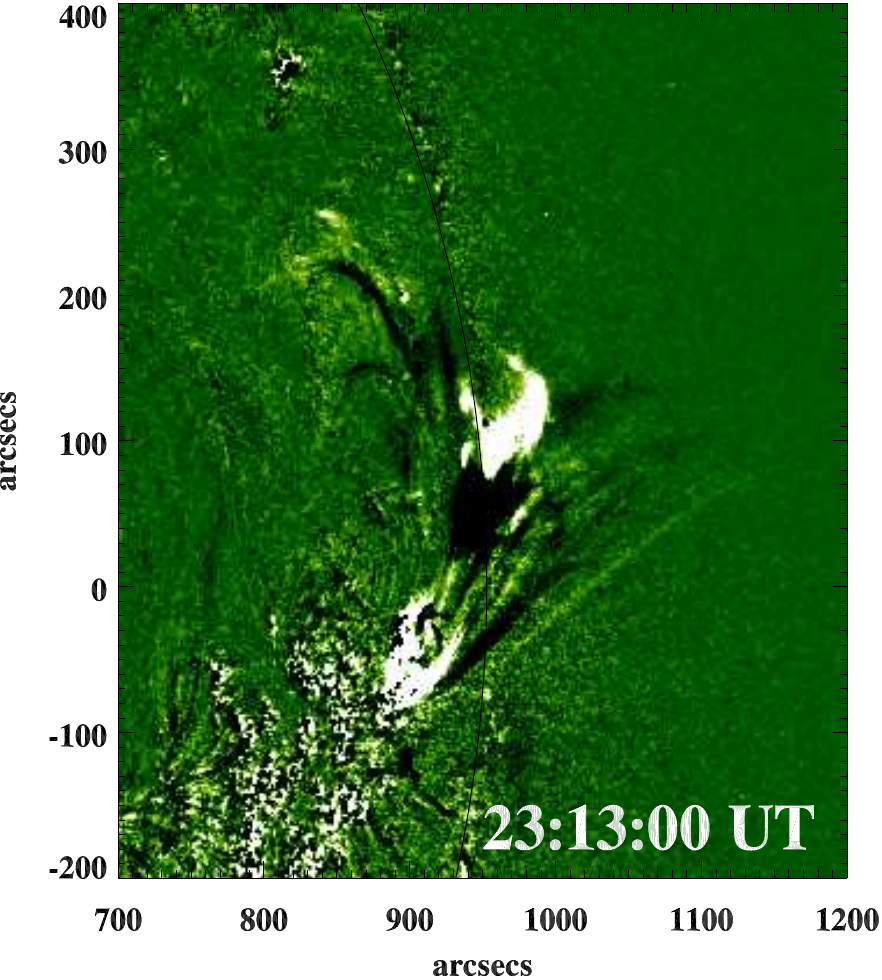}
\includegraphics[width=5cm]{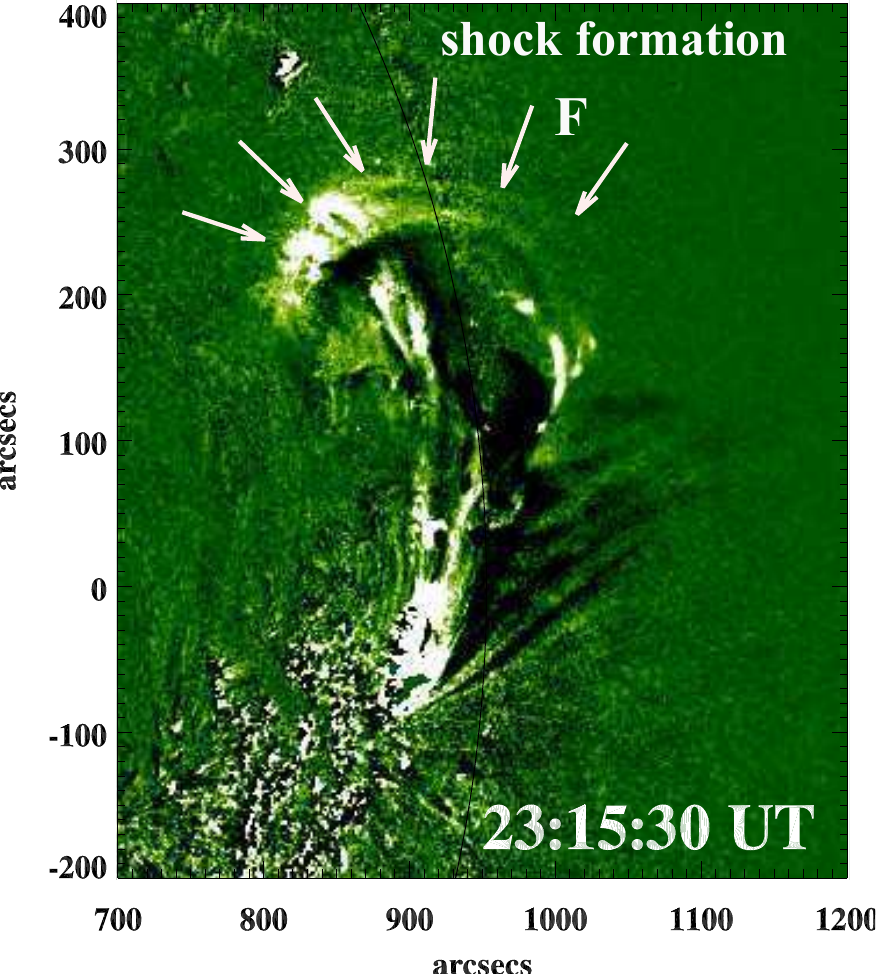}

\includegraphics[width=5cm]{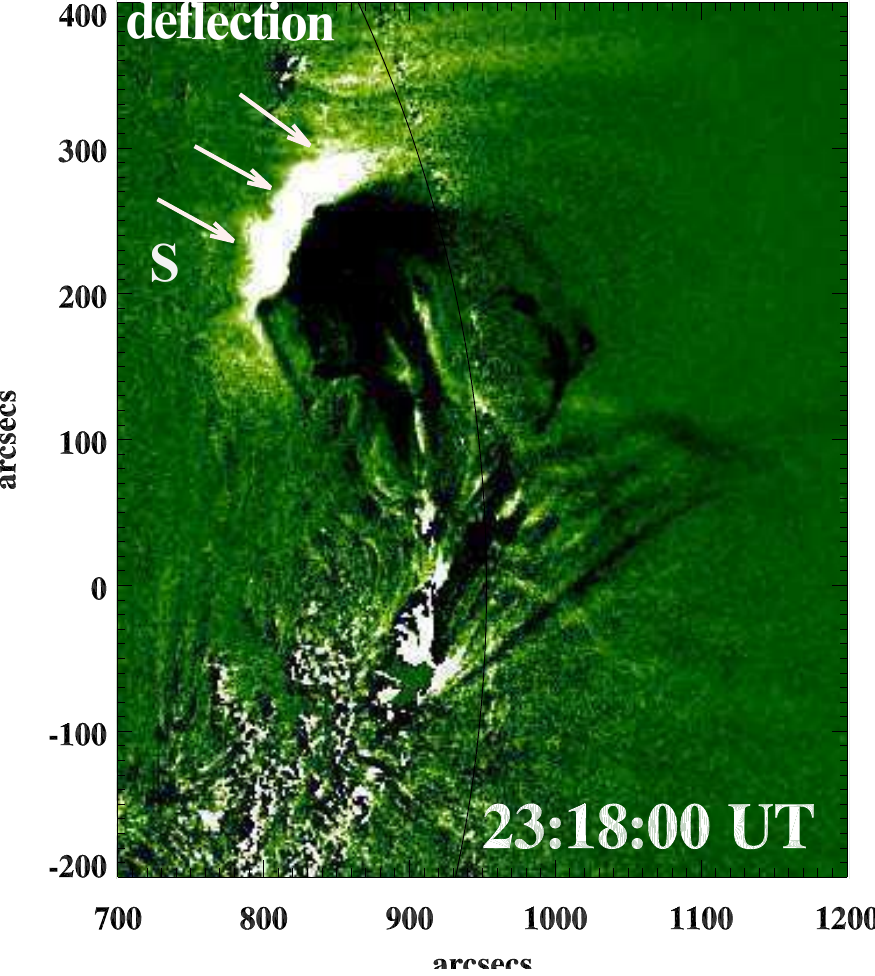}
\includegraphics[width=5cm]{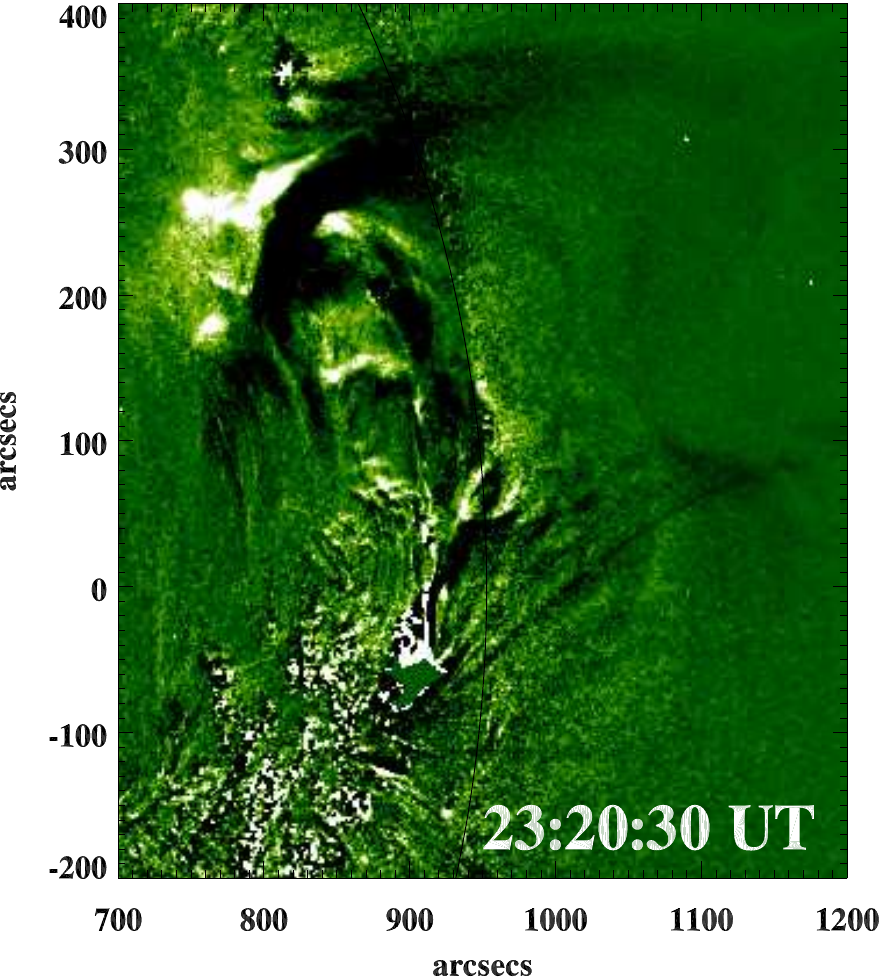}

\includegraphics[width=5cm]{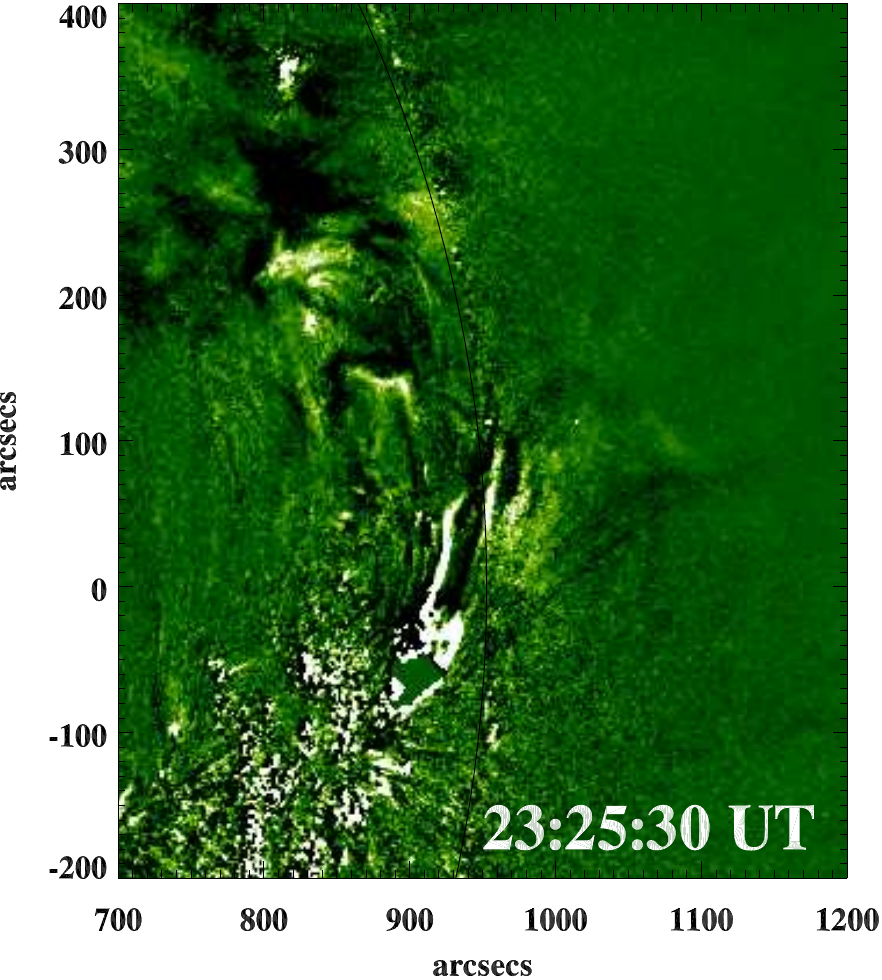}
\includegraphics[width=5cm]{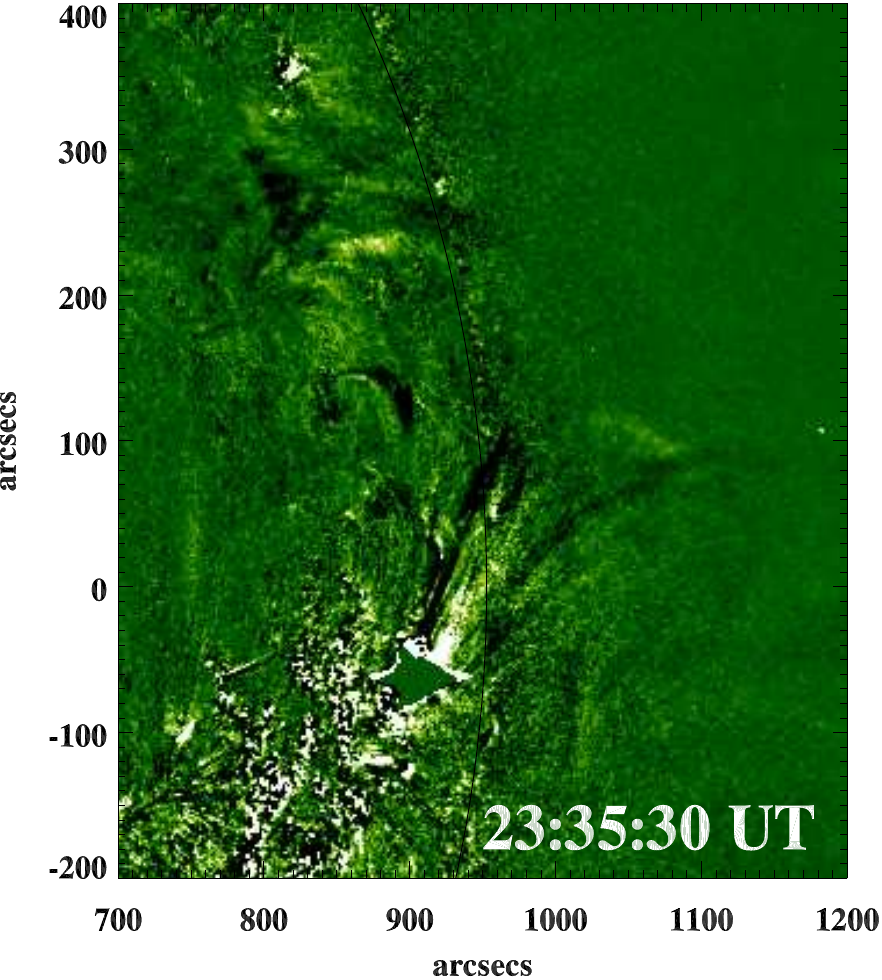}
}
\caption{STEREO-B 195 \AA \ EUV running-difference images showing the eruption of the plasma blob and the formation of EUV wavefronts (indicated by `F' and `S' for the faster and slower front) ahead of it.}
\label{st-b}
\end{figure*}

\begin{figure*}
\centering{
\includegraphics[width=12.0cm]{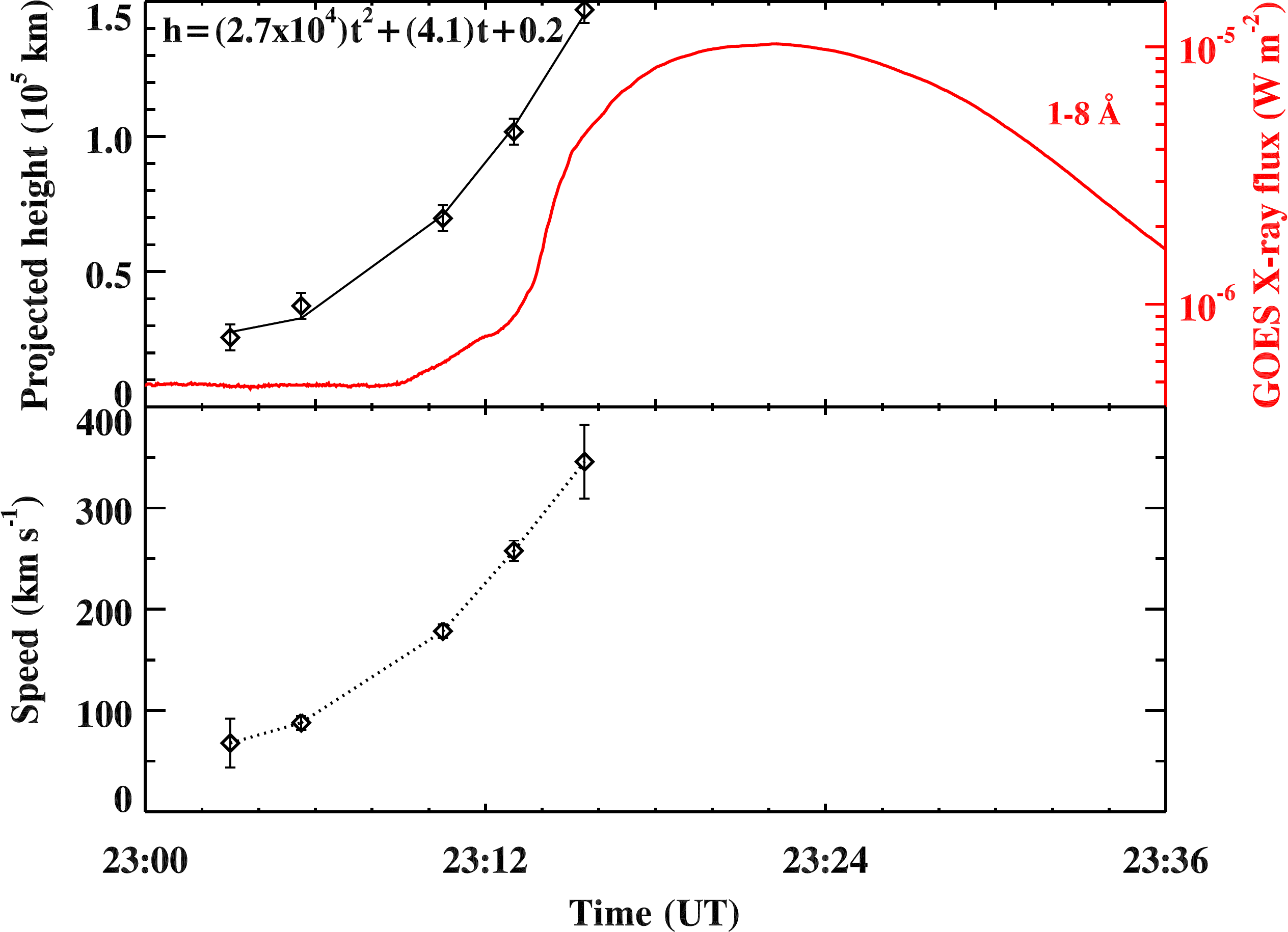}
}
\caption{Top: Blob projected height-time (derived from the STEREO-B 195 \AA \ images) plot with quadratic fit to the measurements. The equation of the fitted curve is h=(2.7$\times$10$^4$)t$^2$+(4.1)t+0.2, where `h' is the height (in km) and `t' is the time (in sec) from 23:03 UT. The red curve shows the GOES soft X-ray flux profile in 1-8 \AA. Bottom: Temporal evolution of the blob speed derived from height-time measurements. The typical error in the height estimate is four pixels (6.4$\arcsec$).}
\label{wave}
\end{figure*}

\begin{figure*}
\centering{
\includegraphics[width=12.0cm]{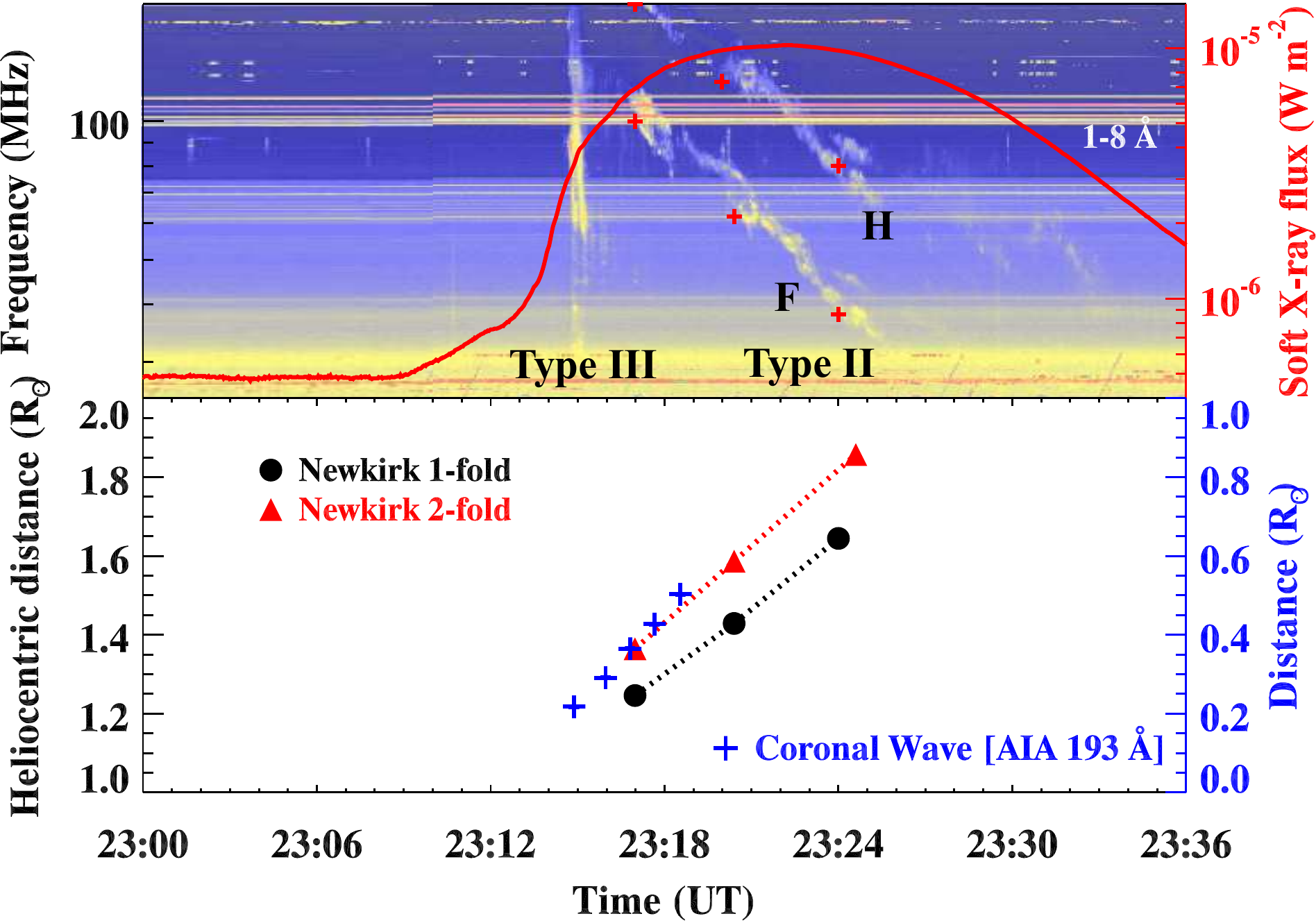}
}
\caption{Top: Radio dynamic spectrum from the Learmonth observatory, Australia in the 25-180 MHz frequency range plotted against the GOES soft X-ray flux in 1-8 \AA. The radio spectrum shows narrow type III and type II radio bursts during the M-class flare. `F' and `H' indicate the fundamental and second-harmonic band emission, respectively. Bottom: Emission heights of type II radio burst (from fundamental band) in the corona estimated from the Newkirk one-fold and two-fold coronal density models. The distance-time profile of the coronal wave is also included.}
\label{spectrum}
\end{figure*}

\begin{figure*}
\centering{
\includegraphics[width=4.5cm]{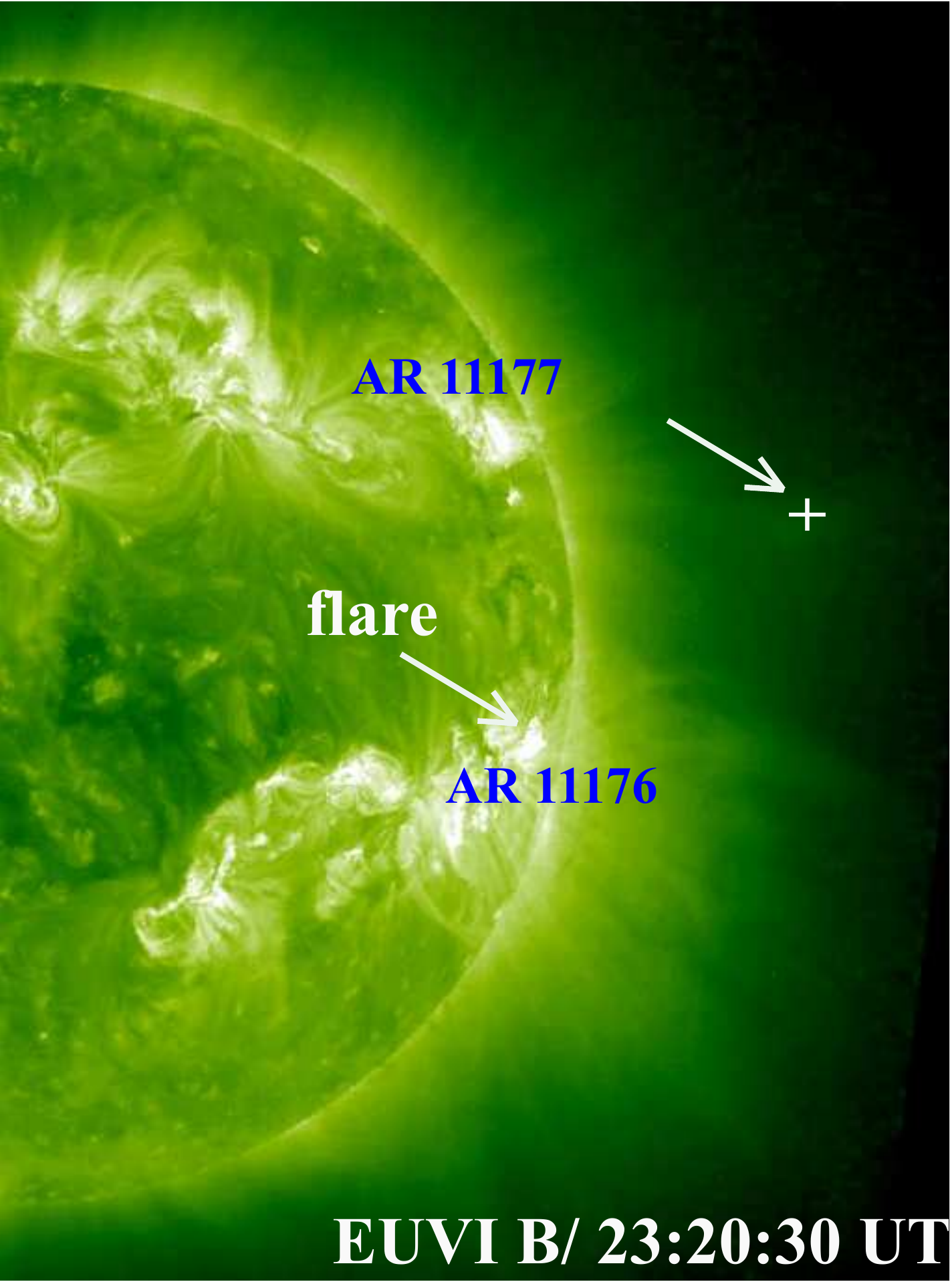}
\includegraphics[width=4.5cm]{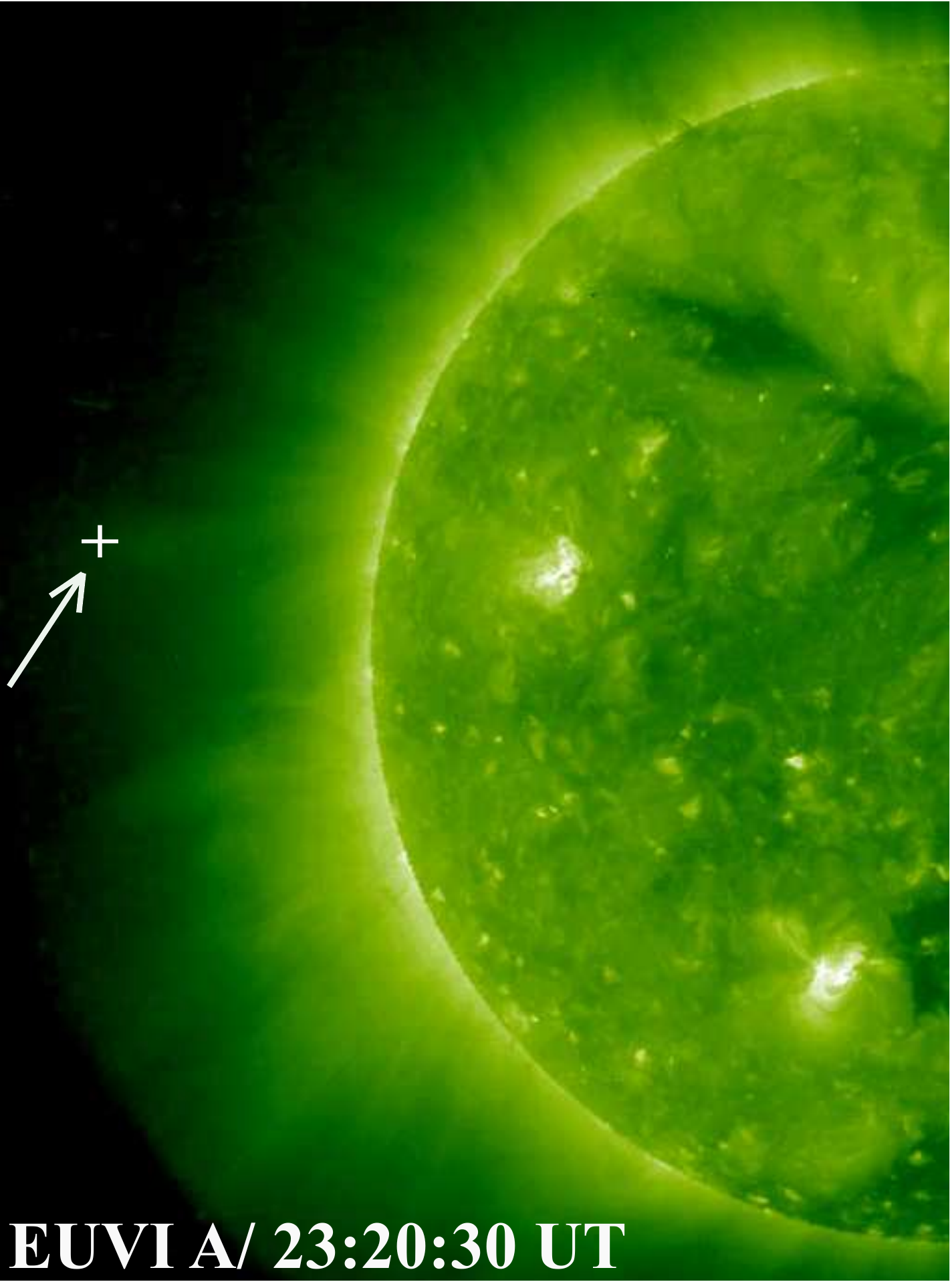}
\includegraphics[width=4.5cm]{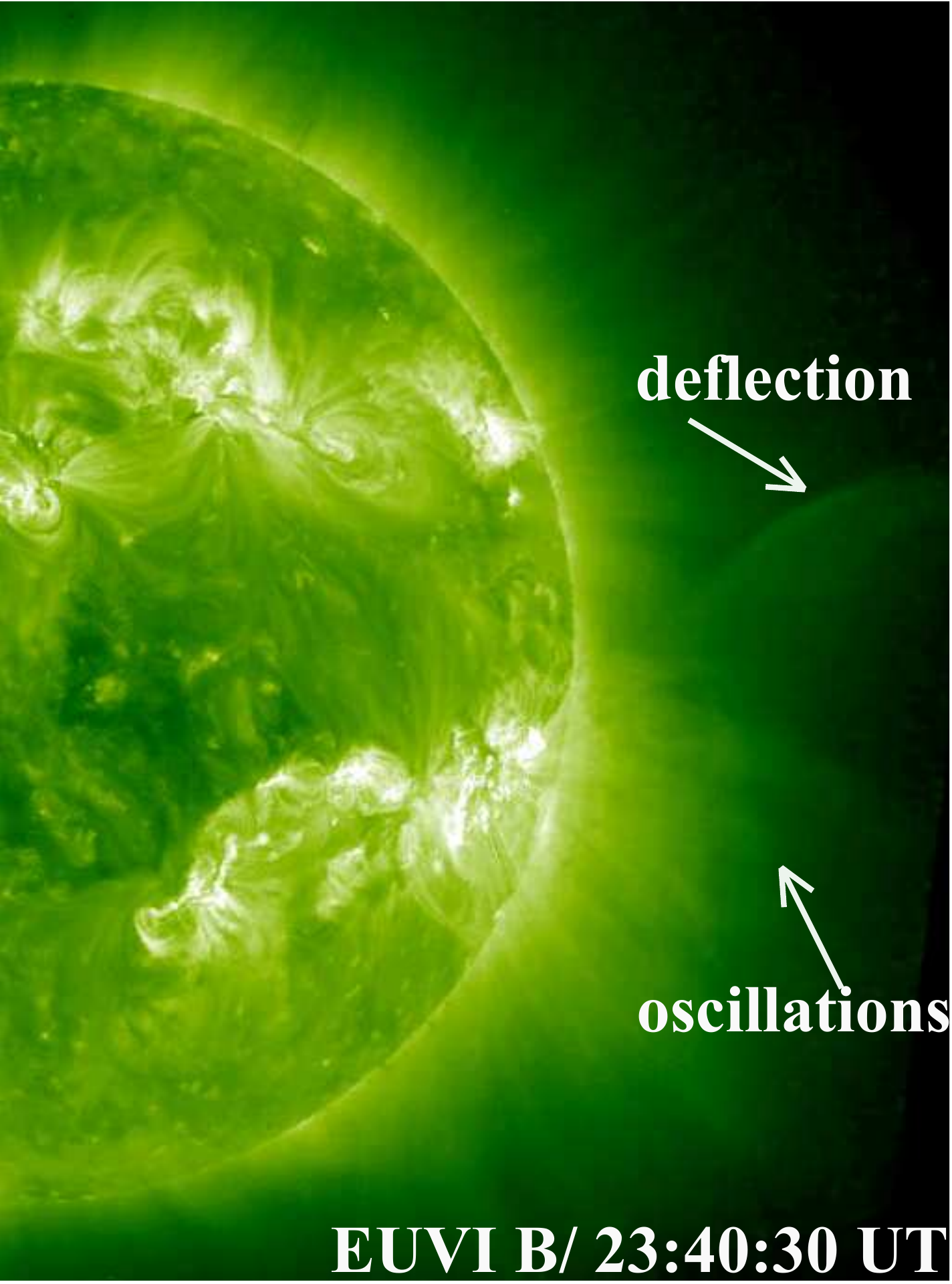}
\includegraphics[width=4.5cm]{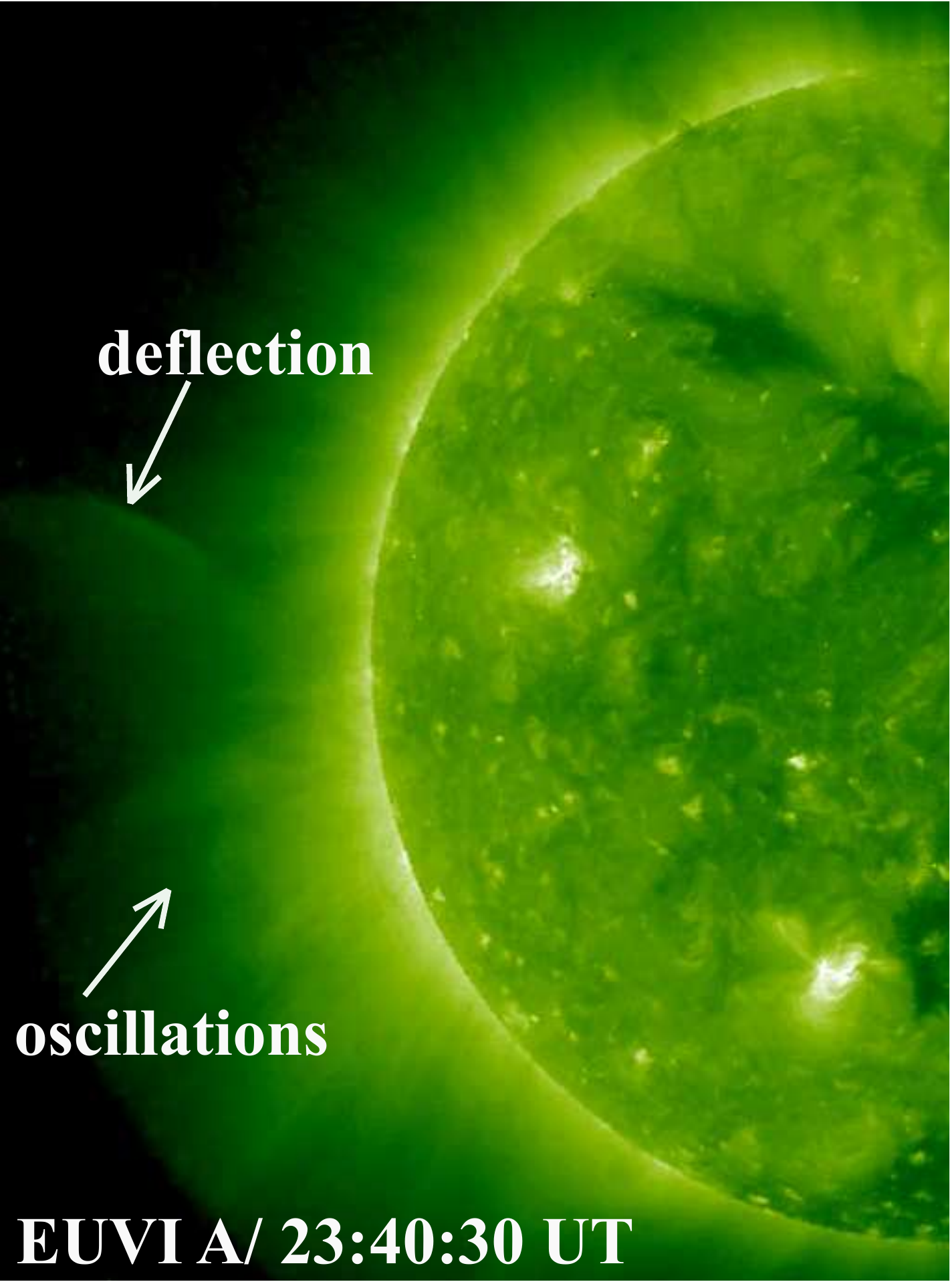}

\includegraphics[width=4.5cm]{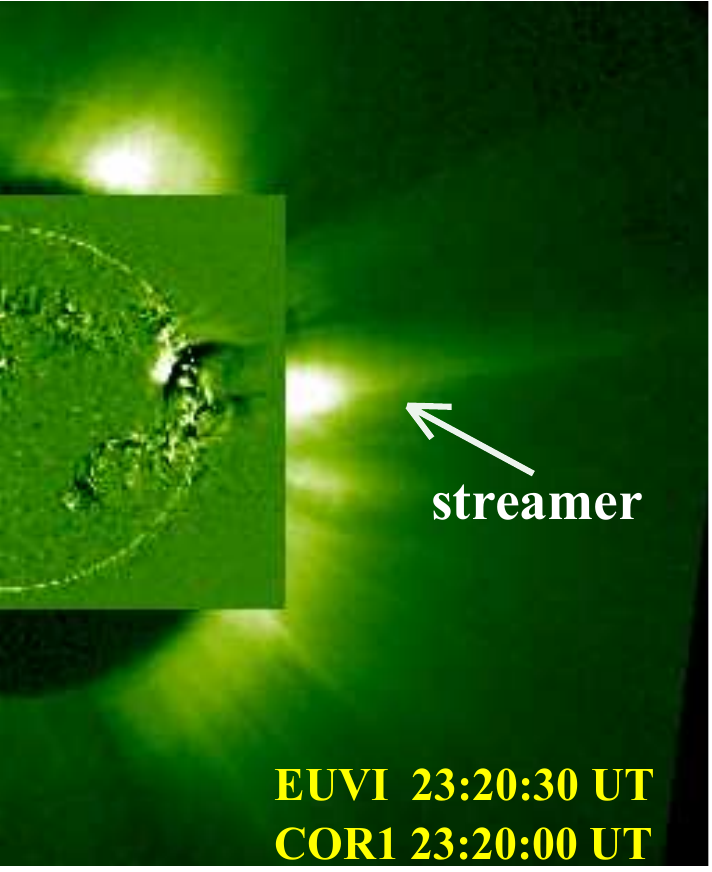}
\includegraphics[width=4.5cm]{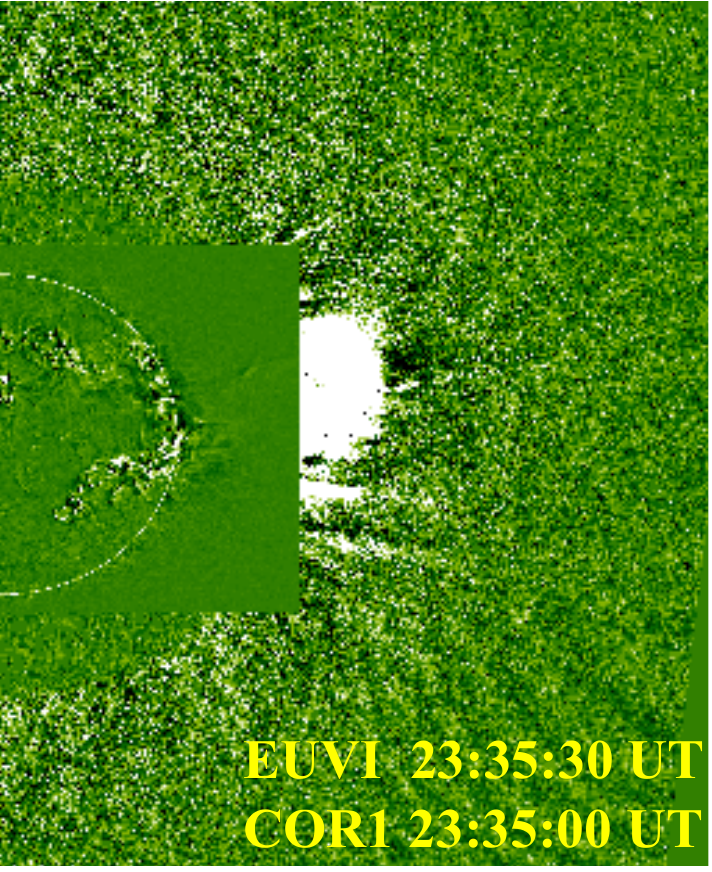}
\includegraphics[width=4.5cm]{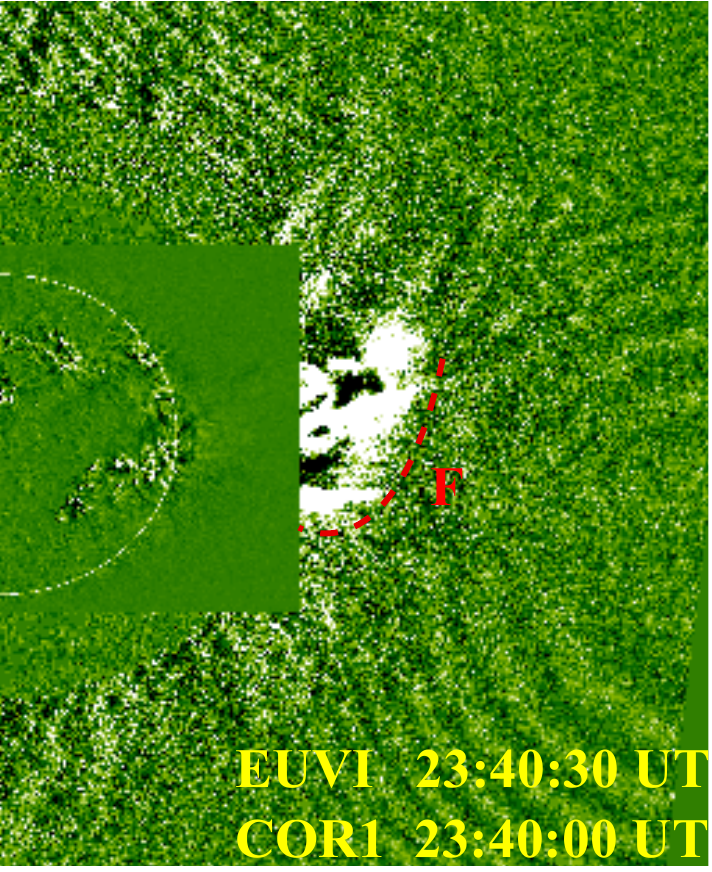}
\includegraphics[width=4.5cm]{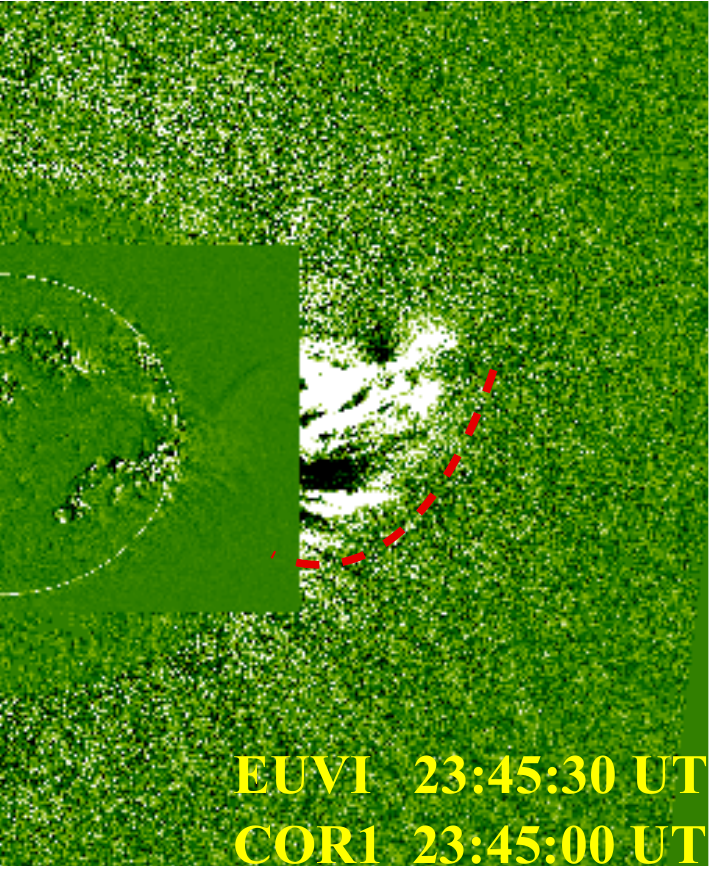}
}
\caption{Top: 195 \AA~ EUV images from STEREO A and B. Bottom: STEREO EUVI and COR1 running-difference composite images. `F' shows the southward propagating front passing through the streamer.}
\label{cor1}
\end{figure*}

Furthermore, we estimated the kinetic energy of the plasma blob from the calculated mass and speed. We assumed a cylindrical shape of the blob with the volume V$_b$ =$\pi$(d/2)$^2$L, where `d' and `L' are the diameter and length of the cylindrical plasma blob. The length and width of the blob are $\sim$70$\arcsec$ and $\sim$30$\arcsec$ (AIA 131 \AA~ image at 23:10:09 UT in Figure \ref{aia131}). The mass of the blob can be estimated by m$_b$ =V$_b$$\times$$\rho$. We adopted an average electron density n$_e$=1.0$\times$10$^9$ cm$^{-3}$, and $\rho$$\sim$n$_e$m$_p$=1.67$\times$10$^{-15}$ g cm$^{-3}$ . Thus, the approximate mass of the blob is m$_b$=3.4$\times$10$^{13}$ g. We took the average and maximum speed of the blob v$_b$$\sim$300 to 540 km s$^{-1}$ from the observation. We estimated the kinetic energy of the blob by K.E.=m$_b$v$_b$$^{2}$/2, which is $\sim$1.5$\times$10$^{28}$ ergs to $\sim$5.0$\times$10$^{28}$ ergs. Using TRACE and AIA observations, the estimated energy of the global fast EUV wave (in other events) is about $\sim$10$^{26}$ ergs and $\sim$10$^{28}$ ergs according to the properties of the kink oscillation of the coronal loop and cavity \citep{nakariakov1999,liu2012}. Therefore, from the estimated energy, it seems that the plasma blob has sufficient energy to trigger the EUV wave. 

It is widely accepted that EUV waves are strongly associated with CMEs \citep{cliver2005}. Without a CME, even M- and X-class flares cannot generate an EUV wave \citep{chen2006}.  Recently, \citet{zheng2012} observed four EUV waves associated with surge activities and suggested that the continuous emergence and cancellation of magnetic flux in the EFR could supply sufficient energy to trigger the surge and associated EUV waves. They suggested that the surges could also generate an initially driven disturbance, which continued to propagate freely after the surge stopped and fell back. Moreover, plasma jets can also generate EUV waves in a similar way as the CMEs drive \citep{vrsnak2008,zhukov2011}. EUV waves associated with the minifilament/jet eruptions have recently  been reported \citep{zheng2011,zheng2012a}, which were interpreted as fast-mode MHD waves. In \citet{zheng2011}, the EUV wave propagates at a uniform velocity of 220-250 km s$^{-1}$ with very little angular dependence. \citet{zheng2012a} observed the speed of the EUV wave from 1100 to 550 km s$^{-1}$ and discussed that the EUV was presumably driven by the rapid expansion of the CME loops, observed in  STEREO/COR1. However, they did not observe  multiple EUV waves in different EUV channels. Our observations clearly show the flux emergence and formation of the EUV waves in front of the plasma blob. However, the peak time of the flare is very close to the wave initiation, but the initial location of the wave was far away from the associated flare site, i.e., in front of the plasma blob. The impulsive acceleration of the blob may be sufficient to trigger the EUV wave. 

From our observation, the acceleration phase duration of the blob is about 12 minutes and the shock formation distance is $\sim$0.23 R$_\odot$ (from the flare center). However, the shock formation time is comparable with the time predicted in the numerical simulation of the piston-driven shock (for 500 km s$^{-1}$ Alfv\'en speed, 100 Mm piston size) by \citet{zic2008}, whereas the shock formation distance is shorter than the predicted value. On the other hand, it is interesting to note that the small-scale plasma blob ejected in a collimated (slightly curved) path can cause a large-scale global EUV wave propagating in a wide angular range across the solar surface. For a piston-driven shock, the shock is strongest and fastest in the forward direction of the piston and has a narrow angular extent around it. In this event, the highest wave speed of $\sim$912 km s$^{-1}$ is in the northwest direction seen in AIA 193 \AA~ channel, while the plasma blob is ejected toward the northeast, where the slower wave ($\sim$341 km s$^{-1}$) is observed. This seems inconsistent with the piston-driven shock scenario. It is more likely if a coronal volume greater than the observed plasma blob is acting as the actual piston. This expanding volume was not clearly detected in the available AIA channels and the observed plasma blob may be only a small portion of it. It is also possible that the plasma blob ejection is the initial trigger that destabilizes the surrounding coronal volume and leads to its eruption as the weak CME, which may be the actual driver of the EUV/shock wave. Indeed, COR1 observations show a weak CME structure associated with the blob eruption. The CME disappeared in the low corona (i.e., in COR1 field of view) and could not propagate further into the SOHO/LASCO and COR2 field of view.

In conclusion, we presented a multiwavelength study of the ejection of a plasma blob associated with an M-class 
flare and the observations of EUV waves (i.e., fast and slow waves, transmitted and secondary waves). These observations revealed the multi-thermal nature of EUV wavefronts, which are observed in different EUV channels. The nature of the EUV waves is not well explored and other explanations are possible. However, more observations and analyses of these eruptive events 
 using high-resolution data from SDO, STEREO, and Hinode can provide a better understanding of the energy build-up/release processes and the nature/driver of EUV waves.
%%%%%%%%%%%%%%%%%%%%%%%%%%%%%%%%%%%%%%%%%%%%%%%%%%%%%%%%%%%%%%%%%%%%%%%%%%%
%%%%%%%%%%%%%%%%%%%%%%%%%%%%%%%%%%%%%%%%%%%%%%%%%%%%%%%%%%%%%%%%
\begin{acknowledgements}
We express our gratitude to the referee for his/her valuable suggestions and help in the proper interpretation of the observations, which improved the manuscript considerably.
SDO is a mission for NASA's Living With a Star (LWS) program. RHESSI is a NASA Small Explorer. We acknowledge the STEREO mission, Fermi GBM, and the Learmonth dynamic radio spectrum data used in this study. PK thanks  P.F. Chen for several helpful discussions. This work is partially supported by the CAWSES-India Program, which is sponsored by ISRO. 
\end{acknowledgements}
%%%%%%%%%%%%%%%%%%%%%%%%%%%%%%%%%%%%%%%%%%%%%%%%%%%%%%%%
\bibliographystyle{aa}
\bibliography{reference}

%%%%%%%%%%%%%%%%%%%%%%%%%%%%%%%%%%%%%%%%%%%%%%%%%%%%%%%%%%%%%%%%%%%%%
%%%%%%%%%%%%%%%%%%%%%fig-1%%%%%%%%%%%%%%%%%%%%%%%%%%%%%%%%%%%
%------------------------------------------------------------------------------------ 

%%%%%%%%%%%%%%%%%%%%%%%%%%fig-2%%%%%%%%%%%%%%%%%%%%%%%%%%%%%%%%%%%%%%%%%%%%%%%%%%%%%%%%%%%%%%%%%
%------------------------------------------------------------------------------------ 
\clearpage 
%%%%%%%%%%%%%%%%%%%%%%%%%%%%%%%%%%%%%%%%%%%%%%%%%%%%%%%%%%%%%%%%%%%%%%%%%%%%%%%%%%%%%%%%%%%
%-----------------------fig3------------------------------------------------------------- 
%%%%%%%%%%%%%%%%%%%%%%%%%%%%%%  
\clearpage 
%%%%%%%%%%%%%%%%%%%%%%%%%%%%%%%fig4%%%%%%%%%%%%%%%%%%%%%%%%%%%%%%%%%%%%%%%%%%%%%%%%%%%%%%%% 
   %------------------------------------------------------------------------------------ 
\clearpage 
%%%%%%%%%%%%%%%%%%%%%%%%%%%%%%%%%%%%%%%%%%%%%%%%%%%%%%%%%%%%%%%%%%%%%%%%%%%%%%%%%%%%%%%%%%% 
%%%%%%%%%%%%%%%%%%%%%%figure-5%%%%%%%%
%------------------------------------------------------------------------------------ 
%%%%%%%%%%%%%%%%%%%%%%%%%%%%%%
%-----------------------------------fig6------------------------------------------------- 

\clearpage 
%%%%%%%%%%%%%%%%%%%%%%%%%%%%%%%%%%%%%%%%%%%%%%%%%%%%%%%%%%%%
%---------------------------------fig7--------------------------------------------------- 

%%%%%%%%%%%%%%%%%%%%%%%%%%%%%%%%%%%%%%%%%%%%%%%%%%%%%%%%%%%%
%%%%%%%%%%%%%%%%%%%%%%%%%%%%%%fig8%%%%%%%
\clearpage 
%%%%%%%%%%%%%%%%%%%%%%%%%%%%%%%%%%%%%%%%%
%%%%%%%%%%%%%%%%%%%%%%%%%%%%%%fig9%%%%%%%%
\clearpage 
%------------------------------------------------------------------------------------
%++++++++++++++++++++++++++++++++++++++++++++++++++fig10

%%%%%%%%%%%%%%%%%%%%%%%%%%%%%%%%%%%%%%%%%%%%%%%%%%%%%%%%%%%%
\clearpage 
%------------------------------------------------------------------------------------fig11 
%%%%%%%%%%%%%%%%%%%%%%%%%%%%%%

%------------------------------------------------------------------------------------fig12 
%%%%%%%%%%%%%%%%%%%%%%%%%%%%%%
\clearpage 
%------------------------------------------------------------------------------------fig13 

%%%%%%%%%%%%%%%%%%%%%%%%%%%%%%%%%%%%%%%%%%%%%%%%%%%%%%%%

%%%%%%%%%%%%%%%%%%%%%%%%%%%%%%%%%%%%%%%%%%%%%%%%%%%%%%%%%
\end{document}